\documentclass{ws-rv975x65}
\usepackage{subfigure}     
\usepackage{ws-rv-van}     
\makeindex
\begin{document}

\chapter[Jet quenching in high-energy heavy-ion collisions]{Jet quenching in high-energy heavy-ion collisions}
\label{ch_jet_quenching}

\author[Guang-You Qin and Xin-Nian Wang]{Guang-you Qin$^1$ and Xin-Nian Wang$^{1,2}$}

\address{$^1$Institute of Particle Physics and Key Laboratory of Quark and Lepton Physics (MOE), Central China Normal University, Wuhan, 430079, China\\
$^2$Nuclear Science Division, MS 70R0319, Lawrence Berkeley National Laboratory, Berkeley, CA 94720
}


\begin{abstract}
Jet quenching in high-energy heavy-ion collisions can be used to probe properties of hot and dense quark-gluon plasma. We provide
a brief introduction to the concept and framework for the study of jet quenching. Different approaches and implementation of multiple
scattering and parton energy loss are discussed. Recent progresses in the theoretical and phenomenological studies of jet quenching
in heavy-ion collisions at RHIC and LHC are reviewed.

\end{abstract}

\body

\section{Introduction}

One of the main goals of ultra-relativistic nucleus-nucleus collisions, such as those performed at the Relativistic Heavy-Ion Collider (RHIC) at Brookhaven National Laboratory (BNL) and the Large Hadron Collider (LHC) at European Organization for Nuclear Research (CERN), is to create the hot and dense quark-gluon plasma (QGP) and study its various novel properties.
QCD jets produced from early stage collisions of beam quarks and gluons from two nuclei play an essential role in studying transport properties of the QGP produced in these energetic collisions.
During their propagation through the hot and dense medium, the interaction between hard jets and the colored medium will lead to parton energy loss (jet quenching) \cite{Bjorken:1982tu,Gyulassy:1990ye,Wang:1991xy}.
There have been many striking experimental signatures of jet energy loss observed at RHIC and the LHC, such as the suppression of inclusive hadron spectra at high transverse momentum ($p_T$) \cite{Adcox:2001jp,Adler:2002xw,Aamodt:2010jd}, the modification of back-to-back hadron-hadron and direct photon-hadron correlations \cite{Adler:2002tq, Adams:2006yt, Adare:2009vd, Abelev:2009gu}, and the modification of reconstructed jets \cite{Aad:2010bu,Chatrchyan:2011sx,Chatrchyan:2012gt,Aad:2014bxa} and jet substructure \cite{Chatrchyan:2012gw, Chatrchyan:2013kwa, Aad:2014wha}, as compared to the expectations from elementary proton-proton collisions.

During the last decades, extensive efforts have been devoted to the study of jet quenching phenomena in ultra-relativistic heavy-ion collisions.
The theoretical study of jet quenching in a dynamically evolving QGP in heavy-ion collisions involves a variety of components, such as the initial jet production spectrum, a realistic description of the density profile of the medium, the propagation of hard jets in the medium, and the hadronization (fragmentation) of jet partons. Production rates for hard jets may be calculated within the framework of perturbative QCD due to the hard scales involved in the jet production.
The medium density profiles are usually obtained from relativistic hydrodynamic calculations which have been very successful in describing the dynamical evolution of the bulk matter and especially in explaining the observed anisotropic collective flows \cite{Heinz:2013th, Gale:2013da, Song:2013gia,Romatschke:2009im,Huovinen:2013wma}. The central topic of jet quenching in heavy-ion collisions is concentrated on the detailed mechanisms for the interaction of energetic partons with the hot and dense nuclear medium, and the manifestation of medium modification of jets in the final state observables.

\section{Jet quenching: general framework}
\label{sec_general_framework}

\begin{figure}
\begin{center}
\includegraphics*[width=10.0cm]{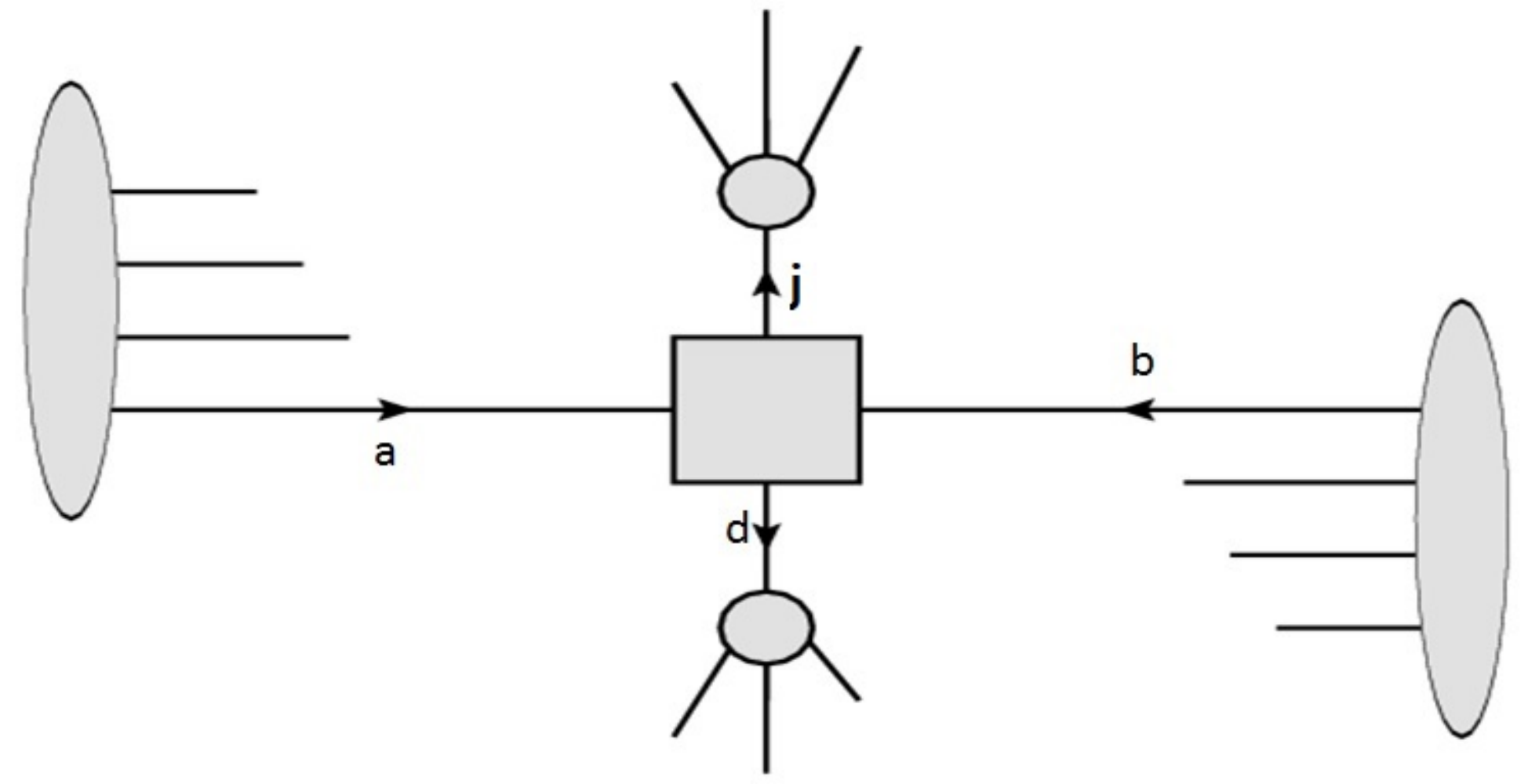}
\caption{(Color online) Schematic illustration of high $p_T$ hadron production in high-energy nuclear collisions which may be factorized into  parton distribution functions, hard partonic scattering cross section and fragmentation functions.
}
\label{fig_jet_production}
\end{center}
\end{figure}

Due to large scales involved, the production of jets and the interaction of jets with dense nuclear matter may be studied within the framework of perturbative QCD.
In perturbative QCD, processes involving large momentum transfer, such as the production of high transverse momentum hadrons, can be described with controlled uncertainties as the convolution of the incoming parton distribution functions (PDFs), hard partonic scattering process, and the final state fragmentation function (FFs), as illustrated Fig. \ref{fig_jet_production}.
The cross section of inclusive hadron production at high $p_T$ can be obtained as follows:
\begin{eqnarray}
d\sigma_{pp \to hX} &&\!\! \approx \sum_{abjd} \int dx_a \int dx_b \int dz_j f_{a/p}(x_a, \mu_f) \otimes f_{b/p}(x_b, \mu_f)
\nonumber\\ &&\!\! \otimes \ d\sigma_{ab\to jd}(\mu_f, \mu_F, \mu_R) \otimes D_{j\to h}(z_j, \mu_F),
\end{eqnarray}
where $x_a = p_a/P_A$, $x_b = p_b/P_B$ are the initial momentum fractions carried by the interacting partons, $z_j = p_h/p_j$ is the momentum fraction carried
by the final observed hadron. $f_{a/p}(x_a,\mu_f)$ and $f_{b/p}(x_b,\mu_f)$ are two parton distribution functions, $d\sigma_{ab\to jd}(\mu_f, \mu_F, \mu_R)$ is the differential cross section for parton scattering process, and $D_{j\to h}(z_j, \mu_F)$ is the fragmentation function for parton $j$ to hadron $h$.
There are three different scales involved in the calculation: $\mu_f$ and $\mu_F$ are factorization scales and $\mu_R$ is the renomalization scale; they are usually taken to be the same $\mu_f = \mu_F = \mu_R$ as a typical hard scale $Q$ involved in the process, such as the hadron $p_T$.
The non-perturbative PDFs and FFs are universal, and their scale dependences obey the Dokshitzer-Gribov-Lipatov-Altarelli-Parisi (DGLAP) evolution equations \cite{Dokshitzer:1977sg, Gribov:1972ri, Altarelli:1977zs}.
They are usually determined by the global fit to the experimental data from elementary collisions, such as $e^+e^-$ experiments, deep inelastic scatterings and proton-proton collisions.

The above formula is written at the leading order (LO) perturbative QCD.
For LO calculation of single inclusive hadron production in proton-proton collisions, one needs LO $2\to2$ hard cross sections, LO resummed PDFs and FFs, and one-loop expression for the strong coupling constant $\alpha_s$.
Since hard cross sections at LO only have $2 \to 2$ scattering processes, there is no dependence on the factorization scale to compensate the scale dependence in resummed PDFs and FFs. This situation can be greatly improved when we go beyond LO and perform full next-to-leading order (NLO) calculation.

At NLO, one needs to calculate real $2 \to 3$ diagrams and $2 \to 2$ one-loop virtual diagrams for hard scattering cross sections (see e.g., \cite{Jager:2002xm}).
Collinear singularities in both real $2 \to 3$ processes and virtual one-loop $2 \to 2$ processes are absorbed in PDFs and FFs at the factorization scales.
Soft infrared singularities in real $2 \to 3$  processes are canceled by virtual one-loop $2 \to 2$ processes.
The ultraviolet poles in virtual diagrams are removed by the renormalization of the coupling constant $\alpha_s$ at the renormalization scale.
Therefore, for NLO calculation of single inclusive hadron production in proton-proton collisions, one needs, NLO hard cross sections , NLO resummed PDFs and FFs, and two-loop expression for the strong coupling $\alpha_s$.

\begin{figure}
\begin{center}
\includegraphics*[width=10.0cm]{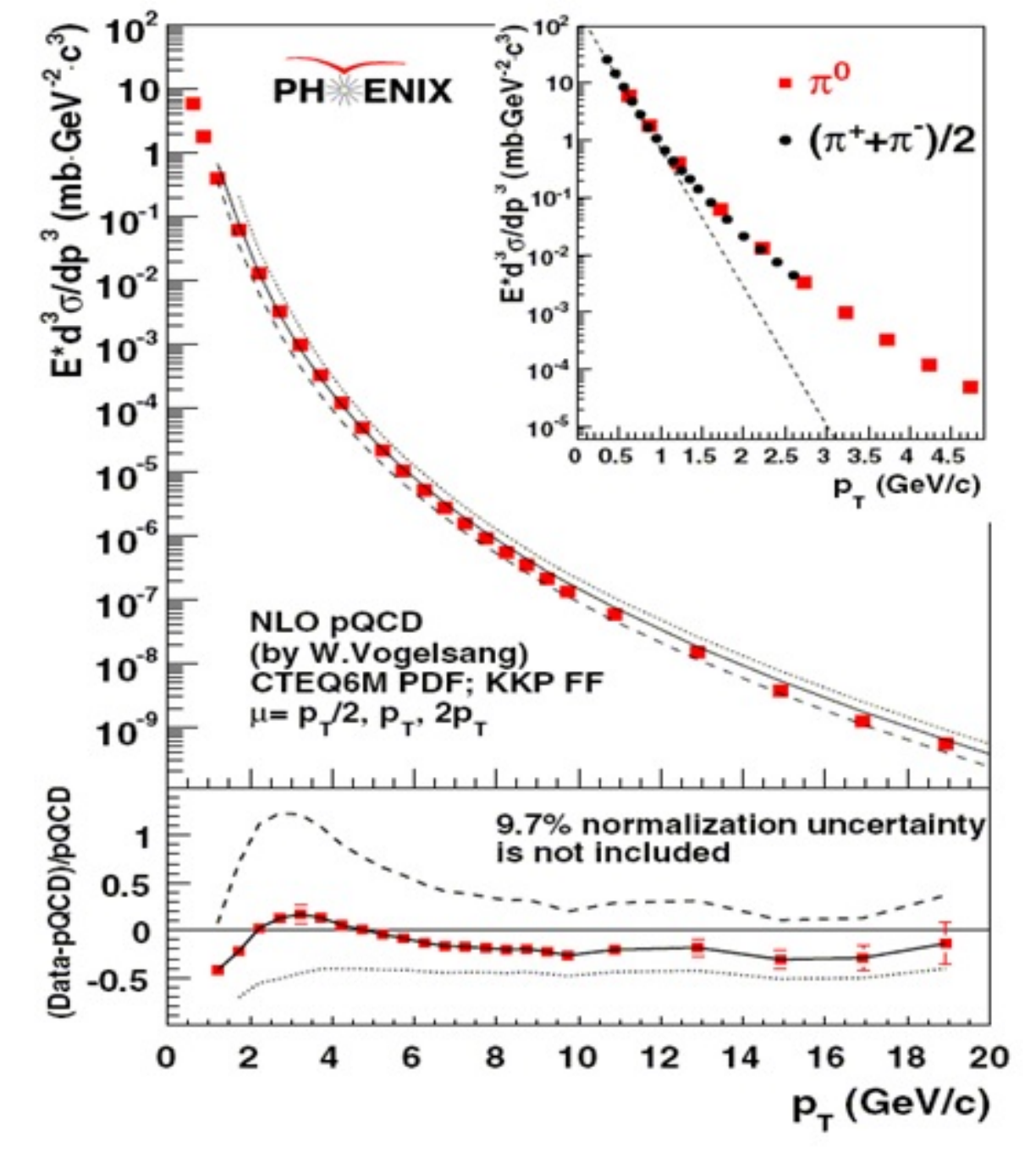}
\caption{(Color online) The cross section for high $p_T$ hadron production in proton-proton collisions at RHIC, compared to NLO perturbative QCD calculations. The figure is taken from Ref. \cite{Adare:2007dg}.
}
\label{fig_PHENIX_hadron_NLO}
\end{center}
\end{figure}

Fig. \ref{fig_PHENIX_hadron_NLO} shows the production cross section for high $p_T$ hadrons in proton-proton collisions at $\sqrt{s_{NN}} = 200$~GeV, compared to NLO pertubative QCD calculations \cite{Jager:2002xm}.
We can see that NLO perturbative QCD can describe the production of single inclusive hadrons at high $p_T$ very well, which indicates that the hard jets in vacuum are under well control on both experimental and theoretical sides.
This serves as the baseline for studying jet-medium interaction and the nuclear modification of jets in high-energy nucleus-nucleus collisions.

Often in many phenomenological studies (especially on jet quenching), LO expression is used with a multiplicative $K$ factor introduced to mimic the NLO corrections to hard jet production.
But a constant $K$-factor usually over-predicts the curvature of single inclusive hadron spectra, especially at small $p_T$.
This may be partially corrected by the incorporation of the intrinsic $k_T$-smearing of partons which originates from initial state radiation.
To perform this, one can introduce the so-called generalized parton distribution function:
\begin{eqnarray}
\tilde{f}(x, k_\perp, Q^2) = f(x, Q^2) g(k_\perp),
\end{eqnarray}
where $g(k_T)$ is the intrinsic $k_T$ smearing function, which is often taken to be Gaussian, $g(k_T) = e^{-k_T^2/\langle k_T^2 \rangle}/(\pi \langle k_T^2 \rangle)$.
After taking into account the intrinsic $k_T$, the single inclusive hadron production can be calculated as:
\begin{eqnarray}
d\sigma_{pp \to hX} &&\!\! \approx  \sum_{abjd} \int dx_a \int dx_b \int dz_j \int d^2k_{T, a} \int d^2k_{T, b} \tilde{f}_{a/p}(x_a, k_{T, a}, Q^2) \nonumber\\ &&\!\! \otimes \ \tilde f_{b/p}(x_b,k_{T, b},Q^2) \otimes d\sigma_{ab\to jd}(Q^2) \otimes D_{h/j}(z_j, Q^2).
\end{eqnarray}
In the above equation, all the scales have been taken to be the same $Q$.
Note that with the inclusion of intrinsic $k_T$, there will also be the modification of kinematics in addition to the integrals $\int d^2 k_{T, a} \int d^2 k_{T, b}$ \cite{Owens:1986mp, Wang:1998ww}.

\begin{figure}
\begin{center}
\includegraphics*[width=10.5cm]{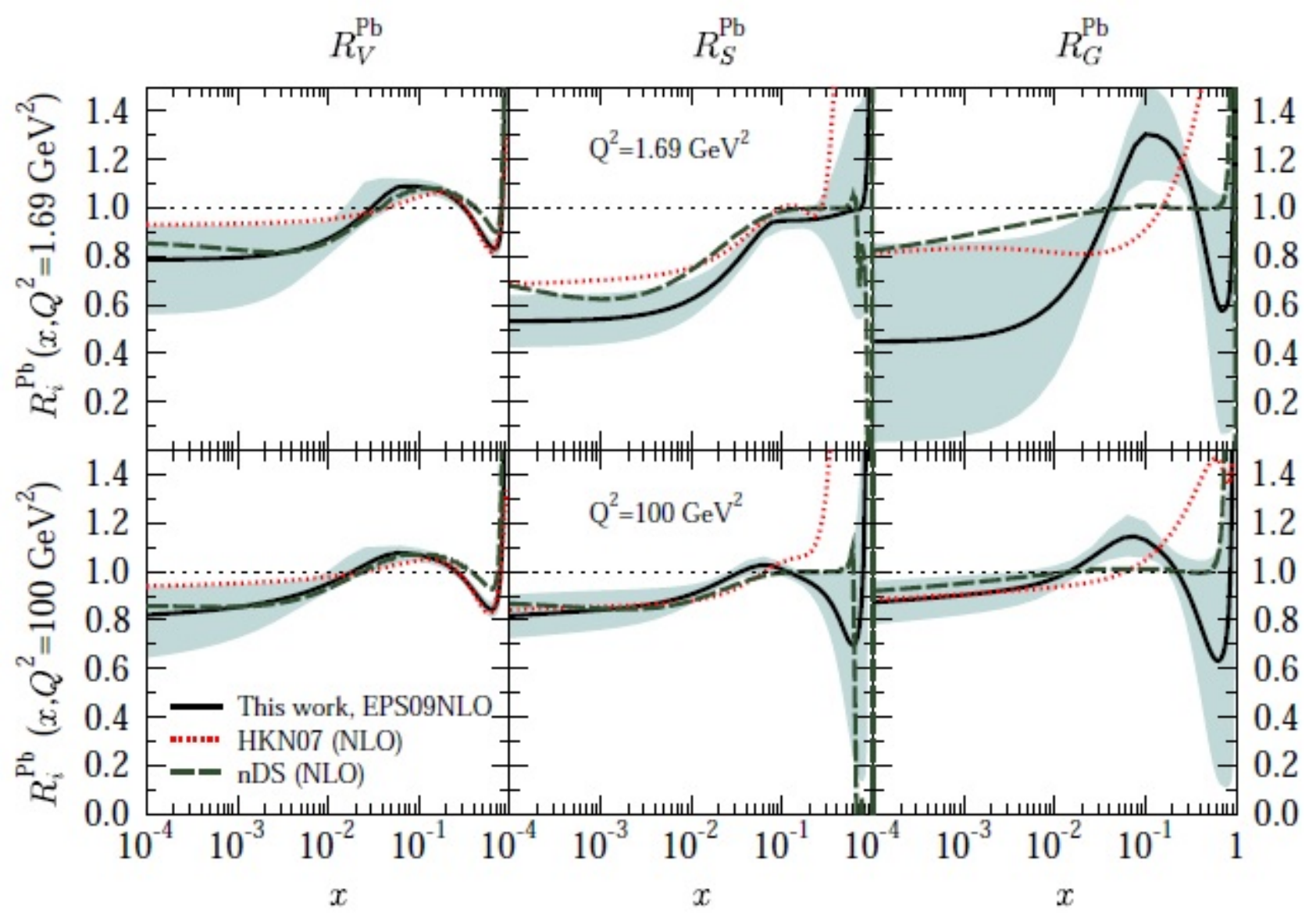}
\caption{(Color online) Nuclear modification factor for the valence and sea quark, and gluon distribution in Pb nucleus at $Q^2 = 1.69$~GeV$^2$ and $Q^2 = 100$~GeV$^2$ from EPS09NLO \cite{Eskola:2009uj}, HKN07 \cite{Hirai:2007sx}, and nDS \cite{deFlorian:2003qf}. The figure is taken from Ref. \cite{Eskola:2009uj}.
}
\label{fig_EPS09_NLO}
\end{center}
\end{figure}

When studying jet energy loss and jet quenching in ultra-relativistic heavy-ion collisions, two types of nuclear effects need to be taken into account.
First, the PDF in nucleus $f_{a/A}$ is different from free proton PDF $f_{a/p}$ used in proton-proton collisions.
Such effect is often called cold nuclear matter (CNM) effect.
To take this into account, one may define the nuclear modification factor  $R_i^A(x, Q^2)$ for the parton distribution function:
\begin{eqnarray}
R_a^A(x, Q^2)  = f_{a/A}(x, Q^2)/f_{a/p}(x, Q^2).
\end{eqnarray}
The PDF nuclear modification factor $R_i^A(x, Q^2)$ is often obtained by the global fit to the data from deep inelastic scattering off a nucleus, proton-nucleus and deuteron-nucleus collisions.
Fig. \ref{fig_EPS09_NLO} shows the nuclear modification factors $R_i^A(x, Q^2)$ for the valence quarks, sea quarks, and gluons in lead (Pb) nucleus at $Q^2 = 1.69$~GeV$^2$ (upper panels) and $Q^2 = 100$~GeV$^2$ (lower panels) from EPS09NLO \cite{Eskola:2009uj}, HKN07 \cite{Hirai:2007sx}, and nDS \cite{deFlorian:2003qf} parameterizations.
One can see that the nuclear modification of parton distribution function generally has four different $x$ regimes: shadowing, anti-shadowing, EMC-effect, and Fermi-motion.

\begin{figure}
\begin{center}
\includegraphics*[width=10.5cm]{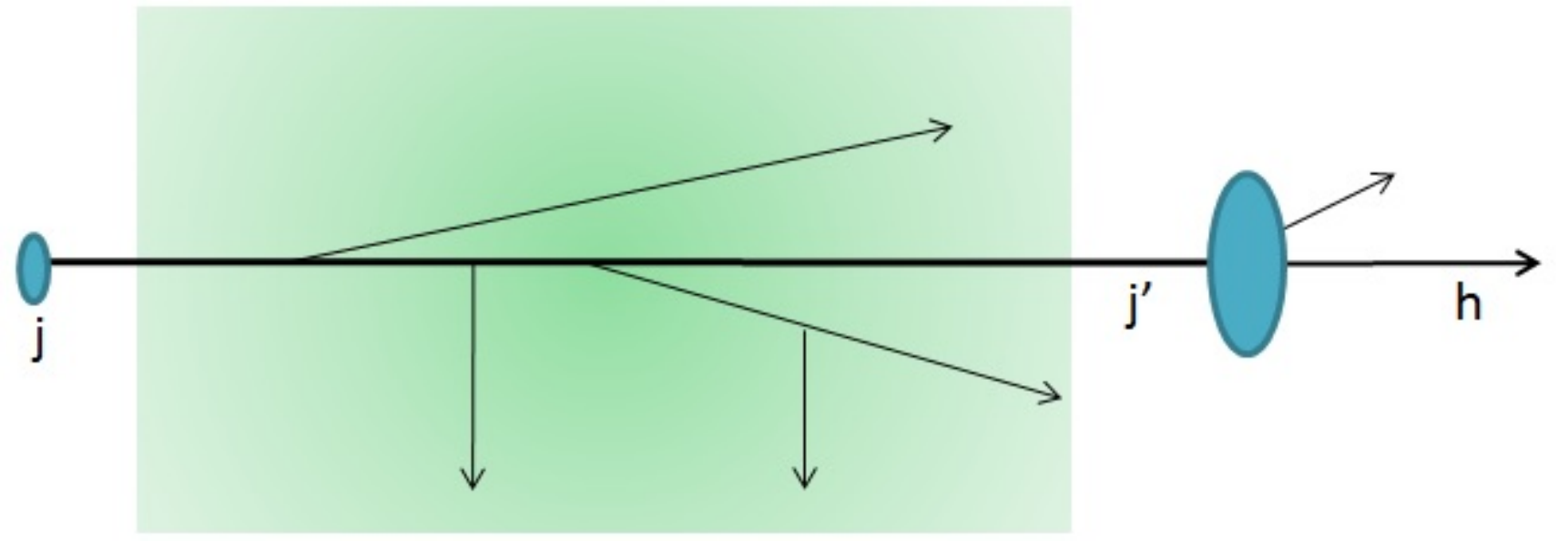}
\caption{(Color online) Schematic illustration of jet modification and parton energy loss in QGP medium before the fragmentation into hadrons.
}
\label{fig_jet_modification}
\end{center}
\end{figure}

The second effect in high-energy nucleus-nucleus collisions is the production of the hot and dense QGP medium.
This is called hot nuclear matter effect and is schematically illustrated in Fig. \ref{fig_jet_modification}.
Partonic jets which are produced from the initial hard scattering processes have to travel through and interact with the hot and dense nuclear matter before fragmenting into final state hadrons which are observed by the detectors.
By taking into account both cold and hot nuclear effects, the above factorized formula for single inclusive hadron production may be modified as
\begin{eqnarray}
d\tilde{\sigma}_{AB \to hX} \approx \sum_{abjj'd} f_{a/A}(x_a) \otimes f_{b/B}(x_b) \otimes d\sigma_{ab\to jd} \otimes P_{j\to j'} \otimes D_{h/j'}(z_{j'}),
\end{eqnarray}
where the additional piece $P_{j \to j'}$ describes the effect of the hard partons $j$ interacting with the colored medium before fragmenting into hadrons.
Here for brevity, the scale dependences of various quantities are not explicitly written.
In phenomenological study, one often defines the so-called medium-modified FF for convenience, by convoluting the vacuum FF with the parton-medium interaction function,
\begin{eqnarray}
\tilde{D}_{h/j}(z_j) \approx \sum_{j'} P_{j\to j'}(p_{j'}|p_j) \otimes D_{h/j'}(z_{j'}).
\end{eqnarray}
This is schematically shown in Fig. \ref{fig_jet_modification}.
One may also define the medium-modified parton scattering cross section by combining the vacuum cross section with the parton-medium interaction function.

The above factorized formula has been widely used in phenomenological studies of jet quenching in ultra-relativistic heavy-ion collisions.
However, currently there is no formal proof of such factorization yet.
The assumption of factorization is motivated by the fact that jet energy loss observed in nucleus-nucleus collisions is a dynamical final state effect which occurs after the hard process. It is also consistent with various phenomenological jet quenching studies.

As has been mentioned, the interaction of hard jets and the dense nuclear medium usually leads to jet energy loss.
One important observable is the suppression of the single inclusive hadron yield at high $p_T$ in nucleus-nucleus collisions as compared to the expectation from proton-proton collisions.
In order to quantify such nuclear modification effect, one may define the nuclear modification factor $R_{AA}$:
\begin{eqnarray}
R_{AA}(p_T, y, \phi_p) = \frac{1}{N_{\rm coll}} \frac{dN_{AA}/dp_Tdyd\phi_p}{dN_{AA}/dp_Tdyd\phi_p},
\end{eqnarray}
where $N_{\rm coll}$ is the number of binary collisions for a given collision centrality class.
Similar nuclear modification quantities may be defined for the production of single inclusive jets, as well as jets/hadrons triggered by jets/hadrons or direct photons.

\begin{figure}
\begin{center}
\includegraphics*[width=7.2cm]{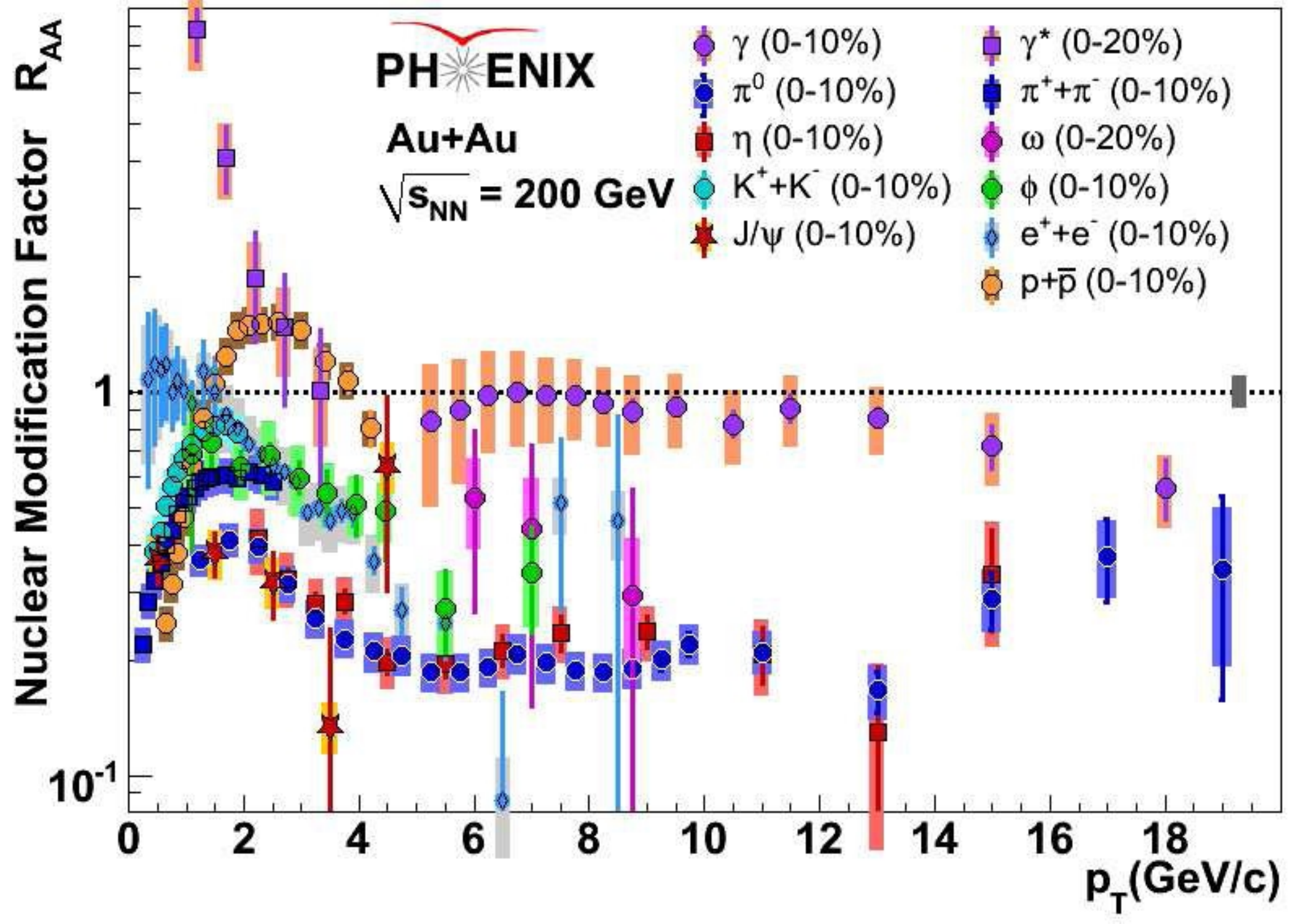}
\includegraphics*[width=5.3cm]{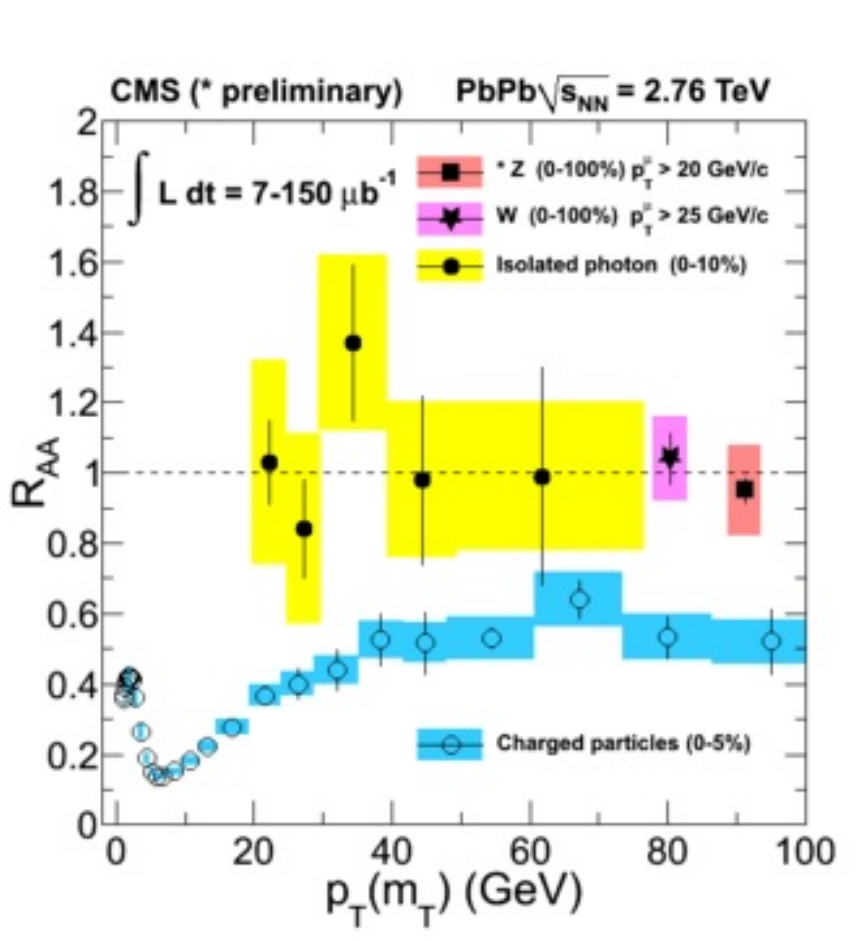}
\caption{(Color online) Left: $R_{AA}$ for identified hadrons, direct photons and non-photonic electrons measured by PHENIX in most central Au+Au collisions at $\sqrt{s_{NN}}=200$~GeV \cite{Sharma:2011zz}. Right: $R_{AA}$ for charged hadrons, direct photons, and Z/W bosons measured by CMS in most central Pb+Pb collisions at $\sqrt{s_{NN}}=2.76$~TeV \cite{CMS:2012aa}.
}
\label{fig_RAA_ID}
\end{center}
\end{figure}

Fig. \ref{fig_RAA_ID} shows the nuclear modification factor $R_{AA}$ as a function of $p_T$ for charged hadrons, direct photons and other identified particles. The left shows PHENIX measurement for most central Au+Au collisions at $\sqrt{s_{NN}}=200$~GeV, and the right shows CMS measurements for most central Pb+Pb collisions at $\sqrt{s_{NN}}=2.76$~TeV.
One can see that the yields of high $p_T$ hadrons are strongly suppressed in nucleus-nucleus collisions compared to those in proton-proton collisions, whereas the nuclear modification factor $R_{AA}$ for high $p_T$ photons is consistent with unity.
Since photons carry no color charge, they only interact with surrounding matter electromagnetically.
Photons, once produced, will fly to the detectors without further rescattering due to the fact that their mean free paths are much larger than the medium size.
This indicates that the observed strong suppression for high $p_T$ hadron production is due to the final state effect, i.e., the interaction of partonic jets with the colored medium which usually causes jets to lose a fraction of their energy.

\section{Parton energy loss formalisms}

\begin{figure}
\begin{center}
\includegraphics*[width=5.9cm]{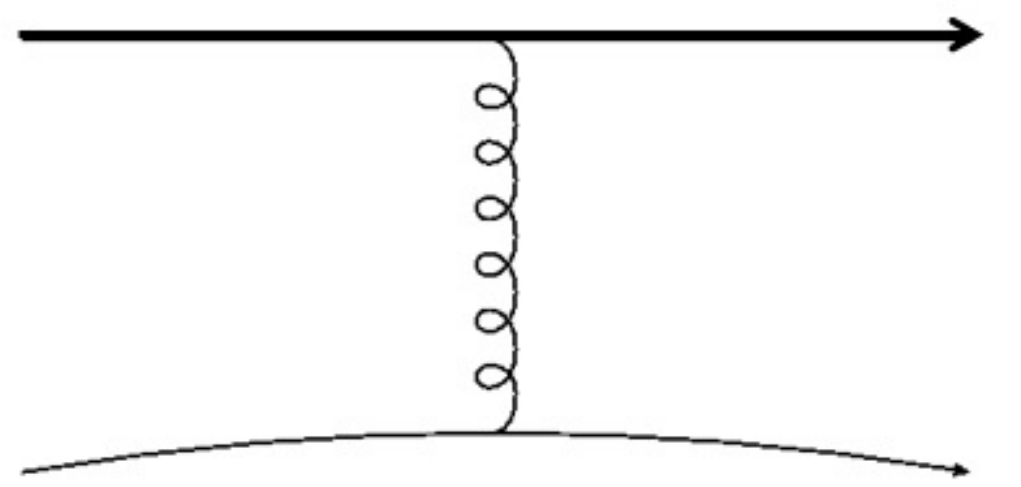}
\includegraphics*[width=6.6cm]{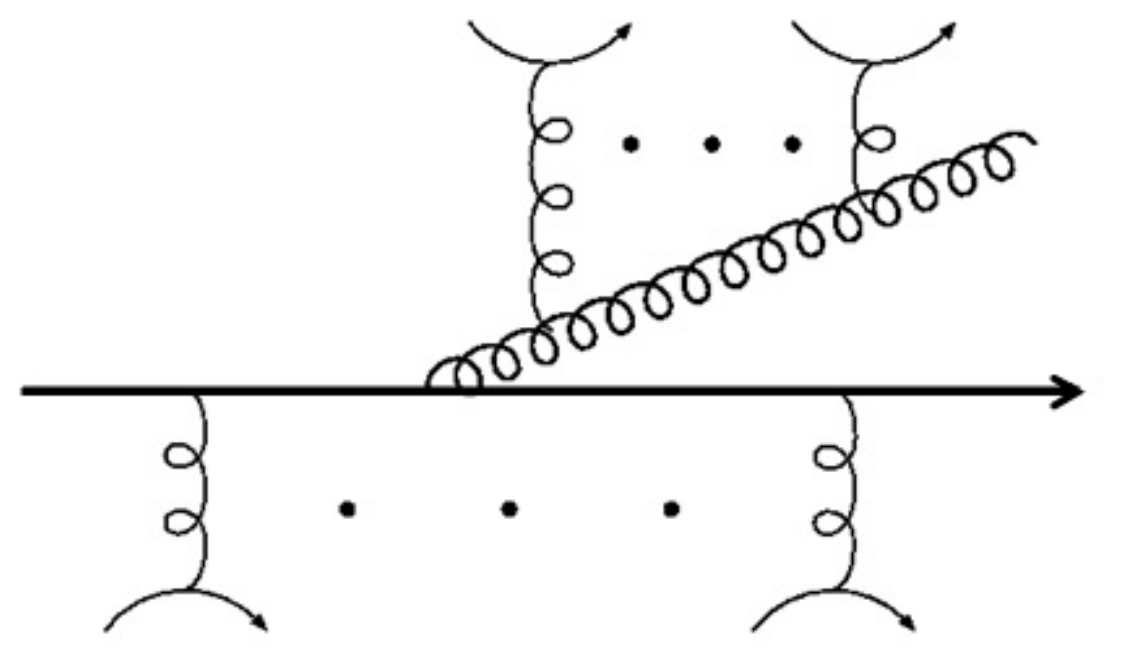}
\caption{(Color online) Typical diagrams for collisional (left) and radiative (right) energy losses of an energetic parton propagating through a hot and dense nuclear medium.
}
\label{fig_coll_rad}
\end{center}
\end{figure}

Hard partonic jets may lose energy via a combination of both elastic and inelastic collisions with the constituents of the hot and dense nuclear medium.
The energy loss of the primary hard parton occurred in $2 \to 2$ elastic collisions is usually called elastic or collisional energy loss.
Multiple scatterings with the medium constituents may induce additional radiation which takes away a fraction of energy from the primary parton; this is called radiative energy loss.
Fig. \ref{fig_coll_rad} shows the typical diagrams for the calculation of jet energy loss originating from collisional (left panel) and radiative (right panel) processes.

\subsection{Radiative energy loss}

Radiative energy loss has been regarded as the most important component in studying parton energy loss in nuclear medium and jet quenching in relativistic heavy-ion collisions.
Even in absence of a medium, high-$p_T$ partons produced from the initial hard collisions between the incoming partons will undergo vacuum splitting processes and reduce their virtuality (offshellness).
The presence of the hot and dense medium will modify the gluon radiation or parton splitting processes as compared to the vacuum case due to the rescatterings of the propagating parton shower with the medium constituents.

One important effect in the study of medium-induced gluon emission or parton splitting processes is the so-called Landau-Pomeranchuk-Migidal (LPM) effect \cite{Landau:1953um,Migdal:1956tc}.
For small angle or collinear radiation, the primary parton and the radiated gluon propagate along similar paths, and a finite time period (the formation time $\tau_f \sim 2\omega/k_\perp^2$, with $\omega$ and $k_\perp$ the energy and transverse momentum of the radiation), is required for the radiation process to complete.
If the formation time $\tau_f$ of the radiation is larger than the mean free path $\lambda$ of the propagating parton, the multiple scatterings experienced by the propagating parton can no longer be treated as independent.
Such quantum interference between successive scatterings caused by the LPM effect will lead to the suppression of the radiation spectrum as compared to the Bethe-Heitler spectrum which assumes incoherent multiple scatterings.
For the case of QCD, since the emitted gluons also carry color charge, their rescatterings with soft gluons in the medium will produce more dominant contribution to the modification of the radiation spectrum.

A number of parton energy loss formalisms have been developed to study medium-induced gluon bremsstrahlung and radiative process in dense nuclear medium, namely, Baier-Dokshitzer-Mueller-Peigne-Schiff-Zakharov (BDMPS-Z) \cite{Baier:1996kr, Baier:1996sk, Zakharov:1996fv}, Gyulassy, Levai and Vitev (GLV) \cite{Gyulassy:1999zd, Gyulassy:2000fs, Gyulassy:2000er}, Amesto-Salgado-Wiedemann (ASW) \cite{Wiedemann:2000za,Wiedemann:2000tf}, Arnold-Moore-Yaffe (AMY) \cite{Arnold:2001ba, Arnold:2002ja} and higher twist (HT) \cite{Guo:2000nz, Wang:2001ifa} formalisms.
In the following, we provide some basic information about different jet quenching formalisms.

{\bf \subsubsection{Single gluon emission}}

In most jet quenching formalisms, the starting point or the central goal is to calculate the single gluon emission kernel (e.g., the differential radiation spectrum $dN_g/d\omega dk_\perp^2 dt$ or its variants).
In Sec. \ref{sec_qhat_rhic_lhc}, we will provide more details about the single gluon emission kernel formula when presenting the phenomenological applications of jet quenching formalisms and the comparison to experimental data.
In this section we only give a brief general description of different formalisms and focus on the the similarity and differences among them.
In the calculation of single gluon emission kernel, various approaches differ in two main aspects: the assumptions about the medium and the treatments of multiple scatterings per emission.
A systematic comparison of different jet energy loss models has been performed in the context of a ``brick" of QGP in Ref. \cite{Armesto:2011ht}. More details may be found there and the references therein.

\vspace{12pt}
\subsubsection*{BDMPS-Z and ASW-MS formalisms}

In the Baier-Dokshitzer-Mueller-Peigne-Schiff-Zakharov (BDMPS-Z) \cite{Baier:1996kr, Baier:1996sk, Zakharov:1996fv} formalism, the calculation of gluon radiation process is formulated in terms of a path-integral in which the scatterings on multiple heavy static colored scattering centers are resummed.
The interference between vacuum and medium-induced radiation in such a path-integral formulation is included later by Wiedemann \cite{Wiedemann:2000za,Wiedemann:2000tf}.
Most analytical and numerical calculations from the BDMPS-Z formalism utilize a saddle point approximation, which assumes that the primary parton interacts with the medium mainly via multiple soft scattering processes. The numerical implementations of the multiple soft scattering version of the BDMPS-Z formalism are based on the work on calculating the quenching weights by Amesto, Salgado and Wiedemann \cite{Salgado:2003gb} (usually denoted as ASW-MS). In such an implementation, the medium is fully characterized by the jet transport coefficient $\hat{q} = d\langle \Delta p_T^2 \rangle /dL$ which characterizes the mean of the transverse momentum squared exchanged between the propagating parton and the colored medium per unit path length.

\subsubsection*{DGLV and ASW-SH formalisms}

The opacity expansion was developed by Gyulassy, Levai and Vitev (GLV) \cite{Gyulassy:1999zd, Gyulassy:2000fs, Gyulassy:2000er} and independently by Wiedemann \cite{Wiedemann:2000za,Wiedemann:2000tf}. Such approach is based on a systematic expansion in terms of the number of scatterings experienced by the propagating parton. The interference between the vacuum and medium-induced radiation is included in this formalism. In most phenomenological calculations, only the leading term ($N=1$) in the expansion is included and usually called single hard scattering limit. The medium effect in this approach is characterized by two parameters: the density $\rho$ of the scattering centers (or the mean free path $\lambda$ of the propagating parton) and the Deybe screening mass $\mu_D$ which is introduced to regulate the infrared behavior of single scattering cross section. The opacity expansion was first developed using heavy static scattering centers to characterize the medium, and the extension to the medium with dynamical scattering centers was later performed by Djordjevic and Heinz \cite{Djordjevic:2008iz}. There are two different model implementations for this approach, namely, DGLV \cite{Djordjevic:2008iz} and the single hard scattering limit of ASW formalism (ASW-SH) \cite{Salgado:2003gb}.

\subsubsection*{AMY formalism}

Arnold-Moore-Yaffe (AMY) approach \cite{Arnold:2001ba, Arnold:2002ja} is a quantum field theory formulation of parton energy loss in a weakly-coupled medium in thermal equilibrium. The medium that the hard propagating parton interacts is formulated as a thermal equilibrium state in the framework of Hard Thermal Loop (HTL) finite temperature field theory. The properties of the colored medium are characterized fully by its temperature. In phenomenological calculations, the model parameter is usually taken to be the strong coupling constant $\alpha_s = g_s^2/(4\pi)$. The AMY formalism was first developed assuming infinite length for the medium, and was recently improved to include the effect of finite medium length by Caron-Huot and Gale \cite{CaronHuot:2010bp}. In this approach, the incoming parton is assumed to be almost on-shell, thus the vacuum radiation of incoming partons is not included, and only the medium-induced effect is taken into account. Since the perturbative description of the thermal QGP medium is applied, the AMY formalism in principle only applies at very high temperature.

\subsubsection*{HT formalism}

Higher-twist parton energy loss formalism was pioneered by Guo and Wang \cite{Guo:2000nz, Wang:2001ifa}. In this approach, the properties of the medium enter into the calculation in terms of higher-twist matrix elements. In phenomenological applications, one usually factorizes out the nuclear PDFs of the incoming stuck parton from the higher-twist matrix elements, and the remainder of the matrix elements describes the interaction between the final state propagating parton and the dense nuclear medium. The formalism was first developed for single scattering and was later extended to include multiple scatterings per emission by Majumder \cite{Majumder:2009ge}. Similar to the GLV formalism, most numerical implementations of the HT approach still make use of single scattering per emission approximation. However, the characterization of the medium in the HT scheme is a little different from GLV. While GLV takes the full functional form for the the single scattering cross section, HT takes the leading moment of the exchanged transverse momentum distribution. Therefore the medium in the HT formalism is solely characterized by the jet quenching parameter $\hat{q}$.
Note that a recent study has shown that the stimulated radiation process may be modified by the longitudinal momentum exchange with the medium constituents (such as the longitudinal drag and diffusion) \cite{Qin:2014mya}.

\vspace{12pt}

Despite many differences among various parton energy loss formalisms as has been mentioned above, there are three main common assumptions that are made in most calculations. The first is the eikonal approximation which assumes both the incoming parton energy $E$ and radiated gluon energy $\omega$ are much larger than the transverse momentum $q_\perp$ exchanged with the medium constituents, i.e., $E, \omega \gg q_\perp$, so the partons are traveling along eikonal trajectories. Additionally, soft radiation approximation is often imposed in some approaches, so that the gluon energy $\omega$ is assumed to be much smaller the primary parton energy $E$, i.e., $x = \omega/E \ll 1$ (Note that such approximation is not made in the AMY calculation).
The second common approximation is the collinear or small-angle radiation approximation which assumes that the transverse momentum $k_\perp$ of the radiated gluon is much smaller than its energy $\omega$, i.e., $k_\perp \ll \omega$.
The third approximation assumes that the momentum transfer between the hard parton and the medium is localized. This means that the mean free path $\lambda$ of the propagating parton is much larger than the Debye screening length $\lambda_D = 1/\mu_D$.
More extensive discussions on the effects of different approximations on the calculation of medium-induced gluon radiation and parton energy loss may be found in Ref. \cite{Armesto:2011ht}.

{\bf \subsubsection{Multiple Gluon Emission}}

\begin{figure}
\begin{center}
\includegraphics*[width=10cm]{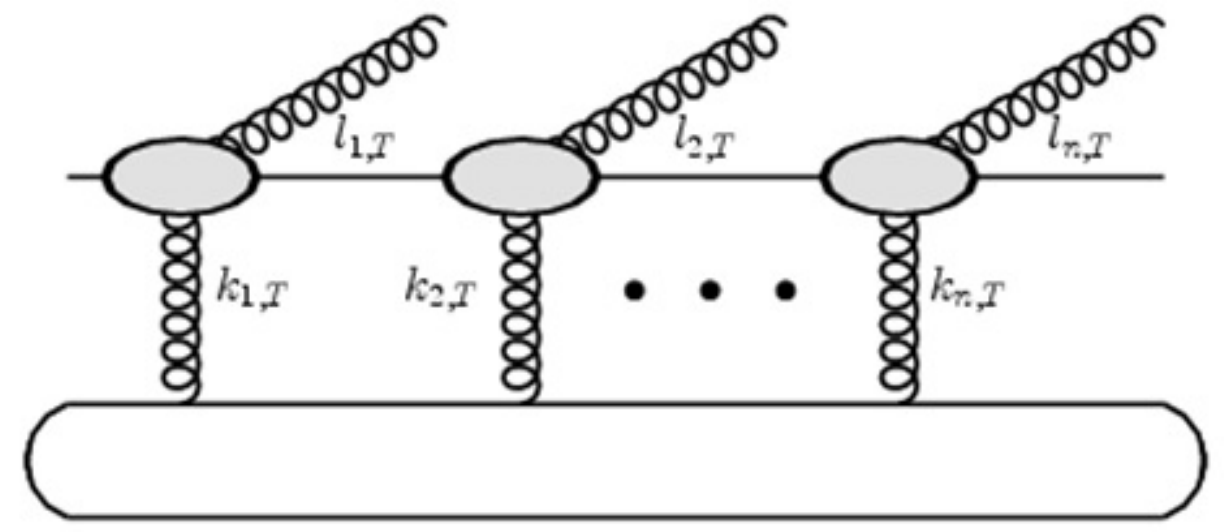}
\caption{(Color online) Schematically illustration of multiple gluon emission reduced to the repeated application of single gluon emission kernel.
}
\label{multiple_gluon emission}
\end{center}
\end{figure}

In general multiple parton splitting may occur in medium, similar to the case of vacuum parton shower.
A full calculation of multiple parton final state would include the interference between different emissions.
There has been some effort made along this direction, such as the calculations of the interference between two emitters (the antenna problem) \cite{Armesto:2011ir}, but the generalization of these results to the case of multiple parton emission is still not clear at present.
In all available phenomenological studies, a common practice to calculate multiple gluon emission is based on the repeated application of the single gluon emission kernel (as schematically illustrated in Fig. \ref{multiple_gluon emission}).
In the following we will present different prescriptions used for the calculation of multiple gluon emission.
\vspace{12pt}

\subsubsection*{Poisson ansatz}

One popular method to obtain multiple gluon emission is the use of Poisson convolution which assumes that the number of emitted gluons in a given path follows a Poison distribution.
The mean of the Poisson distribution is obtained by the integral of single gluon emission kernel, while the energy distribution of each emitted gluon obeys the single gluon emission spectrum.
This method has been widely applied in phenomenological studies using GLV and ASW formalisms \cite{Salgado:2003gb, Renk:2006sx, Wicks:2005gt}.
The central quantity using Poisson ansatz is to calculate the probability distribution $P(\Delta E)$ of parton energy loss, which may be obtained from single gluon emission kernel $dN_g/d\omega$ as follows:
\begin{eqnarray}
P(\Delta E) = \sum_{n=0}^{\infty} \frac{e^{-\langle N_g \rangle}}{n!} \left[ \prod_{i=1}^{n} \int d\omega \frac{dN_g(\omega)}{d\omega} \right] \delta\left( \Delta E - \sum_{i=1}^{n} \omega_i \right),
\label{eq:poisson}
\end{eqnarray}
where $dN/d\omega$ is the spectrum for single gluon emission and  $\langle N_g \rangle = \int d\omega dN_g/d\omega$ is the average number of radiated gluons.
In most jet energy loss calculations that use Poisson ansatz, the degrading of the primary parton energy during its propagation is not dynamically updated after each emission, therefore there could be finite probability for the total radiated energy larger than the incoming parent parton energy.
In practice such finite probability is treated as the probability for jet to lose its total energy.

\subsubsection*{Transport models}

Transport approach is another popular method in the study of parton evolution in medium.
In the AMY formalism, the transport rate equations have been solved to obtain the time evolution of jet energy (momentum) distributions $f(p)=dN(p)/dp$  \cite{Jeon:2003gi, Qin:2007rn}.
The coupled rate equations can be schematically written as:
\begin{eqnarray}
\frac{df(p,t)}{dt} = \int dk \left[ f(p+k,t) \frac{d\Gamma(p+k,k,t)}{dkdt} - f(p,t) \frac{d\Gamma(p,k,t)}{dkdt} \right],
\end{eqnarray}
where $d\Gamma(p,k,t)/dkdt$ is the differential rate for a parton with momentum $p$ to lose momentum $k$ (the parton species may change and the indices for parton species are not explicitly written in the above generic formula).
The above transport equation includes the coupling between quark and gluon distributions of the jets ($f_q$ and $f_g$), and keep track of the change of parton energy during its propagation.
By performing the $\int dk$ integration from $-\infty$ to $\infty$, the above rate equation also takes into account the absorption of thermal partons from the medium, thus preserve the detailed balance.

The transport approach for jet propagation in medium has also been recently implemented in a Linear Boltzmann Transport (LBT) model \cite{Li:2010ts,Wang:2013cia,He:2015pra} which not only follows the propagation of jet shower partons but also keeps track of medium recoil partons so that energy and momentum are conversed throughout the transport processes. Within the LBT model, the propagation of jet shower partons and medium excitation is simulated according to a linearized Boltzmann equation
\begin{eqnarray}
p_1\cdot\partial f_1(p_1)&=&-\int dp_2dp_3dp_4 (f_1f_2-f_3f_4)|M_{12\rightarrow34}|^2 \nonumber \\
 &\times&
(2\pi)^4\delta^4(p_1+p_2-p_3-p_4),
\end{eqnarray}
where $dp_i=d^3p_i/[2E_i(2\pi)^3]$, $f_i=1/(e^{p\cdot u/T}\pm1)$ $(i=2,4)$ are parton phase-space distributions in a thermal medium with local temperature $T$ and fluid velocity $u=(1, \vec{v})/\sqrt{1-\vec{v}^2}$, and $f_i=(2\pi)^3\delta^3(\vec{p}-\vec{p_i})\delta^3(\vec{x}-\vec{x_i}-\vec{v_i}t)$ $(i=1,3)$ are the parton phase-space densities before and after scattering. For elastic scatterings, the amplitudes $|M_{12\rightarrow34}|^2$ are given by the lowest order (LO) pQCD parton scattering processes.  Partons are assumed to propagate along classical trajectories between
two adjacent scatterings. The probability of scattering in each time step $\Delta t$ is determined by,
\begin{equation}
 P_{a}=1-\text{exp}\left[-\Delta t \frac{p \cdot u}{E} \Gamma_{a}(p\cdot u, T)\right],
 \end{equation}
 where $\Gamma_{a}$ is the total scattering rate for parton type $a$ with four-momentum $p$ in the local fluid comoving frame. The backreaction in the Boltzmann transport is accounted for by subtracting initial thermal partons $(p_2)$, denoted as ``negative'' partons, from the final observables after been transported also according to the Boltzmann equation. These negative partons are considered as part of the recoiled partons that are responsible for jet-induced medium excitations \cite{Li:2010ts}. Induced radiation accompanying each elastic scattering is simulated according to the HT approach and multiple gluon emissions within each time step are included according to the Poisson distribution in Eq.~(\ref{eq:poisson}).

\subsubsection*{Modified DGLAP evolution}

HT formalism was developed using perturbative QCD power expansion. It solves the following DGLAP-like evolution equations to obtain the medium-modified fragmentation function $\tilde{D}(z,Q^2)$ \cite{Majumder:2009zu, Qin:2009gw, Majumder:2011uk, Chang:2014fba}:
\begin{eqnarray}
\frac{\partial \tilde{D}(z, Q^2, q^-)}{\partial\ln Q^2} = \frac{\alpha_s}{2\pi} \int \frac{dy}{y} P(y) \int d\zeta^- K(\zeta^-, Q^2, q^-, y) \tilde{D}(z/y, Q^2, q^-y),
\end{eqnarray}
where $q^-$ is the light-cone energy of the hard jet, $y$ the fractional energy of the radiation, and $\zeta^-$ is the location of the jet. $P(y)$ is the vacuum splitting function and $K(\zeta^-, Q^2, q^-, y)$ is the kernel for the parton-medium scattering which induces additional radiation process (see Sec. \ref{sec_qhat_rhic_lhc} for its detailed expression).
In DGLAP-like evolution, the coupling between quark and gluon fragmentation function ($D_q$ and $D_g$) is included, and the change of parton energy during its propagation may be traced as well.
The contribution from vacuum radiation is also implicitly included in the above equation.

\vspace{12pt}

There are some conceptual difference between the above three methods.
AMY rate equations, LBT model and HT DGLAP-like evolution equations have two aspects of improvement as compared to the Poisson convolution used by BDMPS, GLV and ASW formalism,
First, the evolution equations used by the AMY, LBT and HT formalisms keep track of all shower partons while the Poisson convolution only trace the lost energy of the leading parton.
In addition, AMY, LBT and HT include the change of the emission probability due to the degrading of jet energy during its propagation, but the Poisson convolution does not.
Thus the use of Poisson convolution may lead to finite probability to lose more energy than the parent parton (such issue may be fixed by using Monte-Carlo method to calculate the probability distribution $P(\Delta E)$, as done in Ref. \cite{Chang:2014fba}).

In spite of these differences, there is one important point that is not well addressed in all parton energy loss models: the gluon radiation process is not a purely local phenomenon, but has a finite formation time $\tau_f$.
Therefore, the evolution of jet energy loss should be accompanied at the same by an evolution in the coordinate space, i.e., as the jet energy degrades along its propagation, the remaining path length through the medium decreases as well.
Many jet quenching models have taken into account the evolution of the medium and the corresponding change in the local medium information (such as the temperature, the energy/entropy density and the flow velocities), but the combination of the local information and the true finite size effect has not yet been included at present.

\subsection{Collisional energy loss}

Collisional energy loss of hard jets propagating through nuclear matter was first studied by Bjorken in 1982 \cite{Bjorken:1982tu} and then by a few others in recent years \cite{Braaten:1991we,Mustafa:2004dr,Wicks:2005gt,Peshier:2006hi,Djordjevic:2006tw,Qin:2007rn,Schenke:2009ik}.
Many calculations of collisional energy loss were performed in the framework of perturbative finite temperature field theory.
Collisional energy loss may be treated in the same footing as radiative energy loss in the AMY formalism \cite{Qin:2007rn,Schenke:2009ik} and the LBT model \cite{Li:2010ts,Wang:2013cia,He:2015pra}.
In BDMPS-Z, GLV and ASW approaches, the medium is modeled as a collection of heavy static scattering centers, therefore by construction these formalisms neglect the recoil effects and thus conceptually does not allow for collisional energy loss for the propagating hard partons.
In phenomenological application, collisional energy loss in these formalisms is added independently from radiative energy loss.
It is noted that the GLV formalism has been extended to the medium with dynamical scattering centers (DGLV), thus both collisional and radiative energy loss may be formulated in the same setup (DGLV) \cite{Djordjevic:2006tw, Djordjevic:2008iz}.
In the HT formalism, elastic collisions may be added as additional component characterized by the longitudinal drag and diffusion of the propagating hard partons \cite{Majumder:2007hx,Qin:2009gw,Qin:2012fua}.

\begin{figure}
\begin{center}
\includegraphics*[width=6.25cm]{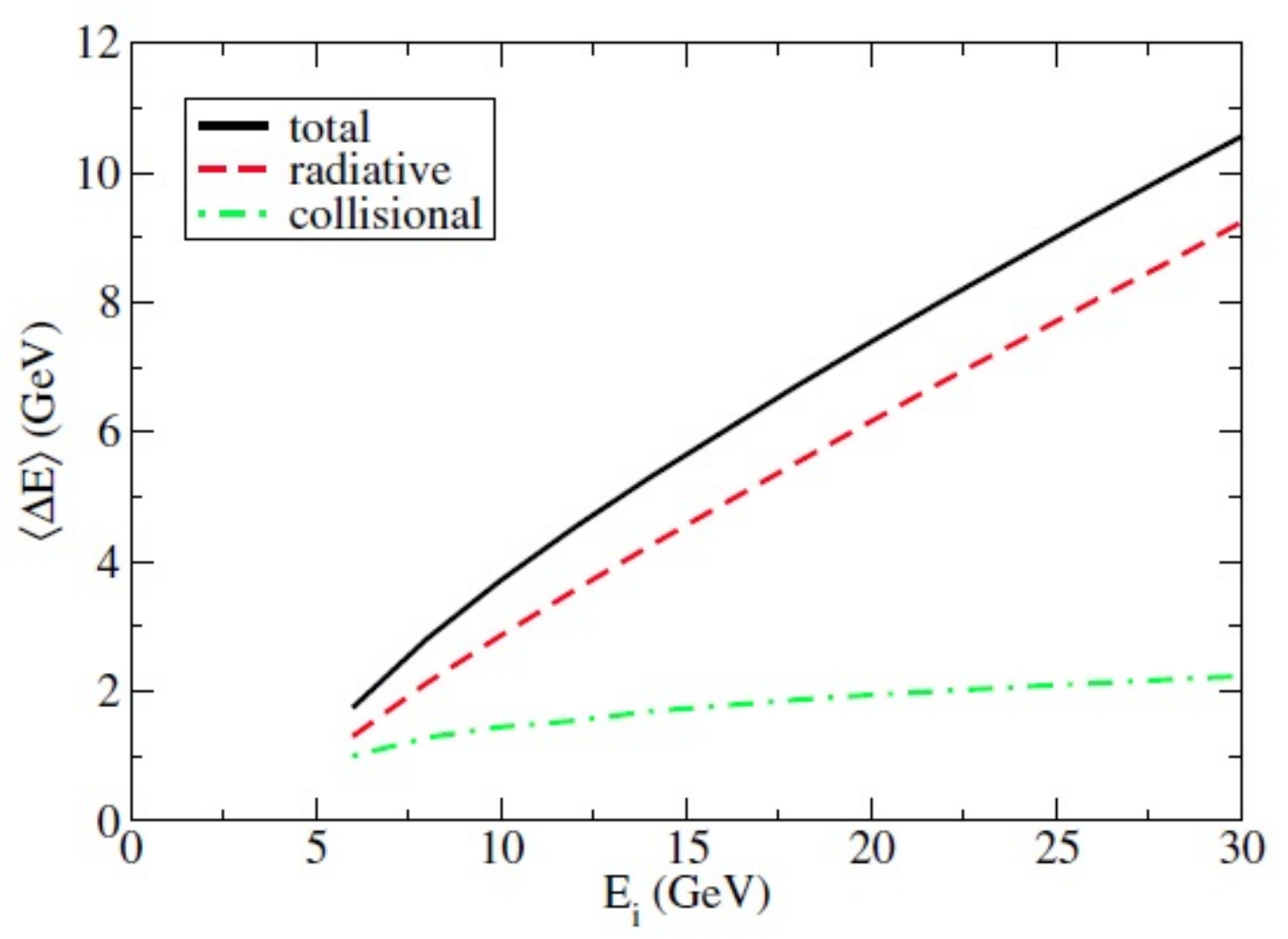}
\includegraphics*[width=6.25cm]{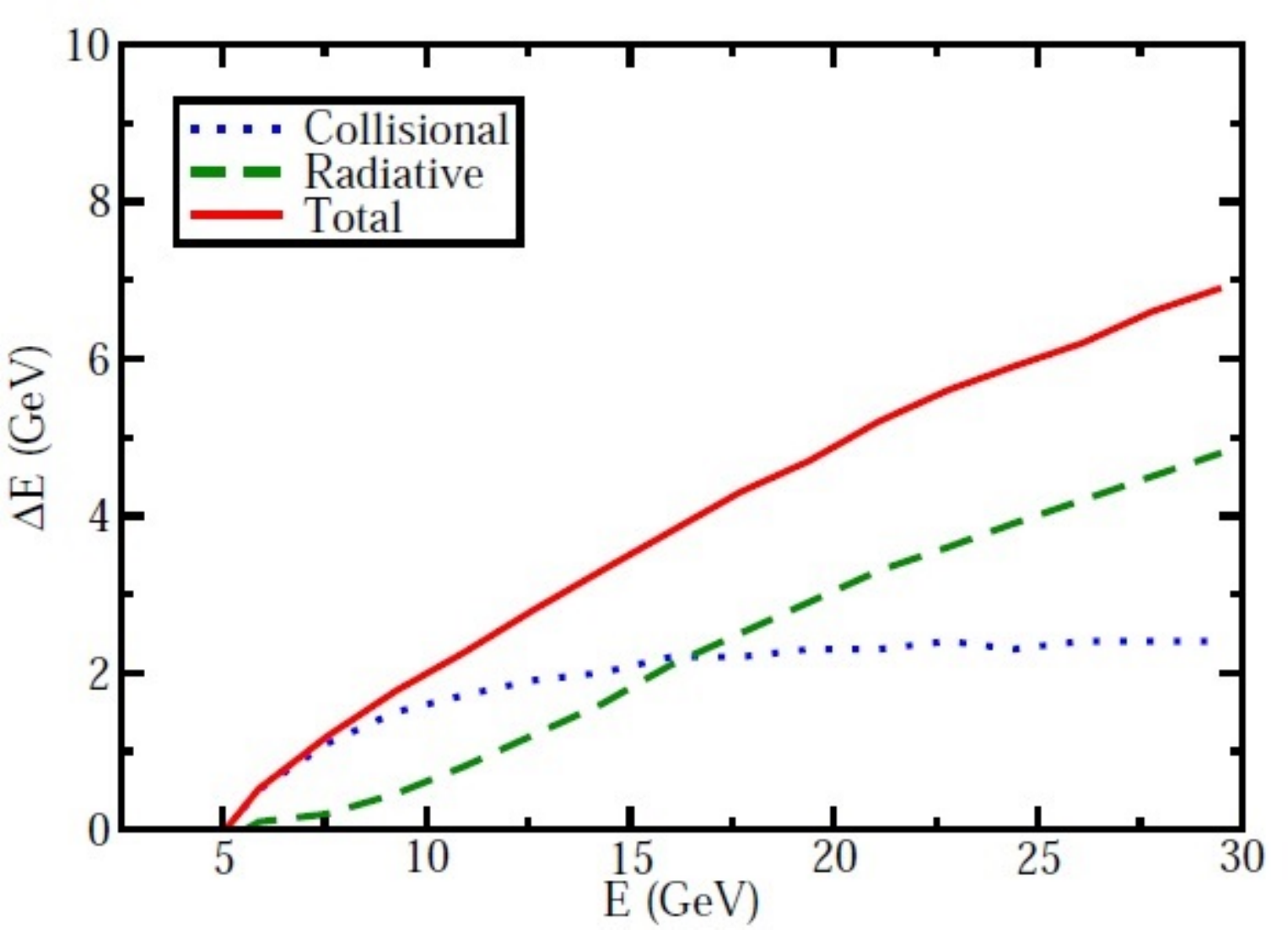}
\caption{(Color online) Left: Average energy loss of a light quark produced at the center of medium passing through the nuclear medium created in most 0-5\% central collisions at RHIC \cite{Qin:2007rn}.
Right: Average energy loss of a bottom quark passing through the nuclear medium created in most 0-7.5\% central collisions at the LHC \cite{Cao:2013ita}.}
\label{fig_Qin_rad_vs_coll}
\end{center}
\end{figure}

Compared to the radiative component of jet energy loss, collisions energy loss is usually considered to be small for light flavor (leading) partons, especially when the energy of the jet is sufficiently high \cite{Wicks:2005gt,Qin:2007rn,Schenke:2009ik}.
However in realistic calculation of nuclear modification factor $R_{AA}$ at RHIC and the LHC energies, collisional energy loss may give sizable contribution and cannot be simply neglected \cite{Wicks:2005gt,Qin:2007rn}.
For heavy quarks, elastic collisions are usually considered as the dominant energy loss mechanism, especially at low and intermediate energy regimes \cite{Moore:2004tg, Mustafa:2004dr}, due to the large finite masses of heavy quarks which tend to reduce the phase space of medium-induced gluon radiation (this is usually called the dead-cone effect \cite{Dokshitzer:2001zm}).
When going to the high-energy regimes where heavy quarks become ultra-relativistic as well, collisional energy loss alone is no longer sufficient for simulating the in-medium evolution of heavy quarks, and radiative contribution becomes more significant \cite{Cao:2012au, Uphoff:2012gb, Abir:2012pu, Nahrgang:2014vza}.
This can be seen from Fig. \ref{fig_Qin_rad_vs_coll} which shows the comparison between radiative and collisional energy loss: the left panel for light partons and the right for bottom quarks.
We can see that for bottom quarks, below around 17~GeV collisional energy loss dominates; above this transition energy, the effect from the finite mass becomes small and radiative components starts to dominate \cite{Cao:2013ita}.
As we will discuss in Sec. \ref{sec_full_jet}, collisional energy loss may play an important role in understanding the energy loss and nuclear modification of full jet in dense nuclear medium \cite{Qin:2010mn}.
It is also an essential ingredient when studying the response of the medium to propagating hard jet \cite{Qin:2009uh,Neufeld:2009ep} (see Sec. \ref{sec_medium_response}).

\section{Recent developments on phenomenology and theory: $\hat{q}$, etc.}
\label{sec_qhat_rhic_lhc}

The main purpose of jet quenching study is to understand the detailed mechanisms of jet-medium interaction and their manifestation in the final state observables.
The ultimate goal is to extract various novel properties of the hot and dense QGP produced in ultra-relativistic heavy-ion collisions.
To achieve this goal, one important strategy is to perform systematic phenomenological studies of jet quenching and compare to a wealth of experimental measurements of final state observables, such as single inclusive hadron production at high $p_T$, the reconstructed jets and their substructures, and hadrons/jets correlated with high-$p_T$ hadrons/jets or direct photons, and so on.

Recently much attention of jet quenching study has been paid to the quantitative extraction of various jet transport coefficients, such as the transverse momentum diffusion rate $\hat{q}$, the elastic energy loss rate $\hat{e}=dE/dt$ and the diffusion of elastic energy loss $\hat{e}_2 = d(\Delta E)^2/dt$, etc \cite{Baier:1996kr, Majumder:2007hx, Qin:2012fua}.
At leading order, these parameters quantify the transverse and longitudinal momenta transferred between the propagating hard partons and the dense nuclear medium via $2\to 2$ elastic collisions.
Since jet transport coefficients can generally be written as the correlations of gluon fields of the dense nuclear medium, they should be able to provide a lot of information about the internal structures of and the interaction nature of the QGP matter produced in high-energy heavy-ion collisions.

In terms of the three-momentum distribution $f(l_q^-, \mathbf{l}_{q\perp}, L^-)$ of the propagating parton, where $l_q^-$, $\mathbf{l}_{q\perp}$ and $L^-$ are light-cone energy, the transverse momentum, and the light cone time (length), the effect of multiple elastic scatterings experienced by the hard parton can be described via the following evolution equation \cite{Qin:2012fua, Abir:2014sxa},
\begin{eqnarray}
\label{drag_diffusion}
\frac{\partial f}{\partial L^-} = \left[ D_{L1}\frac{\partial}{\partial l_q^-} + \frac{1}{2} D_{L2} \frac{\partial^2}{\partial^2 l_q^-} + \frac{1}{2} D_{T2} {\nabla_{l_{q\perp}}^2}  + \cdots \right] f(L^-, l_q^-, \mathbf{l}_{q\perp}).
\end{eqnarray}
The above equation is cut off up to the second order in the gradient expansion of exchanged momentum, and the higher order terms are contained in ``$\cdots$".
The coefficients $D_{L1}$, $D_{L2}$ and $D_{T2}$ are defined in the light-cone coordinate; they are related to $\hat{e}$, $\hat{e}_2$ and $\hat{q}$ as follows:
\begin{eqnarray}
\hat{e} = D_{L1}, & \hat{e}_2 = D_{L2}/\sqrt{2}, & \hat{q} = 2\sqrt{2} D_{T2}.
\end{eqnarray}
Using the initial condition $f(l_q^-, \mathbf{l}_{q\perp}, L^- = 0) =  \delta(l_q^- - q^-) \delta^2(\mathbf{l}_{q\perp})$, the solution for $f(l_q^-, \mathbf{l}_{q\perp}, L^-)$ at a later time $L^-$ can be obtained as,
\begin{eqnarray}
f(l_q^-, \mathbf{l}_{q\perp}, L^-) = \frac{e^{-{(l_q^- - q^- + D_{L1} L^-)^2}/({2D_{L2}L^-})}}{\sqrt{2\pi D_{L2} L^-}}  \frac{e^{-{{l}_{q\perp}^2 }/({2D_{T2}L^-})}}{2\pi D_{T2} L^-}.
\end{eqnarray}
One can see that the parton momentum distribution is a Gaussian, due to the fact that only up to the second order terms are included in the evolution equation.
From the momentum distribution, one may obtain
\begin{eqnarray}
\langle l_q^- \rangle  = q^- - D_{L1} L^-, & \langle (l_q^-)^2 \rangle - \langle l_q^- \rangle^2 = D_{L2} L^-, & \langle l_{q\perp}^2 \rangle  = 2 D_{T2} L^- .
\end{eqnarray}
Therefore, the three terms in Eq (\ref{drag_diffusion}) represent the contributions from the longitudinal momentum loss and diffusion, and the diffusion of transverse momentum of the propagating hard parton.

Particular interest has been paid to the jet quenching parameter $\hat{q}$.
It not only quantifies the transverse momentum squared exchanged between propagating jets and the nuclear medium, but also plays an important role in describing medium-induced radiative energy loss.
The medium modification of radiative process is thought to be dominated by the transverse momentum exchange between the propagating hard jet with the constituents of the dense medium.
The importance of jet quenching parameter $\hat{q}$ has also been emphasized earlier in Ref. \cite{Majumder:2007zh} in which a general relation between the jet parameter $\hat{q}$ and the shear viscosity $\eta$ is derived for weakly-coupled partonic plasmas such as the QGP at sufficiently high temperature.
In a weakly-coupled scenario, the properties of the medium can be described perturbatively on the basis of an appropriate, partonic quasi-particle picture.
Since the shear viscosity to entropy ratio $\eta/s$ saturates in the strong coupling limit as suggested by the Super Yang-Mills (SYM) calculations, while $\hat{q}/T^3$ continues to increase, it is argued that $\hat{q}/T^3$ may serve as a better and more broadly applicable measure of the coupling strength of a QGP than the transport coefficient $\eta/s$.
In addition, a general relation between $T^3/\hat{q}$ and $\eta/s$ is conjectured in Ref. \cite{Majumder:2007zh} as follows:
\begin{eqnarray}
\frac{T^3}{\hat{q}} = \left\{
                                \begin{array}{ll}
                                  \approx \frac{\eta}{s}, & \hbox{weakly-coupled,} \\
                                  \ll \frac{\eta}{s}, & \hbox{strongly-coupled.}
                                \end{array}
                              \right.
\end{eqnarray}
This relation indicates that the precise determination of both quantities ($\hat{q}/T^3$ and $\eta/s$) by systematic phenomenological studies will allow us to quantitatively estimate the interaction strength (nature) of the QGP produced in ultra-relativistic heavy-ion collisions, e.g., when and how a weakly-coupled quark gluon system at sufficiently high temperatures turns into a strongly-coupled fluid at the energies achieved by RHIC and the LHC heavy-ion experiments.

\subsection{Phenomenological studies at RHIC and the LHC}

The study of jet quenching in the dynamically evolving heavy-ion collisions needs a variety of components: the initial hard parton spectrum, a realistic description of the density profiles of the medium, the modeling of jet propagation in dense nuclear medium, and the hadronization (fragmentation) of jets after passing through the medium.
In spite of the complexity involved, tremendous phenomenological studies have been performed in recent years, see e.g., Ref. \cite{Bass:2008rv,Armesto:2009zi,Chen:2010te,Zhang:2007ja,Renk:2008xq,Qin:2009bk,Renk:2011aa,Betz:2014cza,Djordjevic:2014tka,Kang:2014xsa}, in which various jet quenching observables have been investigated utilizing different jet energy loss model calculations.

A significant collaborative effort on the phenomenological study of jet quenching at RHIC and the LHC was recently performed within the framework of JET Collaboration \cite{Burke:2013yra}.
In this survey, within five different existing approaches to parton propagation and medium-induced energy loss in dense medium, systematic phenomenological studies was carried out on the experimental data on the nuclear modification of single inclusive hadrons at large $p_T$ in relativistic heavy-ion collisions at the RHIC and the LHC .
The space-time evolution of the QGP medium used in the study was simulated by the (2+1)-dimensional or (3+1)-dimensional hydrodynamic models which have been constrained by the experimental data on the bulk hadron spectra and anisotropic collective flow.
The goal of this collaborative study is to quantitatively extract the jet quenching parameter $\hat{q}$ and its systematic uncertainties due to model dependence by using the constraint from the experimental data.
The five parton energy loss approaches used for such study are: McGill-AMY \cite{Qin:2007rn}, Martini-AMY \cite{Young:2011ug}, HT-M \cite{Majumder:2011uk}, HT-BW \cite{Chen:2011vt}, DGLV-CUJET \cite{Xu:2014ica}.
In the following we provide some details about different model implementations.

\subsubsection*{DGLV-CUJET model}

The DGLV-CUJET code \cite{Xu:2014ica} was developed at Columbia University as part of JET Collaboration.
It includes both radiative energy loss and elastic energy loss based on DGLV formalism with dynamical scattering centers.
The code takes into account the effects of multi-scale running of the QCD coupling $\alpha_s(Q^2)$ in the DGLV opacity expansion series.

At the first order in the opacity expansion, the medium-induced radiative gluon distribution in the running coupling DGLV (rcDGLV) formalism is given by,
\begin{eqnarray}
x\frac{dN_g}{dx}(\mathbf{r}, \tau) &&\!= \int d\tau \rho_g(\mathbf{r}+\hat{n}(\phi)\tau, \tau) \int \frac{d^2q_\perp}{\pi} \frac{d\sigma}{d^2q_\perp} \int \frac{d^2k_\perp}{\pi} \alpha_s({k_\perp^2}/{x(1-x)})
\nonumber\\ &&\! \times
\frac{12(\mathbf{k}_\perp + \mathbf{q}_\perp)}{(\mathbf{k}_\perp + \mathbf{q}_\perp)^2 + \chi(\tau)} \cdot \left( \frac{\mathbf{k}_\perp + \mathbf{q}_\perp}{(\mathbf{k}_\perp + \mathbf{q}_\perp)^2 + \chi(\tau)} - \frac{\mathbf{k}_\perp}{\mathbf{k}_\perp^2 + \chi(\tau)}\right)
\nonumber\\ &&\! \times
\left( 1-\cos\left[\frac{(\mathbf{k}_\perp + \mathbf{q}_\perp)^2 + \chi(\tau)}{2x_+ E} \right] \right),
\end{eqnarray}
where $x$ and $\mathbf{k}_\perp$ are the fractional energy and transverse momentum of the radiated gluon with respect to the parent parton, $\mathbf{q}_\perp$ is the transverse momentum exchanged with medium constituents in a single scattering.
$\rho_g(\mathbf{r}+\hat{n}\tau, \tau)$ is the local gluon density of the medium, with $\mathbf{r}$ the location of the hard parton, $\hat{n}(\phi)$ the propagation direction in transverse plane, and $\tau$ the evolution time.
The effective differential quark-gluon cross section $d\sigma/d^2q_\perp$ is given by
\begin{eqnarray}
\frac{d\sigma}{d^2 q_\perp} = \frac{\alpha_s^2(q_\perp^2)}{(q_\perp^2 + f_E^2 \mu^2(\tau)) (q_\perp^2 + f_M^2 \mu^2(\tau))},
\end{eqnarray}
where $\mu^2(\tau) = 4\pi \alpha_s(4T^2) T^2$ is the local HTL color electric Debye screening mass squared calculated for a pure gluonic plasma.
In the above equation, the scale $\chi(\tau) = M^2 x_+^2 + f_E^2 \mu^2(T)(1-x_+)/\sqrt{2}$ is utilized to regulate infrared behaviors and control the dead cone and LPM destructive interference effects due to finite quark mass $M$ and a asymptotic thermal gluon mass $m_g = f_E \mu(T) /\sqrt{2}$.
$(f_E, f_M)$ is the HTL deformation parameters which controls the electric and magnetic screening scales related to the HTL.
The vacuum running coupling $\alpha_s(Q^2) = {\rm min}[\alpha_{\rm max}, 2\pi/9\ln(Q^2/\Lambda^2)]$ is used in the calculation.
In CUJET-GLV, there are three main model parameters: ($\alpha_{\rm max}$, $f_E$, $f_M$), and the default values for ($f_E$, $f_M$) are taken to be $(1,0)$.

In CUJET-GLV code,  for each production vertex and propagation direction $(\mathbf{r}, \hat{n})$, the Monte-Carlo integration method is used to evaluate the gluon emission spectra $dN_g/dx$ and average number of gluons $\langle N_g \rangle$. A Poisson ansatz is then utilized to estimate the effect of multiple gluon emission and to obtain the probability distribution of energy loss $P_{\rm rad}(\Delta E|E_0, \mathbf{r}, \hat{n})$,
\begin{eqnarray}
P_{\rm rad}(\Delta E|E_0, \mathbf{r}, \mathbf{n}) = \sum_{n=0}^{\infty} \frac{e^{-\langle N_g \rangle}}{n!} \left[ \prod_{i=1}^{n} \int dx \frac{dN_g}{dx} \right] \delta\left( \Delta E - \sum_{i=1}^{n} x_i E_0 \right).
\end{eqnarray}
The code also computes the elastic energy loss probability $P_{\rm el}(\Delta E|E_0, \mathbf{r}, \hat{n})$, which is convoluted with radiative contribution to get total energy loss distribution $P_{\rm tot}(\Delta E |E_0, \mathbf{r}, \hat{n})$.
The final total energy loss distribution $P_{\rm tot}(\Delta E |E_0, \mathbf{r}, \hat{n})$ is then convoluted with initial parton jet spectra $dN_{pp}/dp_T dy(p_T, y|\mathbf{r},\hat{n})$ to get the final modified parton spectra $dN_{AA}/dp_Tdy(p_T, y|\mathbf{r},\hat{n})$. By averaging over initial jet production configuration (which is determined by Glauber model) and fragmentation (and decay) functions, one gets final spectra and nuclear modification factors for hadrons (or leptons) to be compared with the experimental data.

\subsubsection*{HT-BW model}

In Higher-Twist-Berkeley-Wuhan (HT-BW) model \cite{Chen:2011vt}, the medium modified fragmentation function is given by
\begin{eqnarray}
\tilde{D}_{q\to h}(z_h, Q^2) &&\!= D_{q\to h}(z_h,Q^2)+ \frac{\alpha_s}{2\pi} \int_0^{Q^2} \frac{dl_\perp^2}{l_\perp^2}  \int_{z_h}^1 \frac{dz}{z}
 \nonumber\\&&\! \times
\left[ \Delta\gamma_{q\to qg}(z,l_\perp^2) D_{q\to h}(\frac{z_h}{z}) + \Delta\gamma_{q\to gq}(z, l_\perp^2)D_{g\to h}(\frac{z_h}{z}) \right],
\end{eqnarray}
where $z$ and $l_\perp$ are the momentum fraction and transverse momentum of the daughter partons with respect to the parent parton. The above medium modified fragmentation functions are very similar to the vacuum radiation corrections which lead to vacuum DGLAP evolution equation, except that now the splitting function $\Delta \gamma_{q \to qg}(z, l_\perp^2)$ and $\Delta \gamma_{q \to qq}(z, l_\perp^2)$ depend on the properties of the medium via jet transport coefficient $\hat{q}$.
More specifically, the medium-modified splitting function for a quark jet $\Delta\gamma_{q \to qg}(z, l_\perp^2)$ is given by
\begin{eqnarray}
\Delta\gamma_{q\to qg}(z, l_\perp^2) &&\!= C_A \frac{1}{l_\perp^2} \frac{1+z^2}{1-z} \left(1 - \frac{1-z}{2} \right)
\int dy^- \hat{q}(y^-) 4\sin^2\left[\frac{l_\perp^2 y}{4Ez(1-z)} \right],
\end{eqnarray}
where $\hat{q}(y^-)$ is the transport coefficient for a quark jet at a given location $y^-$.
Note that an extra factor of $1-(1-z)/2$ is to include the corrections beyond the helicity amplitude approximation \cite{Zhang:2003yn}.

In HT-BW calculation, the jet transport coefficient $\hat{q}$ is assumed to be proportional to the local parton density in the QGP phase or the local hadron density in the hadronic phase:
\begin{eqnarray}
\hat{q}(\tau, \mathbf{r}) = \left[\hat{q}_0 \frac{\rho_{QGP}(\tau, \mathbf{r})}{\rho_{QGP}(\tau_0,0)}(1-f_h) + \hat{q}_h(\tau, \mathbf{r}) f_h\right] \times \left(\frac{p\cdot u}{p_0}\right),
\end{eqnarray}
where $\tau$ and $\mathbf{r}$ are the time and the location. $\hat{q}_0$ is the jet transport parameter at the center of the bulk medium in QGP phase at initial time $\tau_0$, and $\rho_{QGP}$ is the parton density in an ideal gas at a given temperature, $f_h$ is the fraction of hadronic phase at any given time $\tau$ and space location $\mathbf{r}$.
The extra factor $(p\cdot u/p_0)$ is to take into account the effect of a dynamically evolving medium on the jet transport coefficient, with $u^\mu$ the local four flow velocity of the fluid.
Assuming a hadron resonance gas, the jet transport parameter in the hadron phase is approximated as:
\begin{eqnarray}
\hat{q}_h = \frac{\hat{q}_N}{\rho_N} \left[\frac{2}{3} \sum_M \rho_M(T) + \sum_B \rho_B(T) \right],
\end{eqnarray}
where $\rho_M$ and $\rho_B$ are the meson and baryon density at a given temperature, and $\rho_N = 0.17~{\rm fm}^{-3}$ is the nucleon density in the center of a large nucleus.
$\hat{q}_N$ is the jet transport parameter for a quark through a large nucleus and taken to be $\hat{q}_N = 0.02~{\rm GeV}^2/{\rm fm}$ from recent studies of deeply inelastic scattering (DIS) compared to HERMES data \cite{Airapetian:2007vu}.

\subsubsection*{HT-M model}

The Higher-Twist-Majumder (HT-M) approach \cite{Majumder:2011uk} is similar to the HT-BW model, but it includes multiple gluon emission through a set of effective modified QCD evolution equations.
The medium-modified fragmentation functions are obtained by solving the DGLAP-like evolution equations with a vacuum splitting kernel plus a medium-induced contribution. This model explicitly assumes the factorization of hard scattering cross section from the final fragmentation function.

In HT-M model, there are two contributions to the medium-modified fragmentation functions: one from vacuum evolution and the other from medium-modified evolution.
For the quark fragmentation function, the vacuum DGLAP evolution equations is given by,
\begin{eqnarray}
\frac{\partial D_{q\to h}(z, Q^2)}{\partial \ln Q^2} = \sum_i \frac{\alpha_s}{2\pi} \int_z^1 \frac{dy}{y} P_{q\to i}(y) D_{i \to h}(\frac{z}{y}, Q^2).
\end{eqnarray}
The medium modified DGLAP-like evolution equations read as follows,
\begin{eqnarray}
\frac{\partial D_{q\to h}(z, Q^2, q^-)|_{\zeta^-_i}^{\zeta^-_f}}{\partial \ln Q^2} &&\!= \sum_i \frac{\alpha_s}{2\pi} \int_z^1 \frac{dy}{y} P_{q\to i}(y) \nonumber\\&&\! \times \int_{\zeta^-_i}^{\zeta^-_f} d\zeta K(\zeta^-, y, Q^2, q^-) D_{i \to h}(\frac{z}{y}, Q^2, q^-y)|_{\zeta^-}^{\zeta^-_f}.
\end{eqnarray}
In the above equations, $P_{q\to i}(y)$ is the regular vacuum splitting function.
The medium modification effect is contained in the scattering kernel $K(\zeta^-, y, Q^2, q^-)$, with $\zeta^-$ the location of the scattering.
$\zeta^-_i$, $\zeta^-_f$ are the original and final locations of the hard parton.
The contribution to the scattering kernel $K(\zeta^-, y, Q^2, q^-)$ from the leading power correction is given by
\begin{eqnarray}
K(\zeta^-, y, Q^2, q^-) = \frac{[\hat{q}_A(\zeta^-) - (1-y)\hat{q}_A/2 + (1-y)^2 \hat{q}_F]}{Q^2} 4\sin^2\left[ \frac{Q^2(\zeta^--\zeta^-_i)}{4q^-y(1-y)}\right]. \ \
\end{eqnarray}
The jet transport coefficient $\hat{q}_A$ and $\hat{q}_F$ are for a gluon jet and a quark jet, and they are related to each other as $\hat{q}_F/\hat{q}_A = C_F/C_A$.

In the HT-M model, the jet quenching parameter $\hat{q}$ is assumed to scale with the entropy density of the dense nuclear medium (either QGP or hadronic phase),
\begin{eqnarray}
\frac{\hat{q}(s)}{\hat{q}_0} = \frac{s}{s_0},
\end{eqnarray}
where $s_0$ is the maximum entropy density achieved at an initial time $\tau_0$ at the center of the medium (produced in most central collisions at RHIC).

When solving both vacuum and medium modified evolution equations, the input fragmentation functions are needed. In HT-M model, the inputs are taken as the vacuum fragmentation functions at the input scale $Q_0^2 = p_T/L$, where $p_T$ is the transverse momentum of the parton and the length factor $L$ is the mean escape length of the hard jet obtained via using the single gluon emission kernel. Such input vacuum fragmentation functions are obtained according to the vacuum evolution equations from $Q_0^2 = 1~{\rm GeV}^2$.

\subsubsection*{McGill-AMY model}

In McGill-AMY model \cite{Qin:2007rn}, the medium-induced radiation and elastic scattering processes are described by the high temperature thermal field theory.
In the model calculation, the first step is to calculate the production of initial stage partons before the formation of thermal QGP.
The second step is to calculate the medium effects on the produced partons which propagate through the hot and dense QGP.
The model solves the a set of coupled transport rate equations for parton momentum distributions to obtain the medium modified parton momentum distributions.
The third step is to convolute the medium modified parton momentum distribution with vacuum fragmentation functions to give final hadron spectra in relativistic heavy-ion collisions.

In McGill-AMY model, the following transport rate equations are solved for the parton momentum distribution $f(p,t) = dN(p,t)/dp$:
\begin{eqnarray}
\frac{df_j(p)}{dt} = \sum_{a,b} \int dk \left[f_a(p+k) \frac{d\Gamma_{a\to j}(p+k,k)}{dkdt} - f_j(p) \frac{d\Gamma_{j\to b}(p,k)}{dkdt} \right],
\end{eqnarray}
where $d\Gamma_{j\to b}(p,k)/dkdt$ is the transition rate for the partonic process $j\to a$, with $p$ the initial parton energy and $k$ the lost energy in the splitting process. The transition rates are calculated in the full leading order thermal QCD with the inclusion of the HTL effects and LPM interference.
In the model calculation,  all the splitting processes are included and all the partons in the shower are kept track of until the partons fragment into hadrons outside the thermal medium.
The energy gained by the propagating parton due to the absorption of the thermal partons is included in the $k<0$ integration part.
The contribution from $2 \to 2$ elastic collisions with the medium constituents are included in a similar way.

In the model, the properties of the thermal QGP medium enter through the local temperature and flow velocity when calculating the parton transition rates. The interaction between the hard parton the the medium is controlled by the HTL resummed elastic collision rates:
\begin{eqnarray}
\frac{d\Gamma_{\rm el}}{d^2 \mathbf{q}_\perp} = \frac{C_s}{(2\pi)^2} \frac{g_s^2 m_D^2 T}{q_\perp^2 (q_\perp^2 + m_D^2)},
\end{eqnarray}
where $g_s = \sqrt{4\pi \alpha_s}$ is the coupling constant, and $m_D^2 = (N_c/3 + N_f/6) g^2 T^2$ is the Debye screening mass squared. The factor $C_s$ is the color factor of the propagating parton. Currently in the calculation, the coupling constant $\alpha_s$ is taken to be constant for all interaction vertices.
The scale dependence of the coupling as well as the finite size effect have not yet been included at present.

\subsubsection*{MARTINI-AMY model}

MARTINI-AMY model \cite{Young:2011ug} is a Monte-Carlo event generator for simulating jet production and the subsequent evolution in relativistic heavy-ion collisions. The production of initial jet partons is obtained by running PYTHIA-8 code from each nucleon-nucleon collision with the spatial distribution of the jet partons following the Glauber geometry.
The propagation of jet partons in the thermal medium is carried out by solving the same coupled rate equations as in McGill-AMY model, but using Monte-Carlo techniques. In addition, Martini-AMY code also includes the scale dependence of the strong coupling $\alpha_s$ for the radiation vertices via $\langle k_T^2 \rangle = (\hat{q} k)^{1/2}$. The parton transition rates also include the finite-size effects via a parameterization of the rates derived in Ref. \cite{CaronHuot:2010bp}.
\vspace{12pt}

\begin{figure}
\begin{center}
\includegraphics*[width=5.4cm,height=3.7cm]{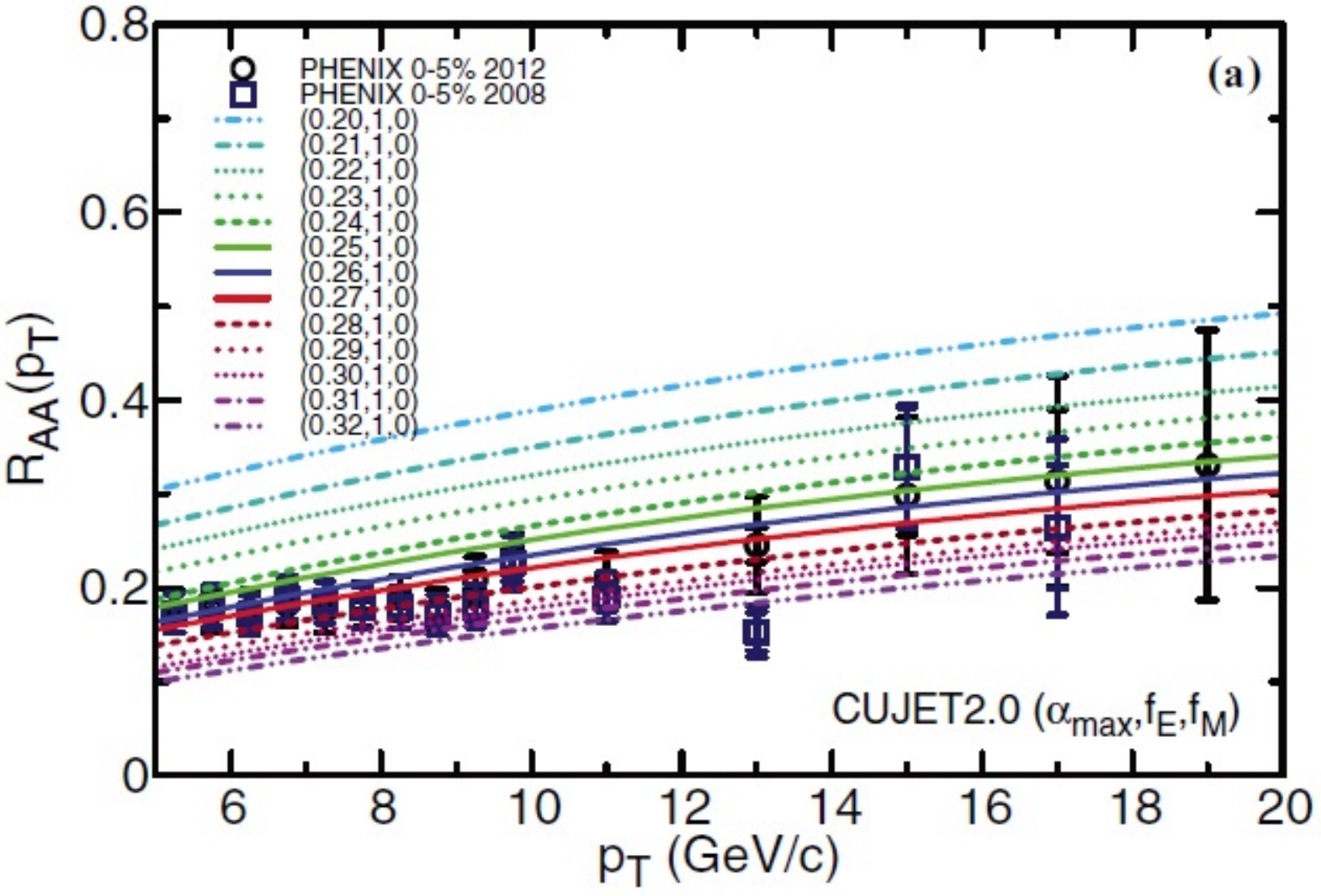}
\includegraphics*[width=5.4cm,height=3.7cm]{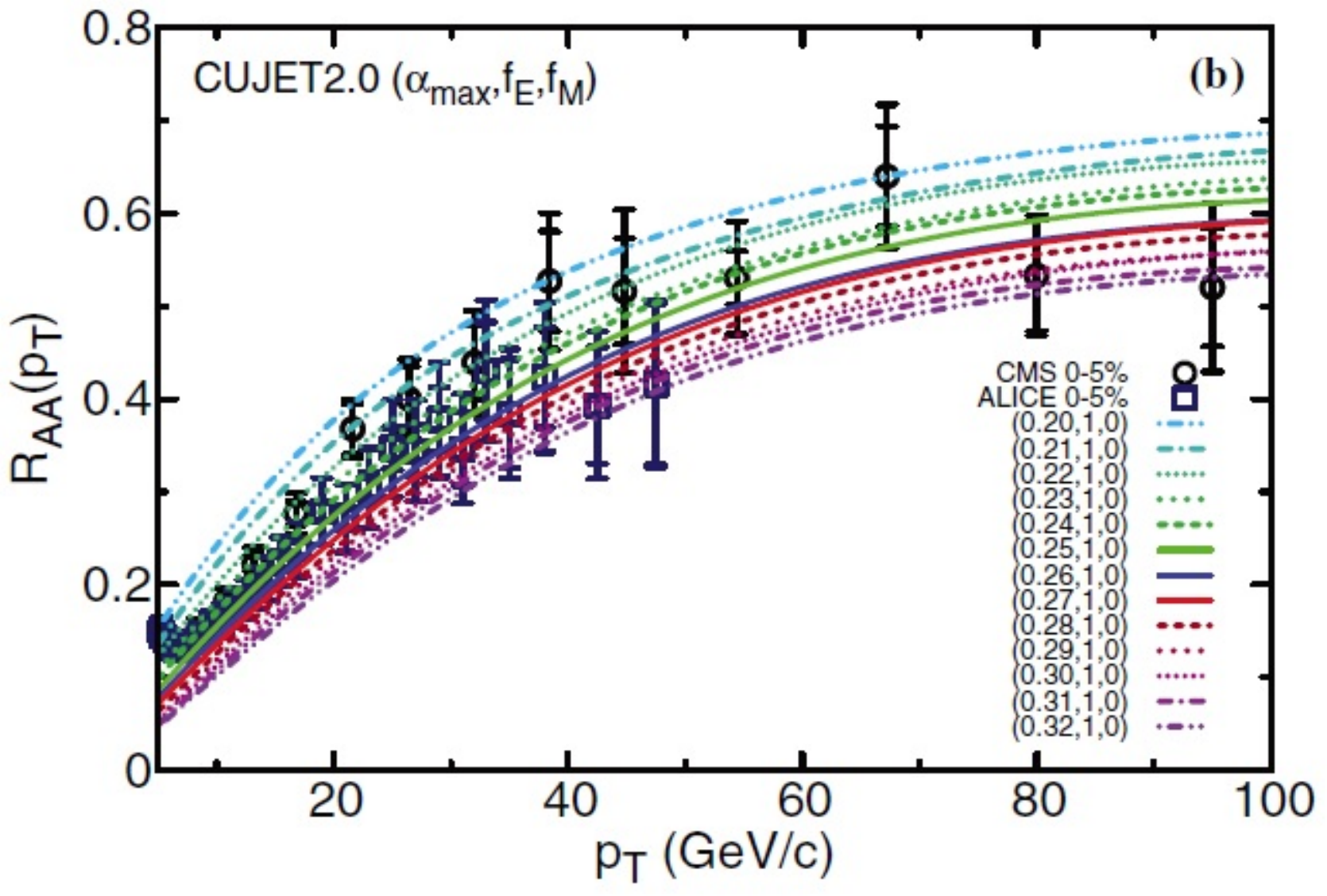}
\includegraphics*[width=5.4cm,height=3.7cm]{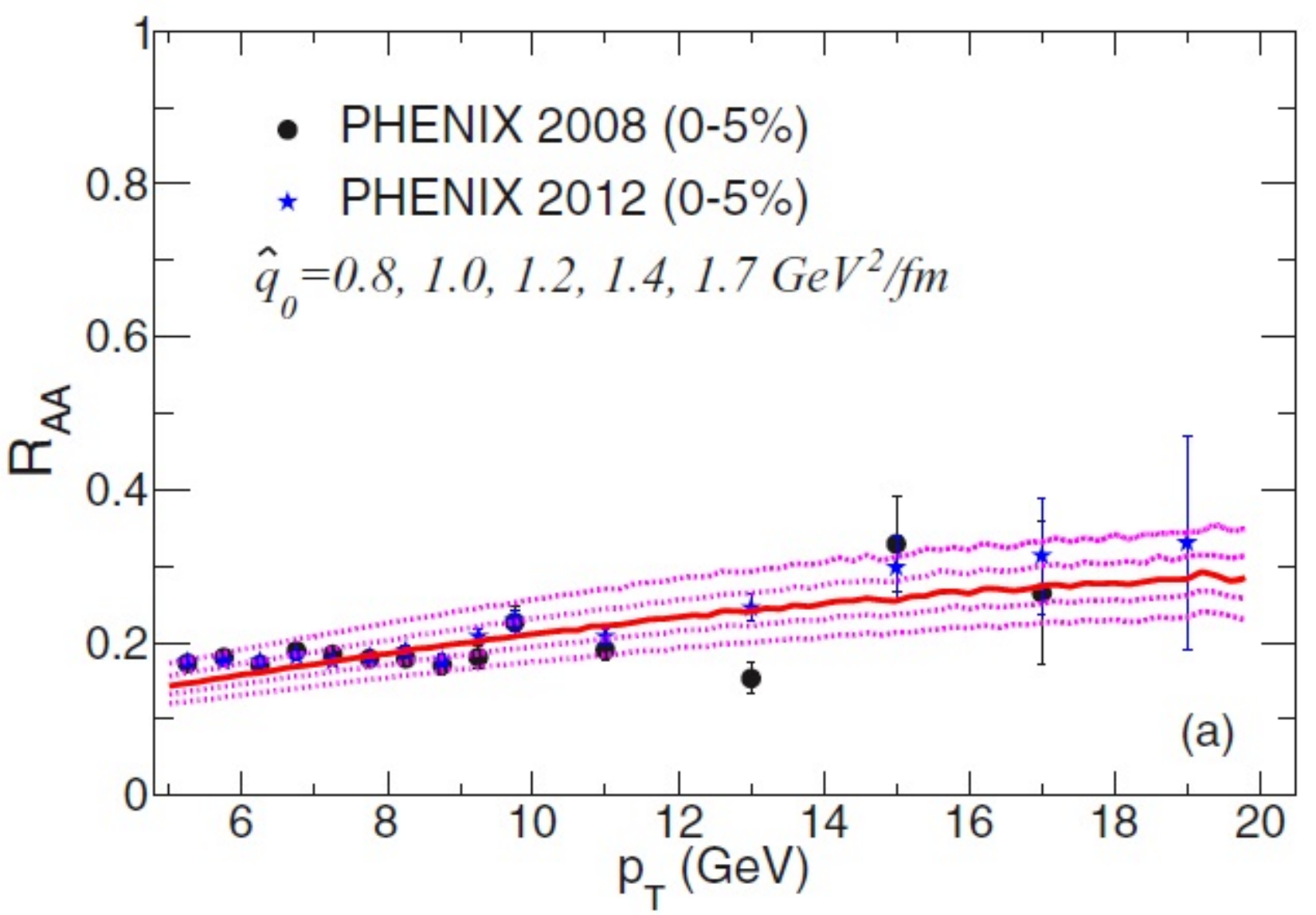}
\includegraphics*[width=5.4cm,height=3.7cm]{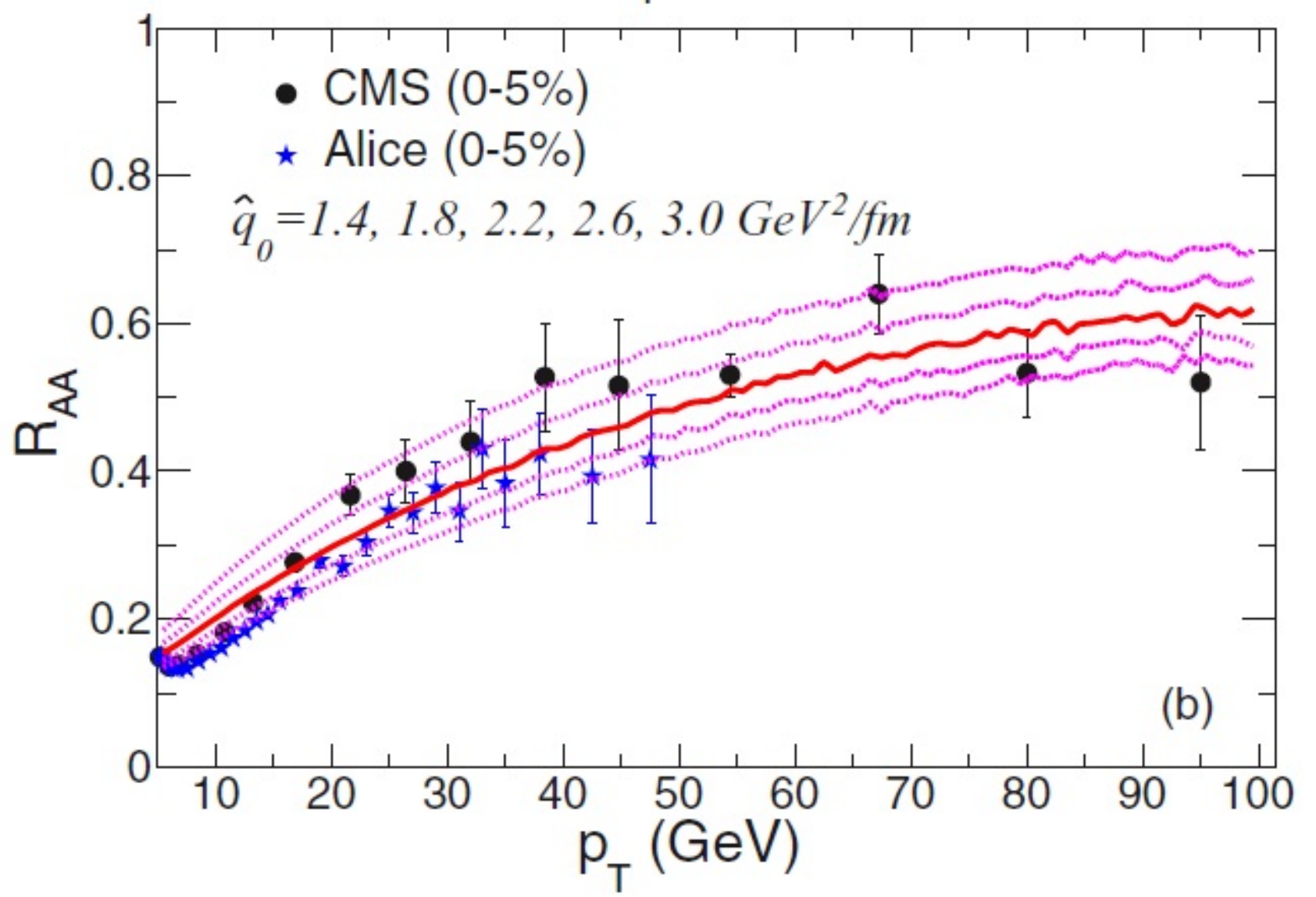}
\includegraphics*[width=5.6cm,height=3.7cm]{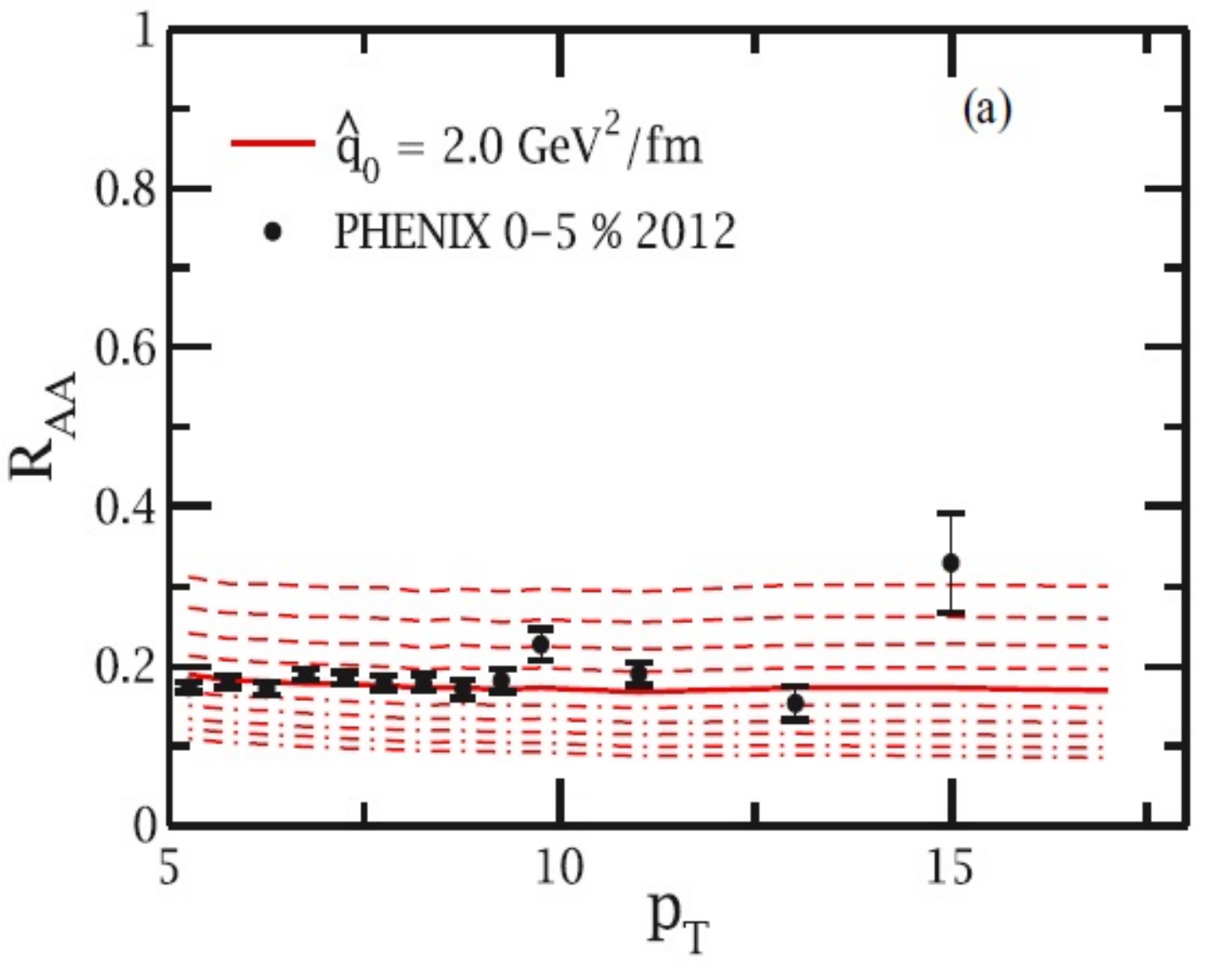}
\includegraphics*[width=5.6cm,height=3.7cm]{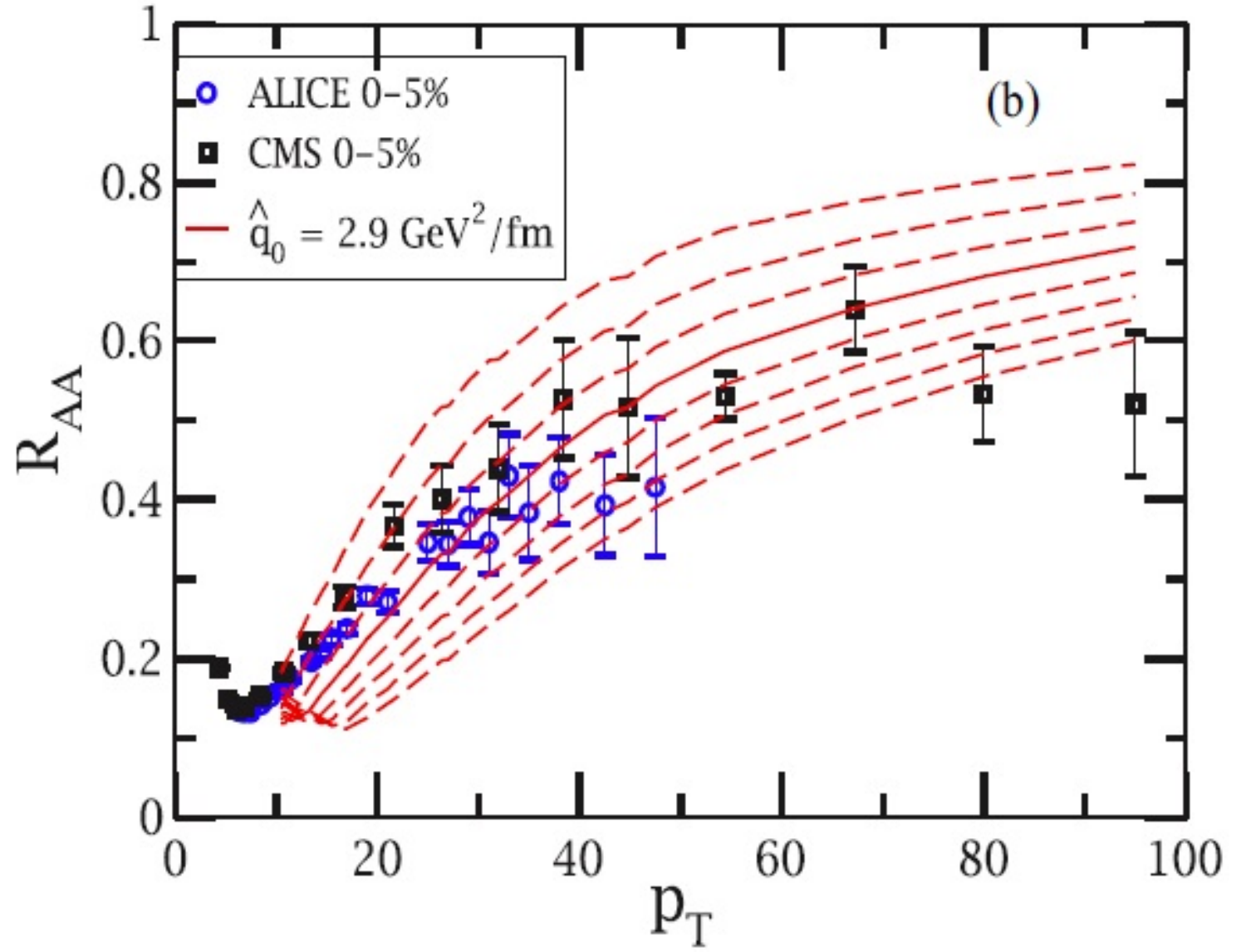}
\includegraphics*[width=5.4cm,height=3.7cm]{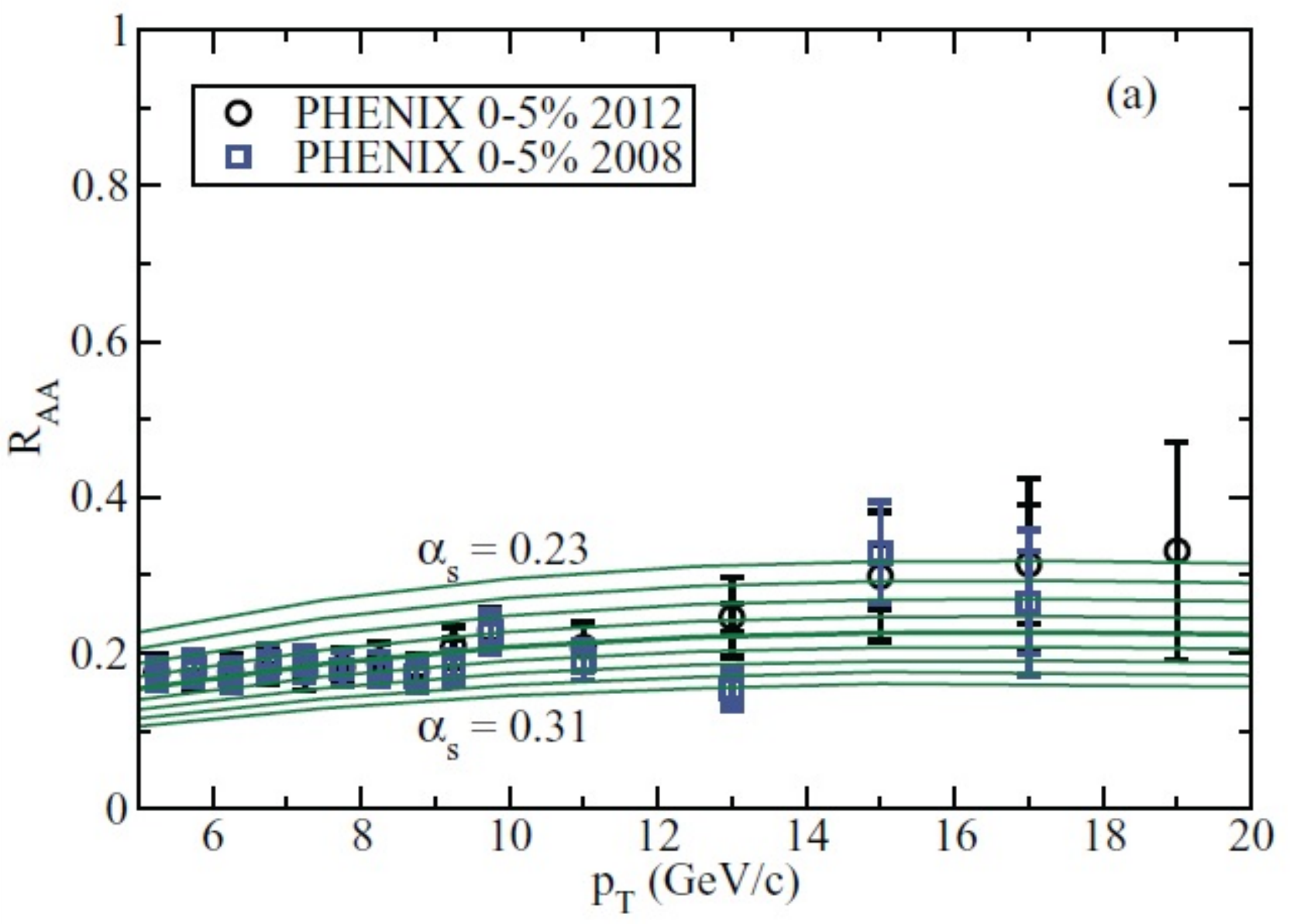}
\includegraphics*[width=5.4cm,height=3.7cm]{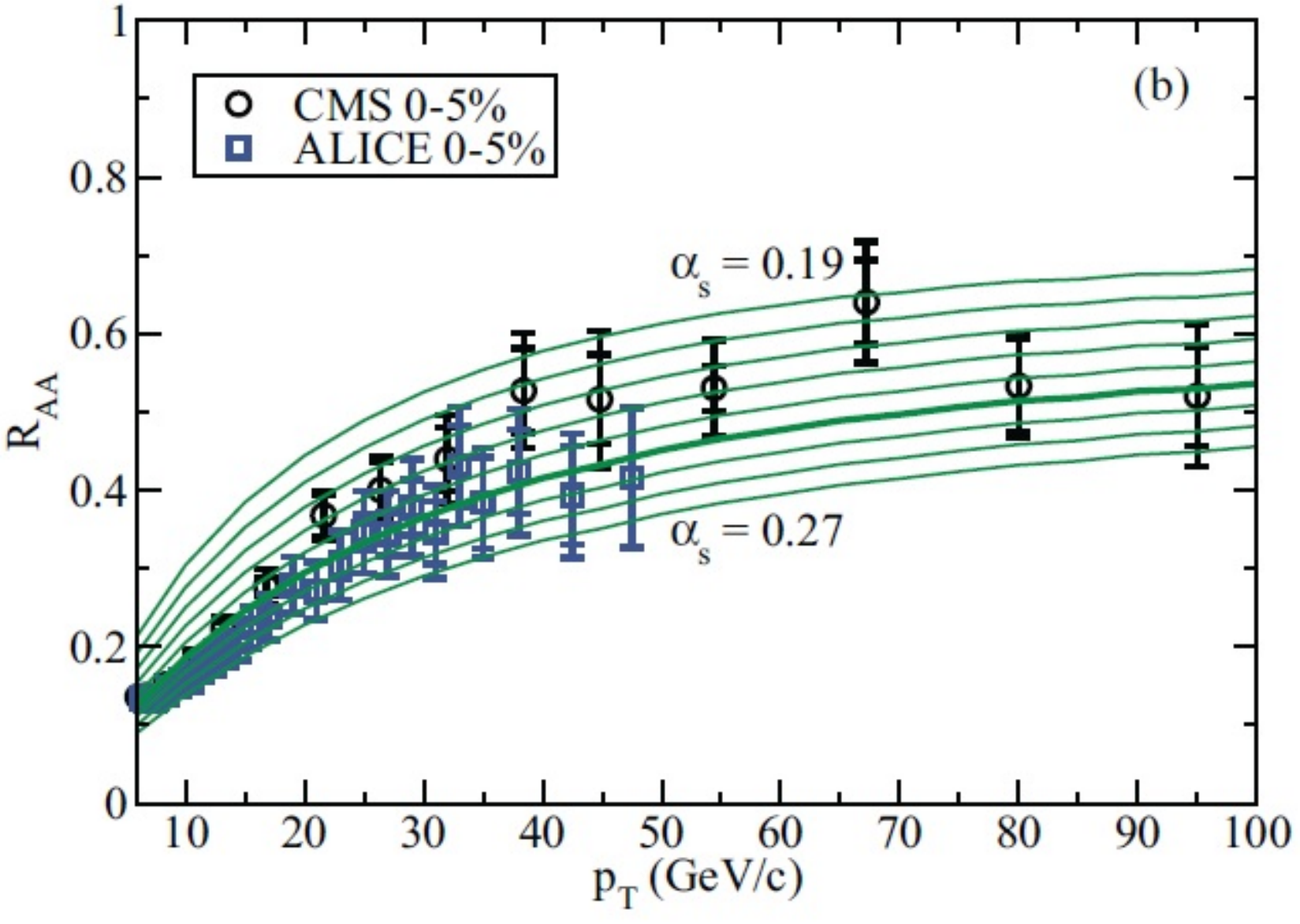}
\includegraphics*[width=5.6cm,height=3.7cm]{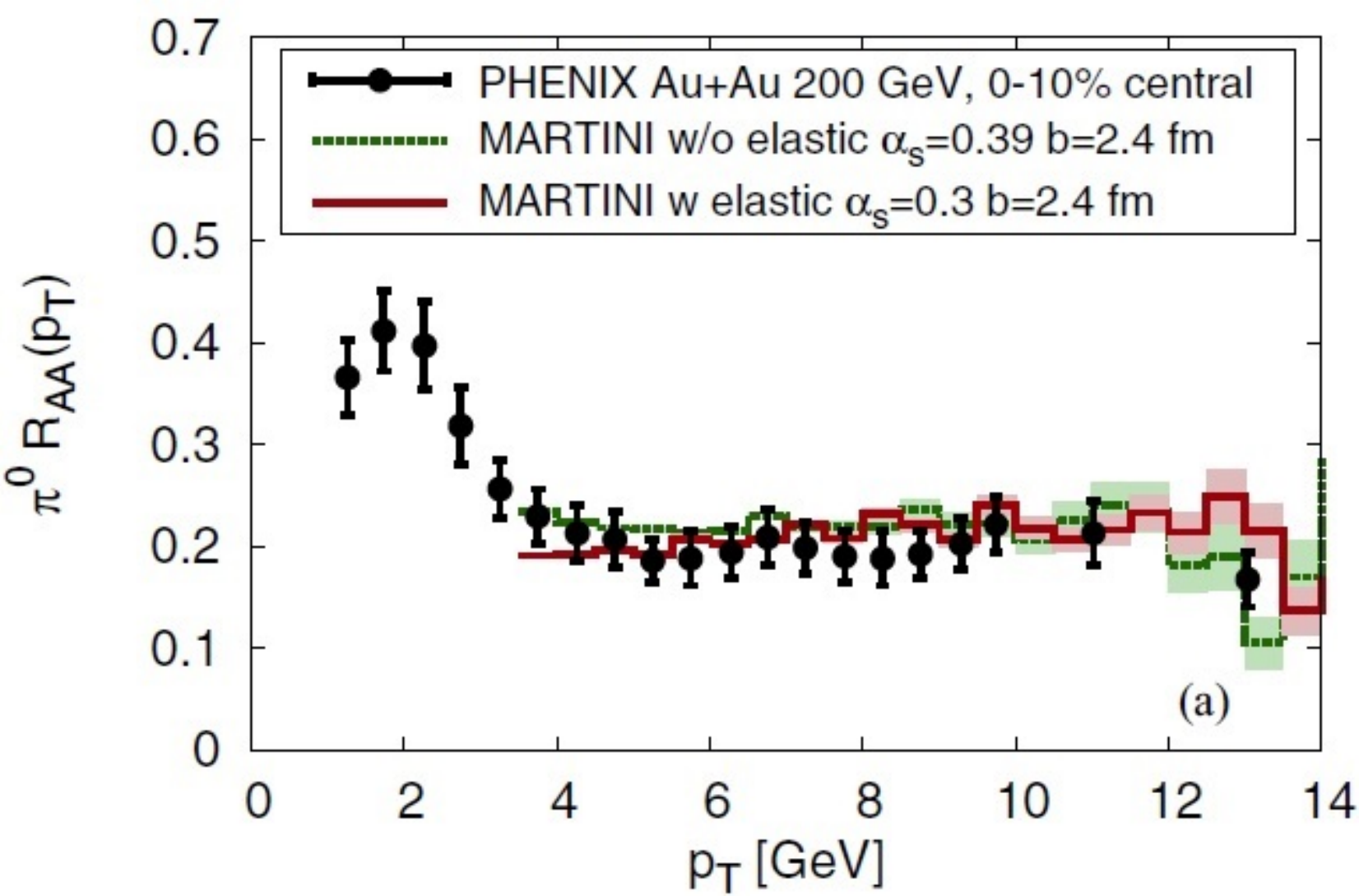}
\includegraphics*[width=5.6cm,height=3.7cm]{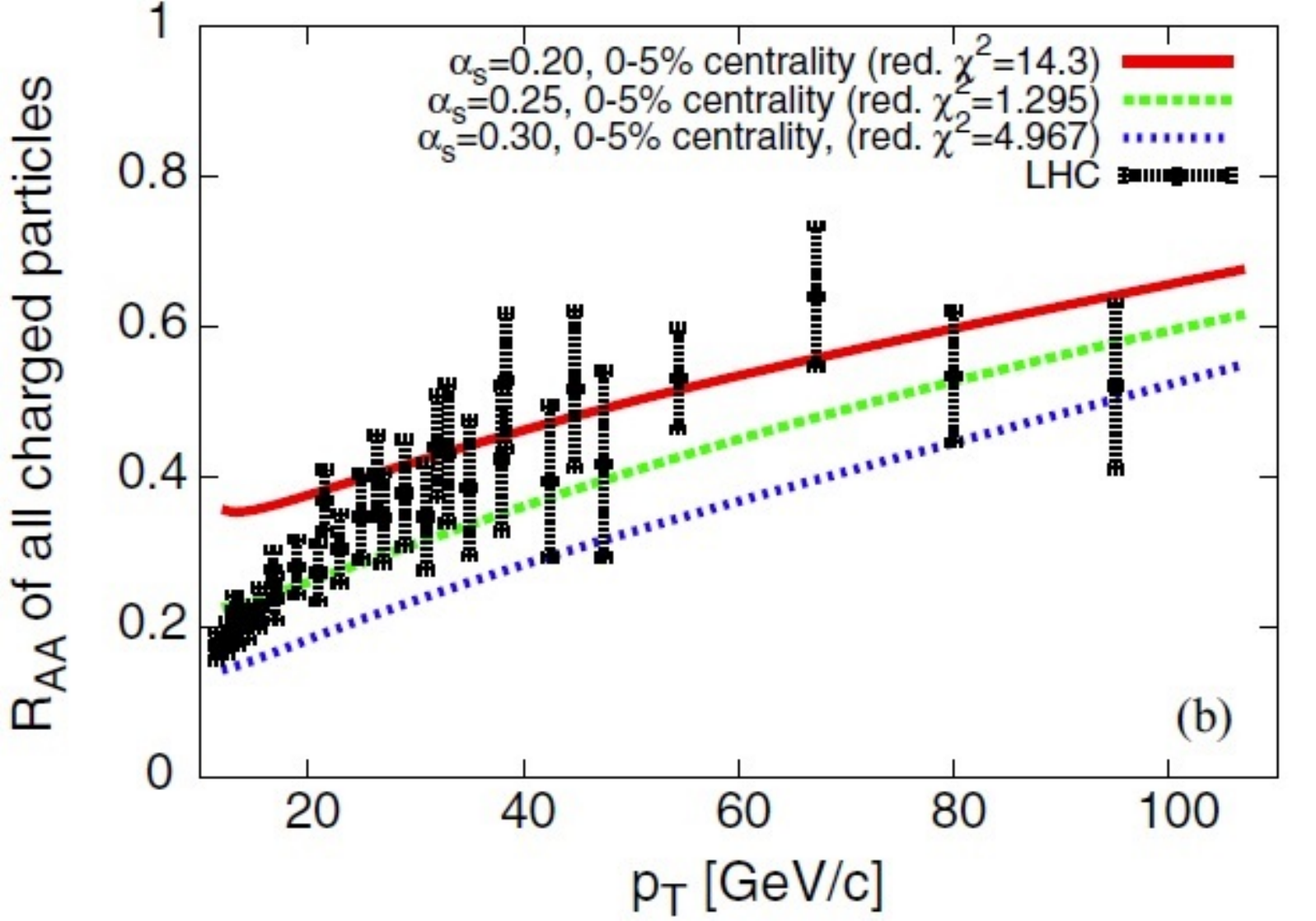}
\caption{(Color online) The nuclear modification factor $R_{AA}$ at both RHIC (left panels) and the LHC (right panels) compared with different model calculations. From top to bottom, the models are: CUJET-GLV, HT-BW, HT-M, McGill-AMY and MARTINI-AMY.
Figures are taken from Ref. \cite{Burke:2013yra} done by JET Collaboration.
}
\label{fig_RAA_RHIC_LHC}
\end{center}
\end{figure}

Within the framework of JET Collaboration, the above five jet quenching models are used to calculate the nuclear modification factor $R_{AA}$ for large $p_T$ single inclusive hadrons in most central collisions at RHIC and the LHC. The results are shown in Fig. \ref{fig_RAA_RHIC_LHC}, where the left panels are for RHIC $R_{AA}$ while the right for the LHC $R_{AA}$. From the top to the bottom are the results from five different models: CUJET-GLV, HT-BW, HT-M, McGill-AMY and MARTINI-AMY. Different curves in each plot are the calculations with different model parameters within each jet quenching approach.
For DGLV-CUJET the model parameters are $(\alpha_{\rm max}, f_E, f_M)$, for two HT calculations jet quenching coefficient $\hat{q}_0$ is the model parameter, and for two AMY calculations, the model parameter is strong coupling $\alpha_s$.
For all the calculations, the realistic ideal/viscous hydrodynamics simulations \cite{Hirano:2002ds, Song:2007ux, Qiu:2011hf} are utilzed for the space-time evolution of the bulk nuclear matter produced in RHIC and the LHC heavy-ion collisions.
The initial time for relativistic hydrodynamics evolution is set as $\tau_0 = 0.6$~fm/c with the initial temperatures set as $T_0 = 346-373$~MeV ($477-486$~MeV) at the center the hot and dense medium produced in most central Au+Au collisions at $\sqrt{s_{NN}} = 200$~GeV at RHIC (most central Pb+Pb collisions at $\sqrt{s_{NN}}=2.76$~TeV at the LHC).

By comparing to the experimental measurements of single inclusive hadron nuclear modification factor $R_{AA}$ at large $p_T$, one may obtain a tight constraint on the model parameters for each jet energy loss model calculation.
In DGLV-CUJET model, the best fits to the experimental data on $R_{AA}$ at RHIC and the LHC give: $\alpha_{\rm max} = 0.28$ for $T < 378$~MeV and $\alpha_{\rm max} = 0.24$ for $378 < T < 486$~MeV.
For HT-BW model, the best fits produce $\hat{q}_0 = 1.2$~GeV$^2$/fm at RHIC, and $\hat{q}_0 = 2.2$~GeV$^2$/fm at the LHC for the quark jet transport coefficients.
For HT-M model, the gluon jet transport coefficients are obtained from the best fits as: $\hat{q}_{0} = 2.0$~GeV$^2$/fm at RHIC, and $\hat{q}_0 = 2.9$~GeV$^2$/fm at the LHC.
For McGill-AMY model, the best fits give $\alpha_s = 0.27$ for RHIC and $\alpha_s = 0.24$ for the LHC.
For Martini-AMY model, $\alpha_s = 0.3$ for RHIC and $\alpha_s = 0.24$ for the LHC are obtained.

\subsection{Quantitative extraction of $\hat{q}$}
\label{sec_JET_qhat}

With the model parameters fixed by the experimental data, one may obtain the effective jet transport coefficient $\hat{q}$ for each model.
For CUJET calculation, it is obtained as follows:
\begin{eqnarray}
\hat{q}(E, T, \alpha_{\rm max}) = \rho_g(T) \int^{q_{\rm max}} dq_\perp^2 q_\perp^2 \frac{d\sigma}{dq_\perp^2},
\end{eqnarray}
where $q_{\rm max}$ is the ultra-violet cutoff and taken to be $q_{\rm max}= 6ET$.
One can see that the jet transport coefficient depends on jet energy $E$ and the medium temperature $T$.
The variation of the temperature $T$ also affects the electric and magnetic screening mass deformation parameters $(f_E, f_M)$.
For McGill-AMY and MARTINI-AMY models, the values of $\hat{q}$ may be obtained as follows:
\begin{eqnarray}
\hat{q} = C_s \alpha_s m_D^2 T \ln(1 + q_{\rm max}^2/m_D^2).
\end{eqnarray}
For HT-BW and HT-M models, the comparisons to experimental data directly give the jet quenching parameter $\hat{q}$.

\begin{figure}
\begin{center}
\includegraphics*[width=10.cm]{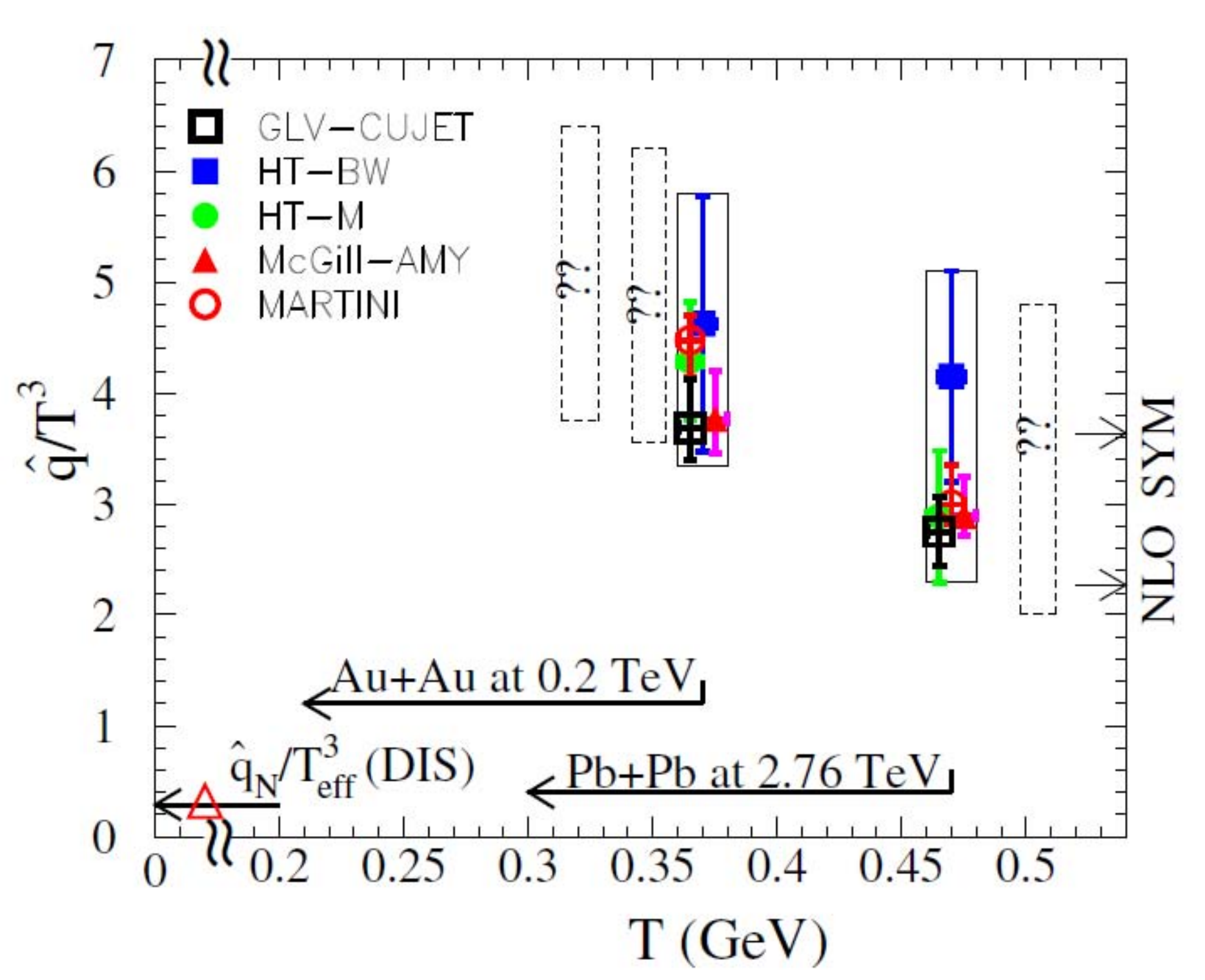}
\caption{(Color online) The extracted values of scaled jet transport parameter $\hat{q}/T^3$ by using single inclusive hadron suppression factor $R_{AA}$ at both RHIC and LHC.
The shown values are for a quark jet with initial energy of $10$~GeV at the center of most central A-A collisions at an initial proper time $\tau_0 = 0.6$~fm/c.
The figure is taken from Ref. \cite{Burke:2013yra} done by JET Collaboration.
}
\label{fig_JET_qhat}
\end{center}
\end{figure}

In Fig. \ref{fig_JET_qhat}, the extracted values for jet quenching parameter $\hat{q}$ scaled by $T^3$ are shown for a quark jet with an energy $E=10$~GeV at the center of most central A-A collisions at an initial proper time $\tau_0 = 0.6$~fm/c at RHIC and the LHC energies (denoted by two solid boxes).
The range of $\hat{q}$ values as constrained by the measured single hadron nuclear modification factors $R_{AA}$ at RHIC and the LHC is obtained as:
\begin{eqnarray}
\frac{\hat{q}}{C_s T^3} = \left\{
                                \begin{array}{ll}
                                  3.5 \pm 0.9, & \hbox{$T \approx 370$ {\rm MeV (at RHIC)},} \\
                                  2.8 \pm 1.1, & \hbox{$T \approx 470$ {\rm MeV (at the LHC)}.}
                                \end{array}
                              \right..
\end{eqnarray}
Translating these values for $\hat{q}$, one obtains: $\hat{q} \approx$ 1.2 and 1.9~GeV$^2$~fm for a quark jet with energy $E=10$~GeV at the highest temperatures reached in the most central Au+Au collisions $\sqrt{s_{NN}} = 200$~GeV at RHIC and Pb+Pb collisions $\sqrt{s_{NN}} = 2.76$~TeV at the LHC.

One can clearly see that the scaled dimensionless quantity $\hat{q}/T^3$ is temperature dependent.
Such temperature dependence can be explored further by extending the current study to the future higher energy Pb-Pb collisions at the LHC and lower energy collisions at RHIC.
The expected values of $\hat{q}/T^3$ in A-A collisions at 0.063~ATeV, 0.130~ATeV and 5.5~ATeV are shown by dashed boxes.
As a comparison, the figure also shows the value of $\hat{q}_N/T_{\rm eff}^3$ in cold nuclei (indicated by the triangle in the figure) as extracted from jet quenching studies in DIS \cite{Majumder:2004pt}, where the value of $\hat{q}_N = 0.02$~GeV$^2$/fm is taken, and the effective temperature is taken assuming an idea quark gas with 3 quarks within each nucleon in a large nucleus.
We can see that the values of $\hat{q}$ in the hot and dense QGP matter are much higher than those in cold nuclei.

The above extracted values of $\hat{q}$ are also compared to the values obtained from a next-to-leading order AdS/CFT calculation \cite{Zhang:2012jd} (shown by two arrows on the right axis of the figure). By utilizing the AdS/CFT correspondence, one may calculate the jet quenching parameter $\hat{q}$ in a $\mathcal{N} = 4$ supersymmetric Yang-Mills (SYM) plasma at the strong coupling limit.
At the leading order (large $N_c$ and large 't Hooft coupling $\lambda$), $\hat{q}$ may be obtained as \cite{Liu:2006ug}:
\begin{eqnarray}
\hat{q}_{SYM}^{LO} = \frac{\pi^{3/2} \Gamma(3/4)}{\Gamma(5/4)} \sqrt{\lambda} T_{SYM}^3,
\end{eqnarray}
where $\lambda = g_{\rm SYM}^2 N_c$. Including the corrections from finite 't Hooft coupling, the jet quenching parameter at next-to-leading order is obtained as \cite{Zhang:2012jd},
\begin{eqnarray}
\hat{q}_{SYM}^{NLO} = \hat{q}_{SYM}^{LO}\left(1 - \frac{1.97}{\sqrt{\lambda}} \right).
\end{eqnarray}
The SYM values shown in the figure are obtained for $\alpha_{SYM} =$~0.22-0.31 and have taken into account different numbers of degrees of freedom in SYM and 3 flavor QCD. One can see that the values of $\hat{q}_{SYM}$ from SYM calculations are within the range of $\hat{q}$ extracted from comparing phenomenological studies to experimental data on $R_{AA}$ at RHIC and the LHC done by JET Collaboration.

In order to fully take into account the systematic uncertainties due to model dependence, more jet quenching calculations (see e.g., Ref.  \cite{Djordjevic:2014tka, Kang:2014xsa, Ficnar:2013qxa, Armesto:2009fj, Zapp:2008gi, Renk:2010zx}) should be included in the future for such systematic phenomenological studies as done by the JET Collaboration.
More experimental observables, besides the single inclusive hadron $R_{AA}$, should be incorporated for analysis so that one can obtain tighter constraints on the modeling of jet quenching and on the quantitative extraction of jet transport coefficients.

\subsection{$\hat{q}$ from Lattice QCD}
\label{sec_LQCD_qhat}

While lattice QCD calcuation cannot provide a full description of the dynamical evolution of heavy-ion collisions and jet transport in the hot and dense nuclear medium, it can provide important guidance for jet quenching studies and many insights for our understanding of jet-medium interaction.
For example, jet transport coefficients (such as $\hat{q}$) can in principle be computed with lattice QCD; this will provide very useful reference for phenomenological jet quenching studies.

The first calculation of jet quenching parameter $\hat{q}$ within the framework of finite temperature lattice gauge theory was performed in Ref. \cite{Majumder:2012sh}.
The operator expression for $\hat{q}$ is quoted here:
\begin{eqnarray}
\hat{q} && \!\! = \frac{4\pi^2 \alpha_s}{N_c} \int \frac{dy^- d^2y_\perp d^2k_\perp}{(2\pi)^3} e^{i\frac{k_\perp^2}{2q^-}y^- - i\mathbf{k}_\perp \cdot \mathbf{y}_\perp} \nonumber\\ && \!\! \times \left\langle {\rm Tr} \left[ t^a F_\perp^{a+\mu}(y^-, \mathbf{y}_\perp) U^\dagger(\infty^-, \mathbf{y}_\perp; 0^-, \mathbf{y}_\perp) T^\dagger(\infty^-, \mathbf{\infty}_\perp; \infty^-, \mathbf{y}_\perp) \right.\right.
\nonumber\\ && \!\! \times \left.\left. T(\infty^-, \mathbf{\infty}_\perp; \infty^-, \mathbf{0}_\perp) U(\infty^-, \mathbf{0}_\perp; 0^-, \mathbf{0}_\perp) t^b F_{\perp \mu}^{b+}(0^-, \mathbf{0}_\perp) \right] \right\rangle,
\end{eqnarray}
where $U$ denotes a Wilson line along the ``$-$" light-cone direction and $T$ denotes a Wilson line along the transverse ``$\perp$" light-cone direction.
Ref. \cite{Majumder:2012sh} utilizes the operator product expansion method, and relates the jet transport parameter $\hat{q}$ as obtained in the physical regime of jet momenta to an infinite series local operators in an unphyical regime of jet momenta via dispersion relations.
Then these local operators are computed in quenched $SU(2)$ lattice gauge theory.
After extrapolating the calculation in quenched $SU(2)$ to the case of $SU(3)$ with $2$ flavors of quarks, $\hat{q}$=1.3-3.3~GeV$^2$/fm is obtained for a gluon jet at a temperature $T=400$~MeV.
This value is within the range of $\hat{q}$ extracted from phenomenological studies done by JET Collaboration as shown in Sec. \ref{sec_JET_qhat}.
In a recent non-perturbative lattice calculation \cite{Panero:2013pla} based on a dimensionally reduced effective theory (electrostatic QCD), which only includes the contribution from soft QGP modes, a value of $\hat{q}$=6~GeV$^2$/fm was obtained at RHIC energies. This is about twice the value obtained from NLO perturbative QCD result \cite{CaronHuot:2008ni}.

\begin{figure}
\begin{center}
\includegraphics*[width=12.5cm]{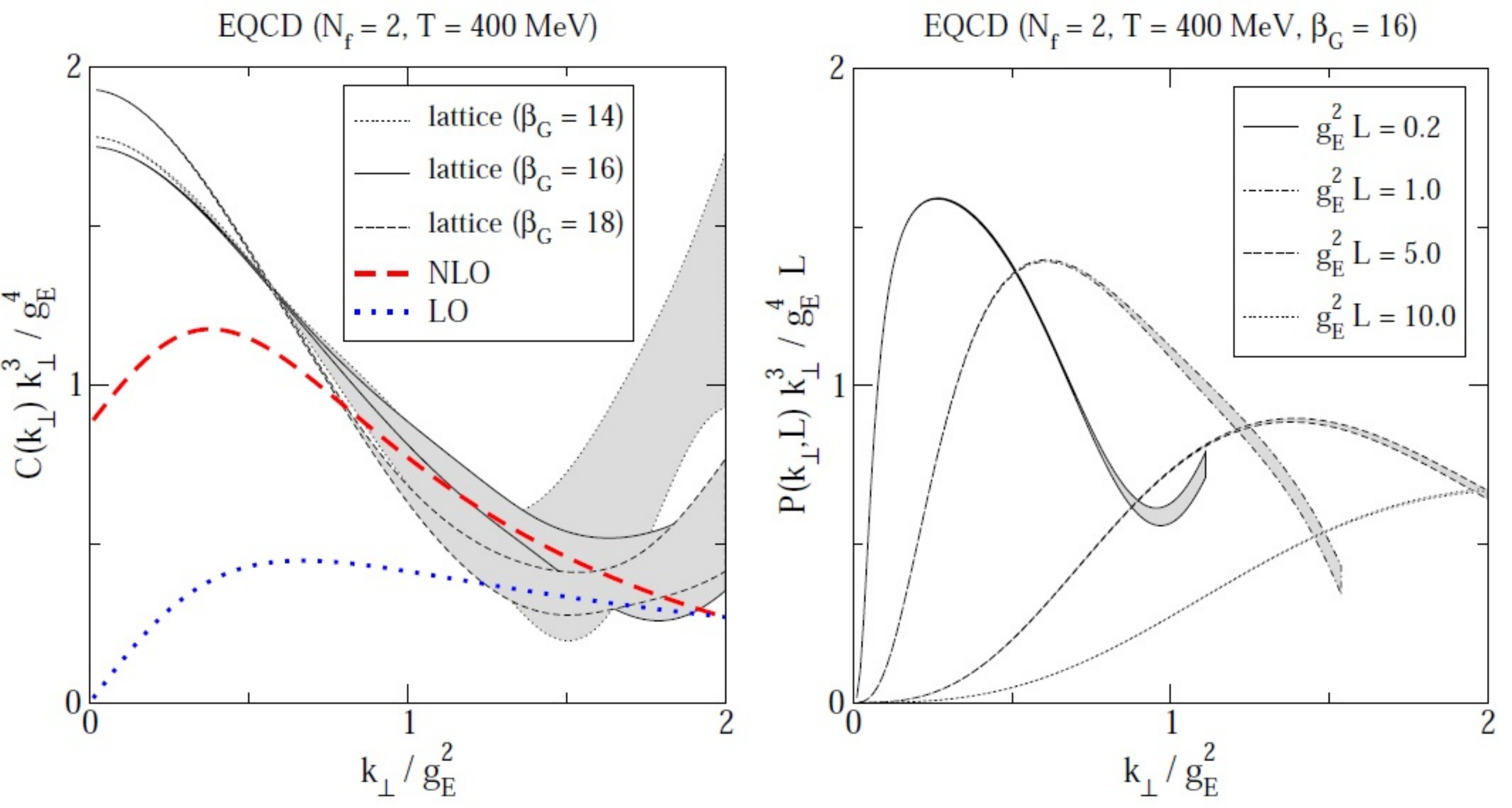}
\caption{(Color online) The transverse collision kernel $C(k_\perp)$ (left) using lattice data from Ref. \cite{Panero:2013pla} compared with the NLO calculation result from Ref. \cite{CaronHuot:2008ni} and the probability distribution $P(k_\perp,L)$ (right).
$g_E^2 \sim g^2T$ is the effective coupling of the dimensionally reduced ¡°EQCD¡± effective theory.
$\beta_G$ is a dimensionless number and related to the lattice spacing $a$ as: $\beta_G = 2N_c/(g^2T a)$.
Figures are taken from Ref. \cite{Laine:2013apa}.
}
\label{fig_EQCD_C(k)_P(k)}
\end{center}
\end{figure}

It should be noted that jet quenching parameter $\hat{q}$ only encodes the second moment of the probability distribution $P(k_\perp, L)$ for the transverse momentum transfer,
\begin{eqnarray}
\hat{q} = \frac{1}{L} \int \frac{d^2k_\perp}{(2\pi)^2} k_\perp^2 P(\mathbf{k}_\perp, L),
\end{eqnarray}
where $L$ is the length that jet traversed.
The probability distribution $P(k_\perp, L)$ satisfies the following evolution equation,
\begin{eqnarray}
\frac{dP(\mathbf{k}_\perp, L)}{dL} = \int \frac{d^2q_\perp}{(2\pi)^2}  C(\mathbf{q}_\perp) \left[P(\mathbf{k}_\perp - \mathbf{q}_\perp) - P(\mathbf{k}_\perp)\right],
\end{eqnarray}
where $C(q_\perp)$ is the transverse collision kernel.
A full distribution of the probability distribution $P(k_\perp)$ or the collision kernel $C(k_\perp)$ will provide more detailed information about jet-medium interaction.
A first-principle calculation of $C(k_\perp)$ and $P(k_\perp, L)$ was carried out in Ref. \cite{Laine:2013apa} and the results are shown in Fig. \ref{fig_EQCD_C(k)_P(k)}.
It is found that the shape of the transverse collision kernel $k_\perp^3 C(k_\perp)$ is consistent with a Gaussian at small transverse momentum $k_\perp$.
As $k_\perp$ becomes large, the calculation breaks down as it only includes the soft modes.

The Gaussian form at small $k_\perp$ for the probability distribution $P(k_\perp)$ is expected in both weakly-coupled and strongly-coupled pictures.
However at large $k_\perp$, the distribution $P(k_\perp)$ takes very different form in these two scenarios: for weakly coupled quark-gluon plasma, the large $k_\perp$ power tail is proportional to $1/k_\perp^2$ , while in the plasma of $\mathcal{N} = 4$ SYM theory in the large-$N_c$ and strong coupling limits, there is no power tail at all \cite{D'Eramo:2012jh}.
So it would be very interesting and useful if lattice calculation as well as phenomenological jet quenching studies can provide the full distribution for the collision kernel or the probability distribution.

\subsection{Renormalization of $\hat{q}$ and NLO developments}
\label{sec_NLO_qhat}

Parton energy loss models based on perturbative QCD techniques have been very successful in studying jet quenching phenomena in relativistic heavy-ion collisions at RHIC and the LHC.
The comparisons between systematic phenomenological studies using various theoretical models and experimental data have led to a tight determination of jet transport coefficients.
However, there still exist strong model dependence due to various approximations applied in the jet quenching calculations.
One important reason is that all the phenomenological jet quenching studies have utilized leading-order parton energy loss formalisms.
In fact, many model differences have been found to originate from to some specific approximations made in the leading order calculation of the gluon radiation spectrum, such as eikonal, soft and collinear approximations \cite{Armesto:2011ht}.

To constrain jet quenching models and better our understanding of jet-medium interaction, the effects of different approximations need to be explored in more detail, e.g., by relaxing these approximations when building parton energy loss formalisms.
The effect of non-eikonal large angle scattering approximation was recently investigated in Ref. \cite{Abir:2013ph} and it is found to be small for jet energies larger than 10-15~GeV.
Ref. \cite{Apolinario:2014csa} re-study the medium-induced radiation by going beyond eikonal approximation and include the finite energy effect in the new formalism.
However, in order to achieve a systematic estimation of the errors made in the leading-order calculation, it is necessary to build a fully next-to-leading order (NLO) framework for studying parton propagation in dense nuclear matter and jet quenching in relativistic heavy-ion collisions.

\begin{figure}
\begin{center}
\includegraphics*[width=12.cm]{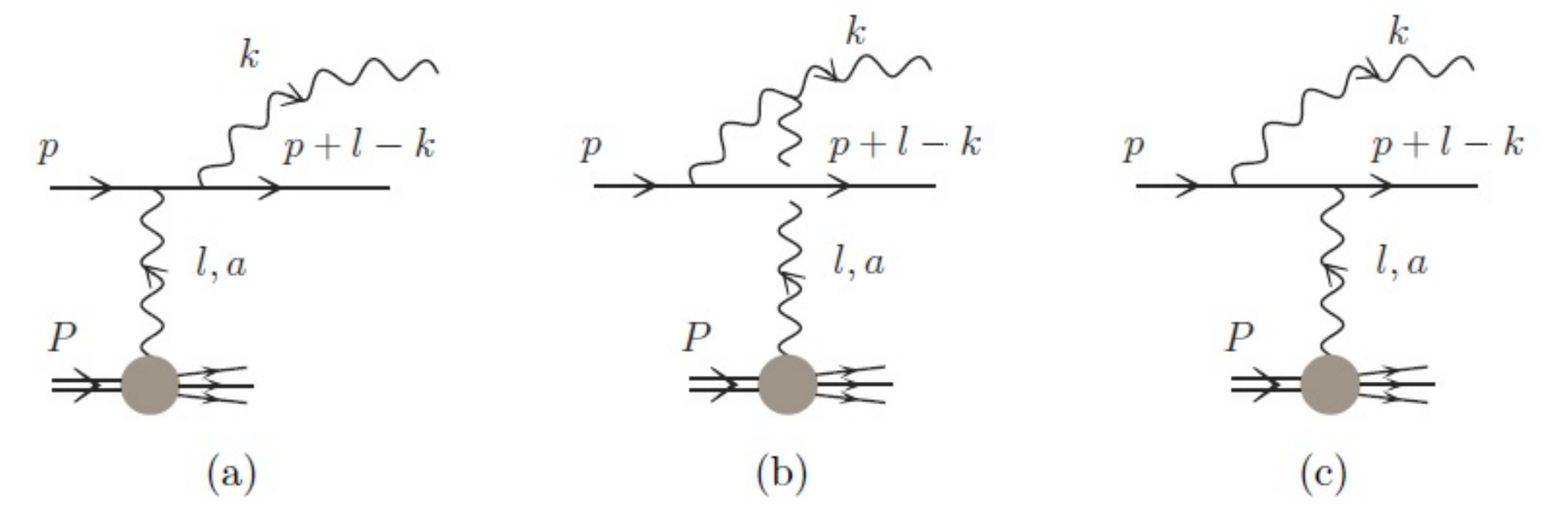}
\caption{(Color online) Induced gluon radiation of a high-energy quark by a single scattering.
}
\label{fig_LMW}
\end{center}
\end{figure}

The first attempt along NLO direction was performed in Ref. \cite{Wu:2011kc, Liou:2013qya}, in which the NLO radiative correction to the transverse momentum broadening is computed.
Consider a high-energy parton propagating through a dense nuclear medium, it experiences both elastic and inelastic collisions with the medium constituents.
Focusing on the transverse momentum broadening for the propagating jet parton, the leading order contribution comes from $2 \to 2$ elastic collisions.
At next-to-leading order, one has to consider the contribution from medium-induced radiation, as shown in Fig. \ref{fig_LMW}.

The calculation was performed in the framework of BDMPS-Z formalism and it was found that the transverse momentum broadening receives a sizable contribution from radiative correction due to the recoil effect of one-gluon emission in the kinetic region of single scattering..
In particular for large media size, the radiative contribution is dominated by the double logarithmic terms.
The double logarithmic correction reads as follows:
\begin{eqnarray}
\langle p_\perp^2 \rangle_{\rm rad} = \frac{\alpha_s N_c}{\pi} \hat{q}_0 L  \int_{\hat{q}_0\tau_0}^{\hat{q}_0 L} \frac{dk_\perp^2}{k_\perp^2} \int_{\frac{\hat{q}_0 \tau_0}{k_\perp^2}}^1 \frac{dx}{x} = \frac{\alpha_s N_c}{2\pi} \hat{q}_0 L \ln^2\left(\frac{L}{\tau_0} \right),
\end{eqnarray}
where $\hat{q}_0$ is the transverse momentum broadening from pure elastic collisions, $L$ is the length of the nuclear matter, and $\tau_0$ is some intrinsic size associated with the constituents of the nuclear matter (which was introduced to regulate the logarithmic divergence in the radiative corrections).
The jet transport coefficient $\hat{q}$ also receives a double logarithmic correction,
\begin{eqnarray}
\hat{q}_{\rm rad} = \frac{\alpha_s N_c}{\pi} \hat{q}_0  \int_{\tau_0}^{L} \frac{d\tau}{\tau} \int_{\hat{q}_0\tau}^{\hat{q}_0 L} \frac{dk_\perp^2}{k_\perp^2} =  \frac{\alpha_s N_c}{2\pi} \hat{q}_0 \ln^2\left(\frac{L}{\tau_0} \right),
\end{eqnarray}
The role of the large double logarithms corrections was further investigated in Ref. \cite{Iancu:2014kga, Blaizot:2014bha}.
It was argued that such double logarithmic correction may be absorbed by a redefinition/renormalization of the jet quenching parameter $\hat{q}$,
\begin{eqnarray}
\hat{q}(\tau) = \hat{q}_0 \left[ 1 + \frac{\alpha_s N_c}{2\pi}\ln^2\left(\frac{\tau}{\tau_0}\right) \right],
\end{eqnarray}
where $\tau$ may be understood as the size of the medium constituents "seen" or probed by the propagating parton.
The infrared cut-off $\tau_0$ can be viewed as a factorization scale in this problem.
With such redefinition of jet quenching parameter, one may write an evolution equation for $\hat{q}(\tau)$ as a function of the scale $\tau$,
\begin{eqnarray}
\frac{\partial \hat{q}(\tau, Q^2)}{\partial \ln \tau} = \int_{\hat{q}_0 \tau}^{Q^2} \frac{dk_\perp^2}{k_\perp^2} \frac{\alpha_s N_c}{\pi} \hat{q}(\tau, k_\perp^2).
\end{eqnarray}
It is further argued that the scale evolution of $\hat{q}$ will render additional medium-length dependence for jet quenching parameter, parton transverse momentum broadening as well as parton energy loss as compared to traditional BDMPS jet energy loss formalism, i.e.,
\begin{eqnarray}
\hat{q}(L) \propto L^{\gamma}\,, \langle p_\perp^2 \rangle \propto \hat{q}_0 L^{1+\gamma} \,, \Delta E \propto \hat{q}_0 L^{2+\gamma},
\end{eqnarray}
where the extra dimension $\gamma$ is found to be $\gamma = 2\sqrt{\alpha_s N_c/\pi}$ for large medium size.

\begin{figure}
\begin{center}
\includegraphics*[width=12.5cm, height=3.0cm]{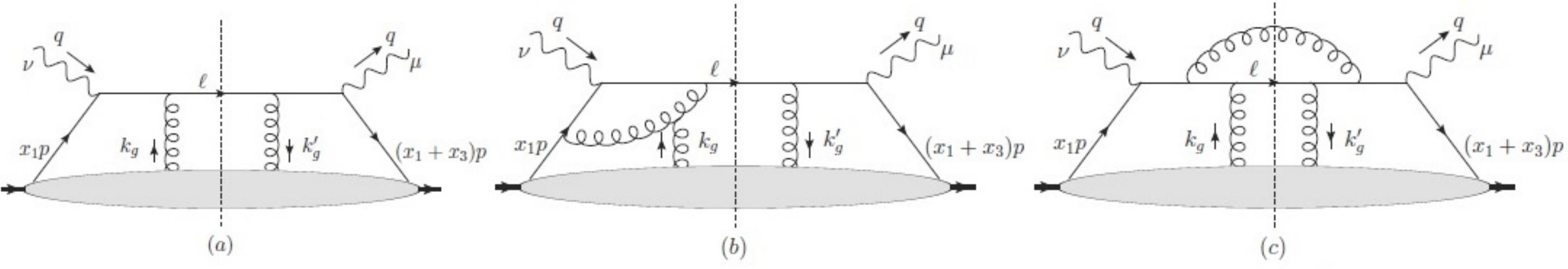}
\caption{(Color online) Some typical diagrams for double scattering contributions to the $k_\perp^2$-weighted differential cross section from leading-order (a), NLO virtual (b), and NLO real (c) processes.
}
\label{fig_Xing}
\end{center}
\end{figure}

The NLO radiative corrections to the transverse momentum broadening and the renormalization of jet quenching parameter $\hat{q}$ have also been studied in Ref. \cite{Kang:2013raa, Kang:2014ela}. This work was performed in the framework of hadron production in semi-inclusive deep-inelastic scattering and lepton pair production in p-A collisions (see Fig. \ref{fig_Xing}).
The cross section weighted with $l_{h\perp}^2$ was computed and a factorization formula was obtained after absorbing the collinear divergence into the redefinitions of nonpertubative correlation functions.
The factorized form for the NLO correction to $l_{h\perp}^2$-weighted cross section reads as:
\begin{eqnarray}
\frac{1}{\sigma_0} \left.\frac{d\langle l_{h\perp}^2 \sigma \rangle}{dz}\right|_{NLO} = T_{qg}(x, 0, 0, \mu_f^2) \otimes H_{NLO}(x, x_B, Q^2, \mu_f^2) \otimes D_{h/q}(z_h, \mu_f^2) + \cdots, \ \ \
\end{eqnarray}
where $T_{qg}$ is the twist-4 quark-gluon correlation function, $D$ is the quark-to-hadron fragmentation function, $H_{NLO}$ is the NLO correction to the hard part.
The twist-4 quark-gluon correlation function $T_{qg}$ is non-perturbative and can not be directly calculated using perturbative QCD.
But its scale dependence may be derived from perturbative QCD, which leads to a new QCD evolution equation (similar to DGLAP equations for PDFs).
The scale dependence of twist-4 quark-gluon correlation function reads:
\begin{eqnarray}
\frac{\partial T_{qg}(x_B, 0, 0, \mu_f^2)}{\partial \ln \mu_f^2} = \frac{\alpha_s}{2\pi} \int_{x_B}^1 \frac{dx}{x} \left[ P_{qg \to qg} \otimes T_{qg} + P_{qg}(\hat{x}) T_{gg}(x, 0, 0, \mu_f^2) \right],
\end{eqnarray}
where $\mathcal{P}_{qg \to qg} \otimes T_{qg}$ is given by
\begin{eqnarray}
\mathcal{P}_{qg \to qg} \otimes T_{qg} &&\!\!= P_{qq}(\hat{x})T_{qg}(x,0,0)  +  \frac{C_A}{2}\left( \frac{4}{(1-\hat{x})_+} T_{qg}(x_B,x-x_B,0) \right. \nonumber\\&&\!\! \left. - \frac{1+\hat{x}}{(1-\hat{x})_+} T_{qg}(x,0, x_B-x) +  T_{qg}(x_B,x-x_B,x-x_B)\right).
\end{eqnarray}
Here $P_{qq}(\hat{x})$ and $P_{qg}(\hat{x})$ are the usual splitting functions in the DGLAP evolution equations, with $\hat{x} = x_B/x$.
From the above new QCD evolution equation, the scale dependence of jet quenching parameter $\hat{q}$ may be in principle obtained if one assumes the nucleus is composed of loosely-bounded nucleons. Neglecting the momentum and spatial correlations of two nucleons inside the nucleus, one may relate the twist-4 quark-gluon correlation function $T_{qg}$ to the jet quenching parameter $\hat{q}$ as follows:
\begin{eqnarray}
T_{qg}(x_B, 0, 0, \mu_f^2) \approx \frac{N_c}{4\pi^2 \alpha_s} f_{q/A}(x_B, \mu_f^2) \int dy^- \hat{q}(y^-, \mu_f^2).
\end{eqnarray}
Unfortunately, the above evolution equation is not closed, and a closed evolution equation can only be obtained for certain kinematic limit.
In the limit of large $x_B$ ($x_B \to 1$), the formation time of the radiated gluon is much large than the size of the nuclear medium, so gluon radiation is suppressed due to the LPM interference effect. In this case, the splitting kernel $\mathcal{P}_{qg \to qg} \otimes T_{qg}$ reduces to the vacuum one $P_{qq}(\hat{x})T_{qg}(x,0,0)$, and one gets scale-independent $\hat{q}$.
For intermediate-$x_B$ regime, one may expand the off-diagonal matrix elements $T_{qg}$ around $x = x_B$ and only keep the leading contribution. Then the scale dependence of $\hat{q}$ can be obtained as follows \cite{Xing:2014qm},
\begin{eqnarray}
\frac{\partial \hat{q}(\mu^2)}{\partial \ln \mu^2} = \frac{\alpha_s}{2\pi} C_A \ln(1/x_B) \hat{q}(\mu^2).
\end{eqnarray}
One may see that the jet transport coefficient $\hat{q}$ is now not only $\mu$-dependent, but also $x_B$-dependent.
The solution of such equation is
\begin{eqnarray}
\hat{q}(\mu^2) = \hat{q}(\mu_0^2) \exp\left[\frac{\alpha_s}{2\pi} C_A \ln(1/x_B) \ln(\mu^2/\mu_0^2) \right].
\end{eqnarray}
This $\mu^2$ dependence of $\hat{q}$ essentially is similar to the violation of scaling behavior of normal PDFs. The $x_B=Q^2/(2ME)$-dependence may be translated to the energy dependence of $\hat{q}$ in the target rest frame, which is consistent with the earlier expectations \cite{CasalderreySolana:2007sw}.

\section{Full jet evolution and modification}
\label{sec_full_jet}

The study of full jets and their nuclear modification can provide many insights into the mechanisms of jet-medium interaction and parton energy loss in ultra-relativistic heavy-ion collisions.
The basic idea of reconstructing full jets is to recombine final state hadron fragments and to infer the information about the original hard partons and the medium effects.
Since both leading and sub-leading fragments of full jets are included in the reconstructed full jets, they are more differential than single inclusive observables.
Therefore they are expected to have the ability to put more stringent constraints on various approximations and assumptions made in jet energy loss calculations, which will be very helpful to eliminate the large systematic uncertainties in the quantitative extraction of the jet transport parameters.
One of the biggest challenges in studying reconstructed full jets in ultra-relativistic heavy-ion collision is the contamination from fluctuations of large bulk background, as compared to elementary proton-proton collisions.
The development of various experimental techniques in recent years have now enabled us to disentangle the reconstructed full jets from fluctuating background.
The availability of many full jet related experimental observables have triggered a lot of theoretical activities as well.
The goal of full jet study is to understand how full jets interact with medium constituents and lose energy, and how the medium effects manifest in the final state observables.

\subsection{Full jet studies at RHIC}

\begin{figure}
\begin{center}
\includegraphics*[width=6.4cm]{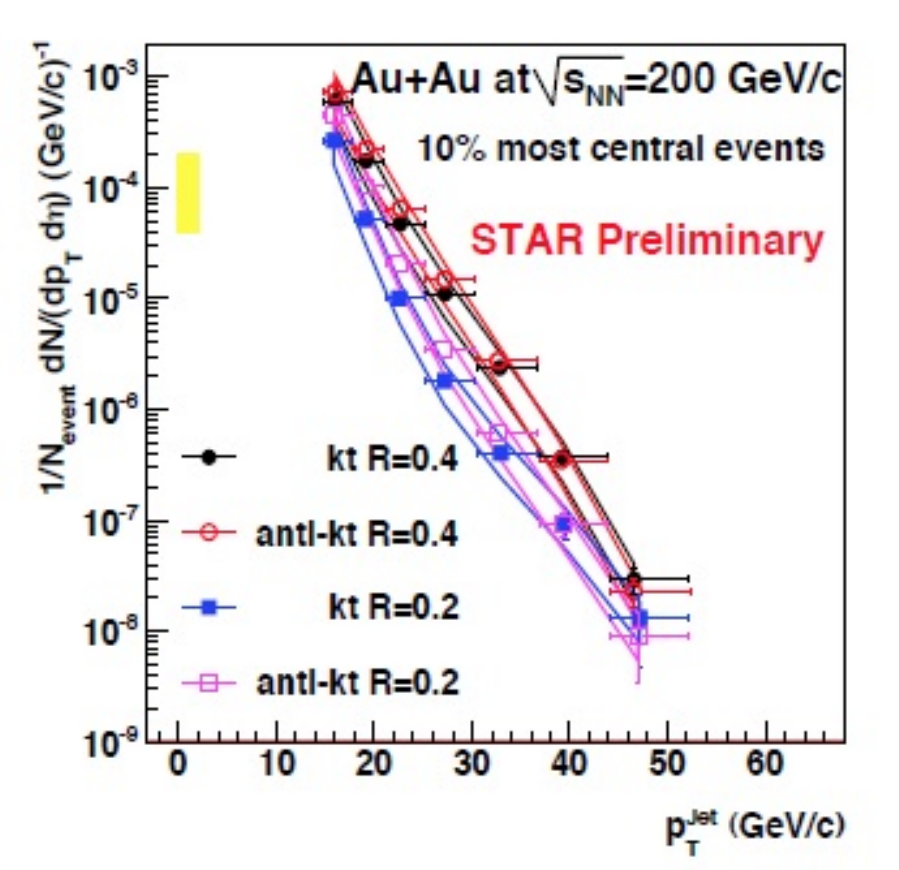}
\includegraphics*[width=6.1cm]{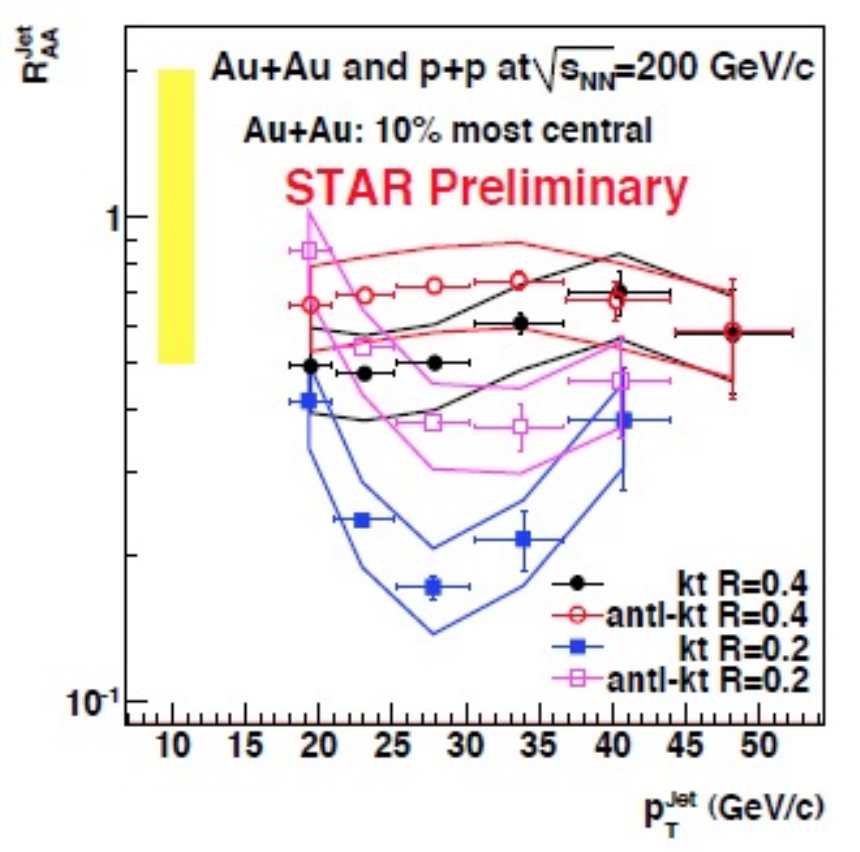}
\caption{(Color online) The cross sections (left) and the nuclear modification factor $R_{AA}$ for inclusive jet production in Au-Au collisions at $\sqrt{s_{NN}} = 200 $GeV measured by STAR Collaboration.
Figures are taken from Ref. \cite{Ploskon:2009zd}.
}
\label{fig_star_full_jet}
\end{center}
\end{figure}

Early studies of reconstructed full jets in relativistic heavy-ion collisions were performed in Au-Au and Cu-Cu collisions at RHIC by both STAR and PHENIX collaborations \cite{Ploskon:2009zd, Lai:2009zq}.
Many interesting results have been obtained about reconstructed full jets and their medium modification, even though with large uncertainties.
For example, Fig. \ref{fig_star_full_jet} shows the measurements of the cross sections (left) and the nuclear modification factor $R_{AA}$ (right) for inclusive jet production in Au-Au collisions at $\sqrt{s_{NN}} = 200 $GeV.
The inclusive jet nuclear modification factor $R_{AA}^{\rm jet}$ is defined as follows:
\begin{eqnarray}
R_{AA}^{\rm jet} = \frac{dN_{AA}^{\rm jet}/{dE_T dy}}{N_{\rm coll} dN_{pp}^{\rm jet}/{dE_T dy}}.
\end{eqnarray}
The dependence on different full jet reconstruction algorithms ($k_T$ and anti-$k_T$) has been explored and turned out to be quite strong.
Such strong dependence may originate from different responses to the fluctuating background in heavy-ion collisions when different reconstruction algorithms are utilized.
Despite large uncertainties and strong dependence on jet reconstruction algorithms, the results tend to indicate that there are substantial nuclear modification of full jets in heavy-ion collisions as compared to proton-proton collisions.
Especially, the strong dependence of jet production and their nuclear modification on the jet resolution parameter $R$ in heavy-ion collisions indicates that full jets are broadened due to their interaction with the hot and dense nuclear medium.

\begin{figure}
\begin{center}
\includegraphics*[width=6.6cm, height=4.5cm]{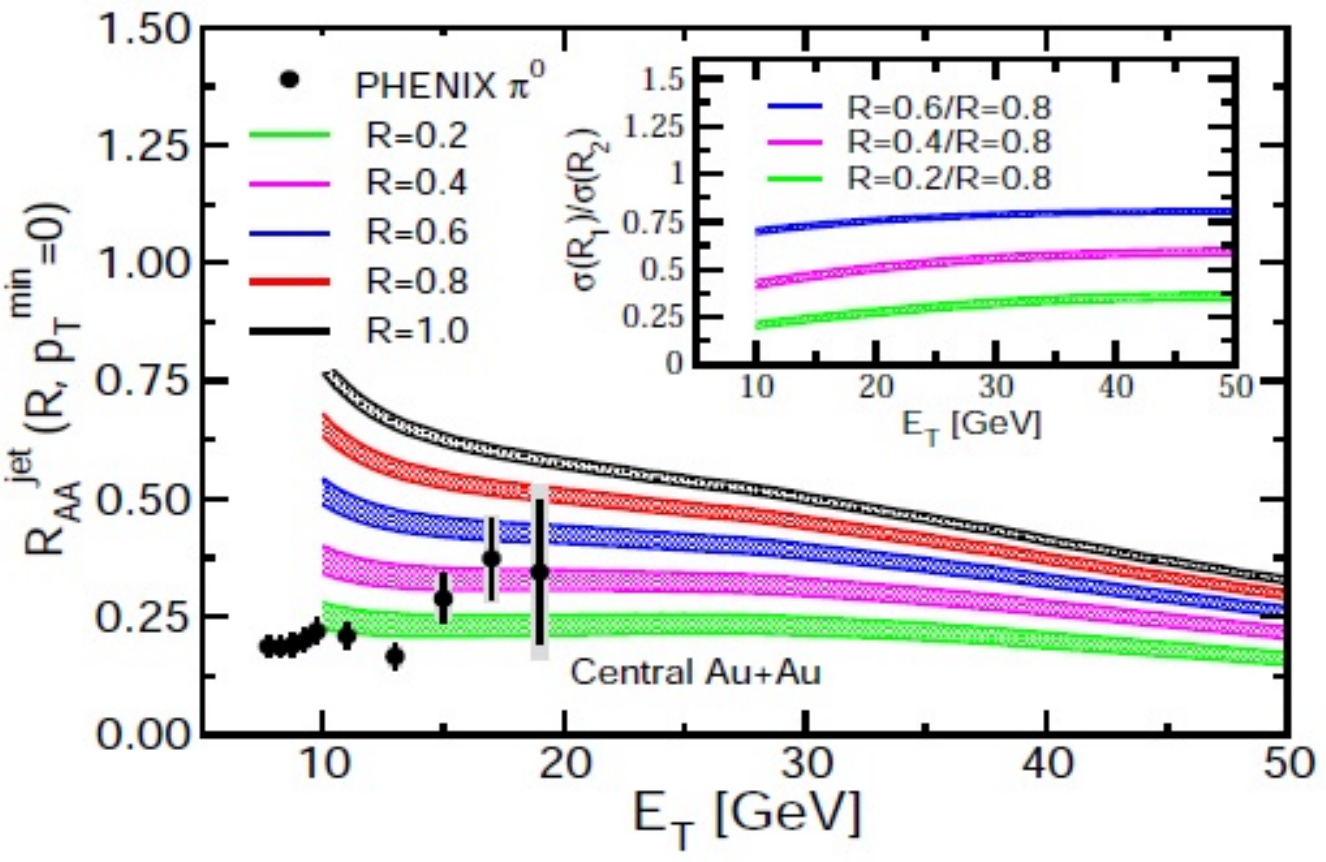}
\includegraphics*[width=5.9cm, height=4.5cm]{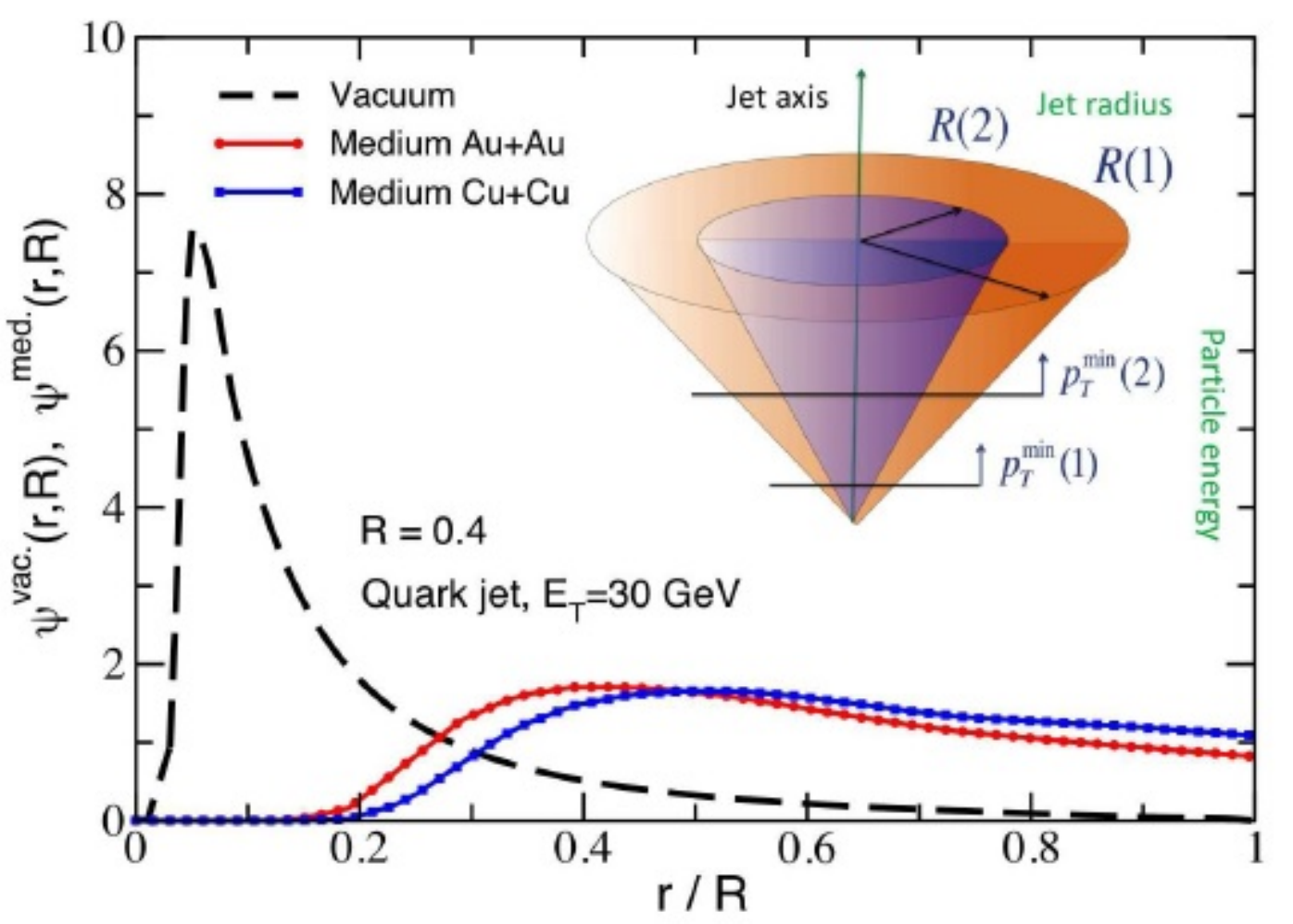}
\caption{(Color online) Left: The nuclear modification factor $R_{AA}^{\rm jet}$ for different cone radii $R$ in $b=3$~fm Au-Au collisions at RHIC. The inserts show the ratios of jet cross sections for different $R$.
Right: The differential jet shape in vacuum compared to the medium-induced jet shape for a quark with $E_T = 30$~GeV in Au+Au and Cu+Cu collisions at $\sqrt{s_{NN}} = 200 $GeV. The insert shows a method for studying the substructure of full jets.
Figures are taken from Ref. \cite{Vitev:2009rd}.
}
\label{fig_Zhang_Vitev}
\end{center}
\end{figure}

Earlier theoretical studies of full jets and their medium modification also supported the broadening of full jets as observed in heavy-ion collisions at RHIC.
In Ref. \cite{Vitev:2009rd}, the medium effect on full jets were studied for Au-Au and Cu-Cu collisions at RHIC energies using the framework of GLV formalism.
First, with the assumption of Poisson ansatz for the medium-induced gluon radiation, the differential radiation spectrum $dN_g/d\omega dr$ is utilized to calculate the probability distributions $P_q(\epsilon, E)$ and $P_g(\epsilon, E)$ for high-energy quarks and gluons which lose fraction of energy $\epsilon$.
The same differential radiation spectrum is also utilized to calculate the fraction of lost energy that is redistributed inside the jet cone:
\begin{eqnarray}
f_{q, g}^{\rm in} = \frac{\int_0^R dr \int_{p_T^{\rm min}}^{E_T}d\omega \frac{dN_g}{d\omega dr} }{\int_0^{R_\infty} dr \int_{p_T^{\rm min}}^{E_T}d\omega \frac{dN_g}{d\omega dr}},
\end{eqnarray}
where $p_T^{\rm min}$ is the momentum cut for the particles $i$ that constitute the jet, and $R$ is the jet cone radius.
One expects that as $R \to R_{\infty}$ and $p_T^{\rm min} \to 0$, the fraction $f_{q, g}^{\rm in} = 1$, i.e., the final state medium effects to inclusive jet cross section vanishes.
The medium-modified jet cross section per binary nucleon-nucleon scattering can be calculated as follows:
\begin{eqnarray}
\frac{d\sigma^{AA}(R)}{dE_Tdy} = \sum_{q, g} \int_0^1 P_{q, g}(\epsilon, E) \frac{1}{[1-(1-f_{q,g}^{\rm in}) \epsilon]^2} \frac{d\sigma_{q, g}^{\rm CNM}}{d^2E_T' dy},
\end{eqnarray}
where $(1-f_{q,g}^{\rm in})\epsilon$ represents the fraction of the lost energy from the parent parton outside the jet cone $R$,
and $E_T' = E_T / [1-(1-f_{q,g}^{\rm in}) \epsilon]$ is initial parton energy before interacting with the hot and dense QGP medium.
Note that in the above expression, $p_T^{\rm min} = 0$ is used.

The result for inclusive jet $R_{AA}^{\rm jet}$ is presented in Fig. \ref{fig_Zhang_Vitev} (left panel) for central Au-Au collisions collisions at $\sqrt{s}_{NN} = 200$~GeV at RHIC.
The bands represent the uncertainties due to a $\sim 20\%$ change in the rate of parton energy loss relative to the default value.
The nuclear modification factor for leading $\pi^0$ production is also shown for comparison.
One can see clearly the strong dependence of inclusive jet $R_{AA}$ on the jet cone size $R$: there is more suppression for inclusive jet for smaller jet radius.
This indicates full jets is broadened due to additional medium-induced radiation when interacting with the hot and dense nuclear matter created in nucleus-nucleus collisions.
The inserted plot shows the ratio of nuclear modification factor for different cone radii in order to reduce the the cold nuclear matter (CNM) effect which may contribute about $1/2$ of the observed suppression for jet energies $E_T > 30$~GeV.

The broadening effect of full jets in heavy-ion collisions may be investigated using jet shape observables.
The differential jet shape function is defined as:
\begin{eqnarray}
\psi(r/R) = \frac{d}{dr} \left[\frac{\sum_i E_{T,i}\theta(r-R_{i, jet})}{\sum_i E_{T,i}\theta(R-R_{i, jet})} \right],
\end{eqnarray}
where the sum over $i$ runs over all the particles inside the full jet.
By definition, the differential jet shape function is normalized to unity, $\int_0^R \psi(r, R) dr = 1$.
The right plot in Fig. \ref{fig_Zhang_Vitev} shows the differential jet shape function in vacuum compared to medium-induced jet shape function in Au-Au and Cu-Cu collisions at RHIC.
In the plot, the medium-induced jet shape function is directly calculated by using differential gluon bremsstrahlung spectrum with proper normalization: $\psi^{\rm med} \propto dN_g/d\omega dr$.
One can clearly see a characteristic large-angle distribution away from jet axis for medium-induced radiation.
It should be noted that the final medium-modified jet shape function has contributions from both vacuum radiation and medium-induced radiation.
It may be calculated as follows:
\begin{eqnarray}
\psi^{\rm tot}(r/R) &&\!\!= \sum_{q, g} \int_0^1 d\epsilon P_{q, g}(\epsilon, E) \frac{\chi_{q, g}(R, E_T, E_T')}{[1 - (1-f_{q, g}^{\rm in})]^3} \nonumber\\
&& \times \left[ (1-\epsilon) \psi_{q, g}^{\rm vac}(r/R, E_T') + \epsilon f_{q, g}^{\rm in}  \psi_{q, g}^{\rm med}(r/R, E_T) \right],
\end{eqnarray}
where $\chi_{q, g} = \frac{d\sigma_{q, g}^{CNM}/d^2 E_T' dy}{d\sigma^{AA}/d^2E_T dy}$.
The first term in the bracket denotes the vacuum contribution, and the second is the contribution from medium-induced radiation.
It is instructive to look at the first moment of the jet shape distribution, or the mean jet width, which may be obtained as follows:
\begin{eqnarray}
\langle r/R \rangle = \int_0^1 d(r/R) (r/R) \psi(r/R).
\end{eqnarray}
For jets with transverse energies $E_T = 30$~GeV and cone size $R=0.4$, one may calculate the broadening effect in terms of the change of mean jet width. For central Au-Au collisions, $\Delta\langle r/R \rangle = \langle r/R \rangle_{\rm tot} - \langle r/R \rangle_{\rm vac} = 0.283-0.271=0.012$, which is less than $5\%$ increase.
This results indicates even with striking suppression pattern for full jets as shown by jet $R_{AA}$, a very modest growth is seen for the change of mean jet width.
More detailed information may be obtained if the full distribution of jet shape function and its medium modification are analyzed (see below).

\subsection{Full jets at the LHC era}

\begin{figure}
\begin{center}
\includegraphics*[width=13.0cm]{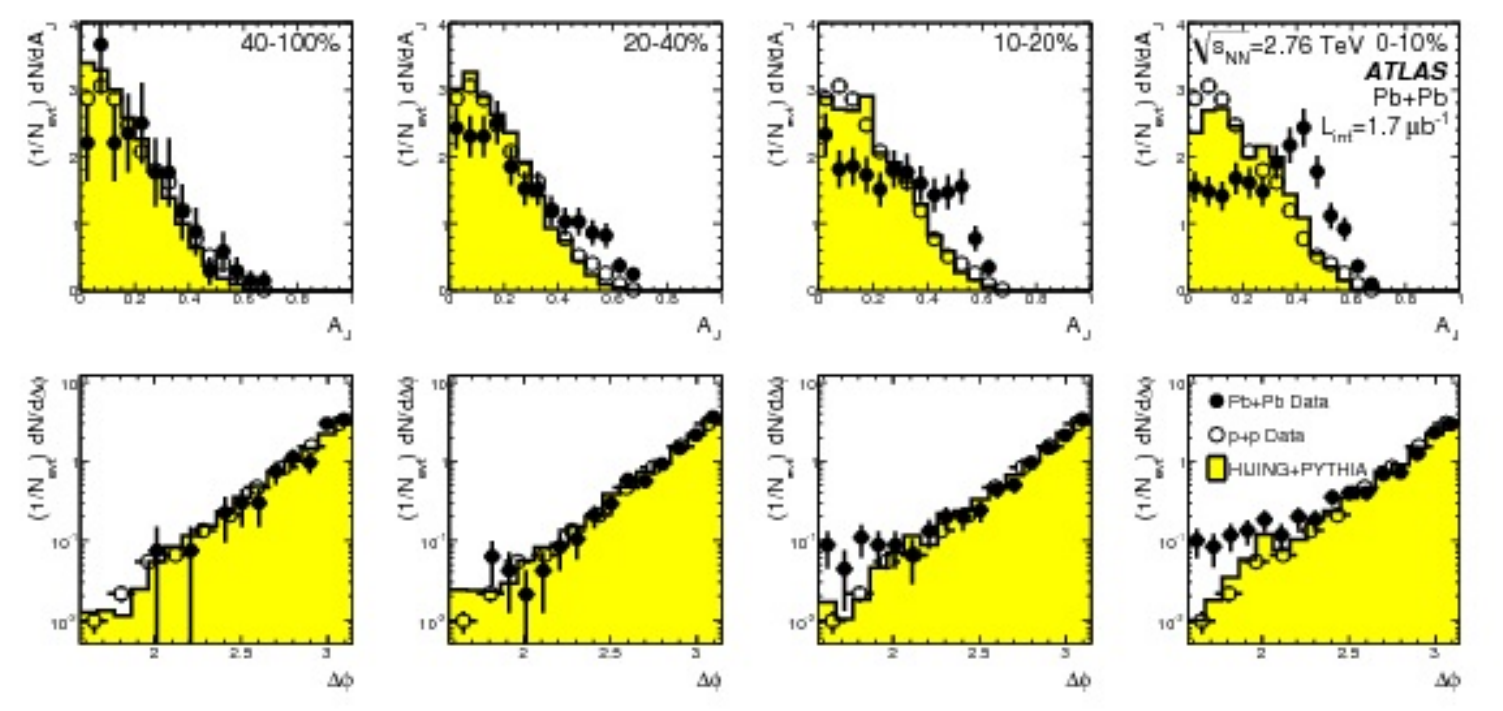}
\caption{(Color online) ATLAS 2010 measurements of dijet asymmetry distributions (top panels) and azimuthal angle $\Delta \phi$ distribution (bottom panels) for data (points) and unquenched HIJING+PYTHIA (solid yellow histograms) for different centralities (left to right from peripheral to central). Proton-proton data for 7~TeV is shown as open circles.
Figures are taken from Ref. \cite{Aad:2010bu}.
}
\label{fig_ATLAS_dijet}
\end{center}
\end{figure}

With the launch of the LHC, the center of mass energy of the collisions has been increased by more than a factor of 10 compared to the top collision energies at RHIC.
The large kinematics accessed at the LHC have enabled us to investigate the properties of jets and the effects of hot and dense nuclear medium on the propagation of jets with transverse energies over a hundred GeV. One of the most interesting channels for studying the effect of jet-medium interaction is the correlated back-to-back jet pairs.
The energy loss of full jets can be studied by measuring the imbalance/asymmetry of the transverse energies between two correlated jets (and the centrality dependence). One may define the energy asymmetry factor $A_J$ for the correlated jet pairs as
\begin{eqnarray}
A_J = \frac{E_{T, 1} - E_{T, 2}}{E_{T, 1} + E_{T, 2}},
\end{eqnarray}
where $E_{T, i}(i = 1, 2)$ denotes the transverse energy of the leading and sub-leading jets, respectively.
The relative azimuthal angle $\Delta \phi = |\phi_1 - \phi_2|$ between two correlated jets can provide information about the degree of deflection that jet experience after passing through the hot and dense QGP medium.
The transverse energy asymmetry $A_J$ and the azimuthal angle $\Delta \phi$ of back-to-back jet pairs have been measured by both ATLAS and CMS Collaborations.
ATLAS selected events with the leading jets having $E_{T,1} > 120$~GeV/c and the subleading jets having $E_{T, 2} > 100$~GeV/c, while for CMS $E_{T,1} > 50$~GeV/c and $E_{T,2} > 25$~GeV/c.
The results from ATLAS measurements are shown in Fig. \ref{fig_ATLAS_dijet}, with the top panels for dijet asymmetry distributions and the bottom panels for azimuthal angle $\Delta \phi$ distribution.

One of the most spectacular results from such measurements is strong modification on dijet transverse energy imbalance distribution in Pb-Pb collisions, as compared to proton-proton collisions, while the the distribution of their relative azimuthal angles is largely unchanged.
The nuclear modification of dijet asymmetry strongly depends on the collision centrality: the modification is larger in more central collisions.
The transverse energy imbalance between full jets and the correlated high $p_T$ direct photons in Pb-Pb collisions have also been measured at the LHC \cite{Chatrchyan:2012gt}, and very similar results have been obtained.
These observations indicated that the subleading jets may experience a significant amount of energy loss after traveling through the hot and dense nuclear matter created in relativistic heavy-ion collisions.
These measurements have triggered a lot of theoretical interests on the investigation of parton shower evolution and full jet energy loss in hot QCD matter.
Various jet energy loss based models have been utilized to explain the observed transverse energy imbalance between dijets and photon-jet pairs \cite{Qin:2010mn,CasalderreySolana:2010eh,Lokhtin:2011qq,Young:2011qx,He:2011pd,Renk:2012cx,CasalderreySolana:2012ef,Blaizot:2013hx,Dai:2012am,Qin:2012gp,Wang:2013cia}.

\begin{figure}
\begin{center}
\includegraphics*[width=12.5cm,height=4.5cm]{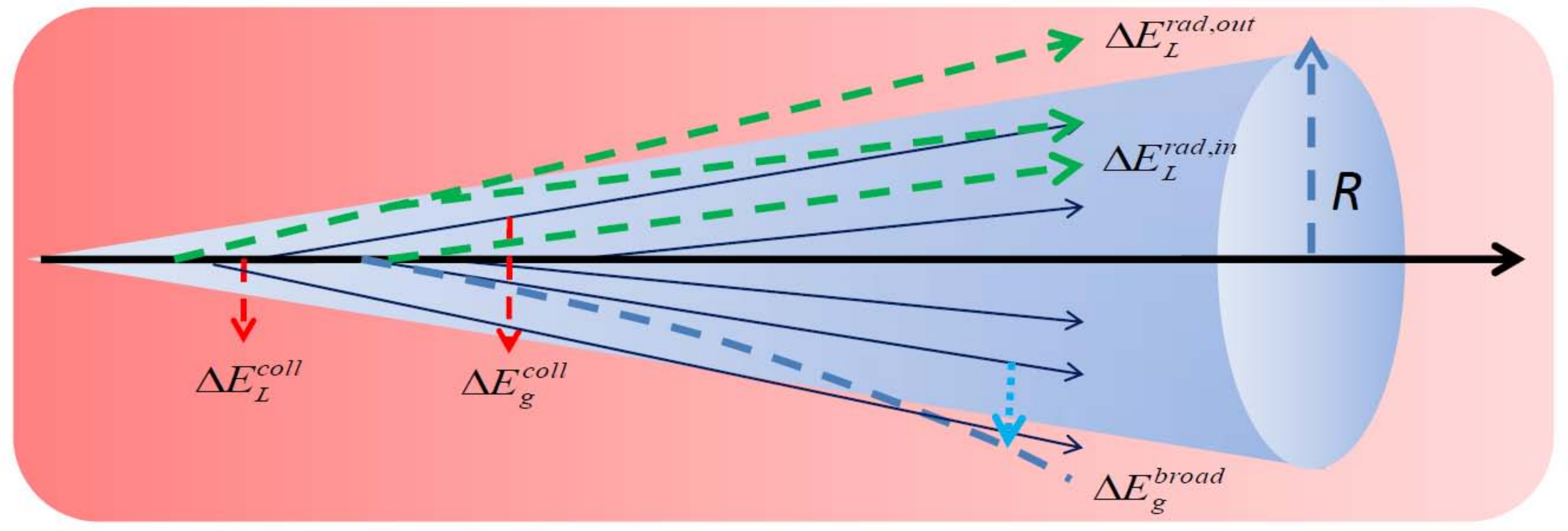}
\caption{(Color online) A schematic illustration of the evolution of full jet, and different medium-induced processes that contribute to full jet energy loss and modification in a quark-gluon plasma.
}
\label{fig_full_jet_evolution}
\end{center}
\end{figure}

The evolution of full jet in a quark-gluon plasma may be schematically illustrated in Fig. \ref{fig_full_jet_evolution}, where the thick solid arrow through the center of jet cone represents the leading parton of the jet, and other lines represent the accompanying radiated gluons.
One can see that compared to leading hadron observables, a few additional ingredients have to be taken into account when studying the evolution and modification of full jets in hot and dense nuclear medium.
One needs to include the interaction of the medium with both the leading parton as well as with the radiated gluons.
The radiative gluons may also get deflected by interacting with medium constituents, and some of them may be kicked out of the jet cone and contribute to the energy loss of full jet.
Therefore, the total energy loss of energy from the jet cone is the sum of energy loss incurred by the leading parton and the accompanying radiated gluons, as well as the gluons that are scattered out of jet cone.

Based on the above picture, Ref. \cite{Qin:2010mn} performed the first quantitative analysis of physical processes that are responsible for full jet energy loss and the observed large di-jet energy asymmetry. The following transport equation is solved for the distribution of radiated gluons:
\begin{eqnarray}
\frac{df_g(\omega, k_\perp, t)}{dt} = -\hat{e}\frac{\partial f_g}{\partial \omega} + \frac{1}{4} \hat{q} \nabla_{k_\perp}^2 f_g + \frac{dN_g^{\rm rad}}{d\omega dk_\perp^2 dt},
\end{eqnarray}
where $f_g(\omega, k_\perp, t)$ is the double differential distribution of the accompanying gluons of the full jets.
In the above equation, the first and second terms describe the evolution of radiated gluons which may transfer energy into the medium by elastic collisions with medium constituents and accumulate transverse momentum in the process.
The sizes of these two effects are controlled by two transport coefficients: elastic energy loss rate $\hat{e}$ and the rate of exchanging transverse momentum squared $\hat{q}$.
The last term in the equation represents the additional gluon radiation induced by the interaction of primary hard parton with the medium.
These medium-induced radiative gluons, after produced, may also lose energy and accumulate transverse momentum during their propagation.

The above transport equation is combined with the higher-twist formalism which provides the rate for the medium-induced gluon radiation \cite{Wang:2001ifa, Majumder:2009ge},
\begin{eqnarray}
\frac{dN_g^{\rm rad}}{d\omega dk_\perp^2 dt} = \frac{2\alpha_s}{\pi} \frac{xP(x) \hat{q}(t)}{\omega k_\perp^4} \sin^2 \left( \frac{t-t_i}{2\tau_f} \right),
\end{eqnarray}
where $P(x)$ is the vacuum splitting function with $x=\omega/E$ denoting the energy fraction of radiated gluons, and $\tau_f = 2Ex(1-x)/k_\perp^2$ the formation time for the radiated gluons. The transport coefficient $\hat{q}$ depends on the evolution time because of the changing position of the primary parton.

After the above evolution equation for the radiated gluons is solved, one may obtain the information about the full jets after propagating through the medium.
From the energy and transverse momentum distribution of the accompanying gluons, the total energy of these gluons contained inside the jet cone can be obtained as follows:
\begin{eqnarray}
E_g(t_f, R) = \int_R d\omega dk_\perp^2 \omega  f_g(\omega, k_\perp^2, t_f),
\end{eqnarray}
where the subscript $R$ means that the integration is taken inside the jet cone, i.e., $k_\perp/\omega < R$.
One may calculate the final energy of the leading parton after passing through the medium,
\begin{eqnarray}
E_L(t_f) = E_L(t_i) - \int dt \hat{e}_L(t) - \int dt d\omega dk_\perp^2 \omega \frac{dN_g^{\rm rad}}{d\omega dk_\perp^2 dt}.
\end{eqnarray}
The total energy inside the jet cone is the sum of the leading parton energy and the energy of the radiated gluon inside the jet cone:
\begin{eqnarray}
E_{\rm jet}(t_f, R) = E_L(t_f) + E_g(t_f, R).
\end{eqnarray}
Then one may calculate the reduction of the full energy,
\begin{eqnarray}
\Delta E_{\rm jet}(t_f, R) = E_{\rm jet}(t_i, R) - E_{\rm jet}(t_f, R),
\end{eqnarray}
where $E_{\rm jet}(t_i, R)$ is the initial energy contained within the jet cone.
The final expression for the energy loss from the jet cone may be obtained as follows:
\begin{eqnarray}
\Delta E_{\rm jet}(t_f, R) &&\!\!= \int dt d\omega dk_\perp^2 \omega \frac{dN_g^{\rm rad}}{d\omega dk_\perp^2 dt} + \int \hat{e}_L(t) dt \nonumber\\ &&\!\! - \int_R d\omega dk_\perp^2 \omega \left[ f_g(\omega, k_\perp^2, t_f) - f_g(\omega, k_\perp^2, t_i) \right],
\end{eqnarray}
where the first two terms represent the radiative and collisional energy loss of the leading hard parton.
To calculate the loss of the energy from jet cone, one has to subtract the energy carried by the radiated gluons inside the jet cone (which is represented by the last two terms) from the leading parton energy loss.
Therefore the energy of the original full jet is now decomposed into several parts:
\begin{eqnarray}
E_{\rm jet} = E_{\rm in} + E_{\rm lost} = E_{\rm in} + E_{\rm out, rad} + E_{\rm out, brd} + E_{\rm coll, th}.
\end{eqnarray}

\begin{figure}
\begin{center}
\includegraphics*[width=11.0cm,height=6.5cm]{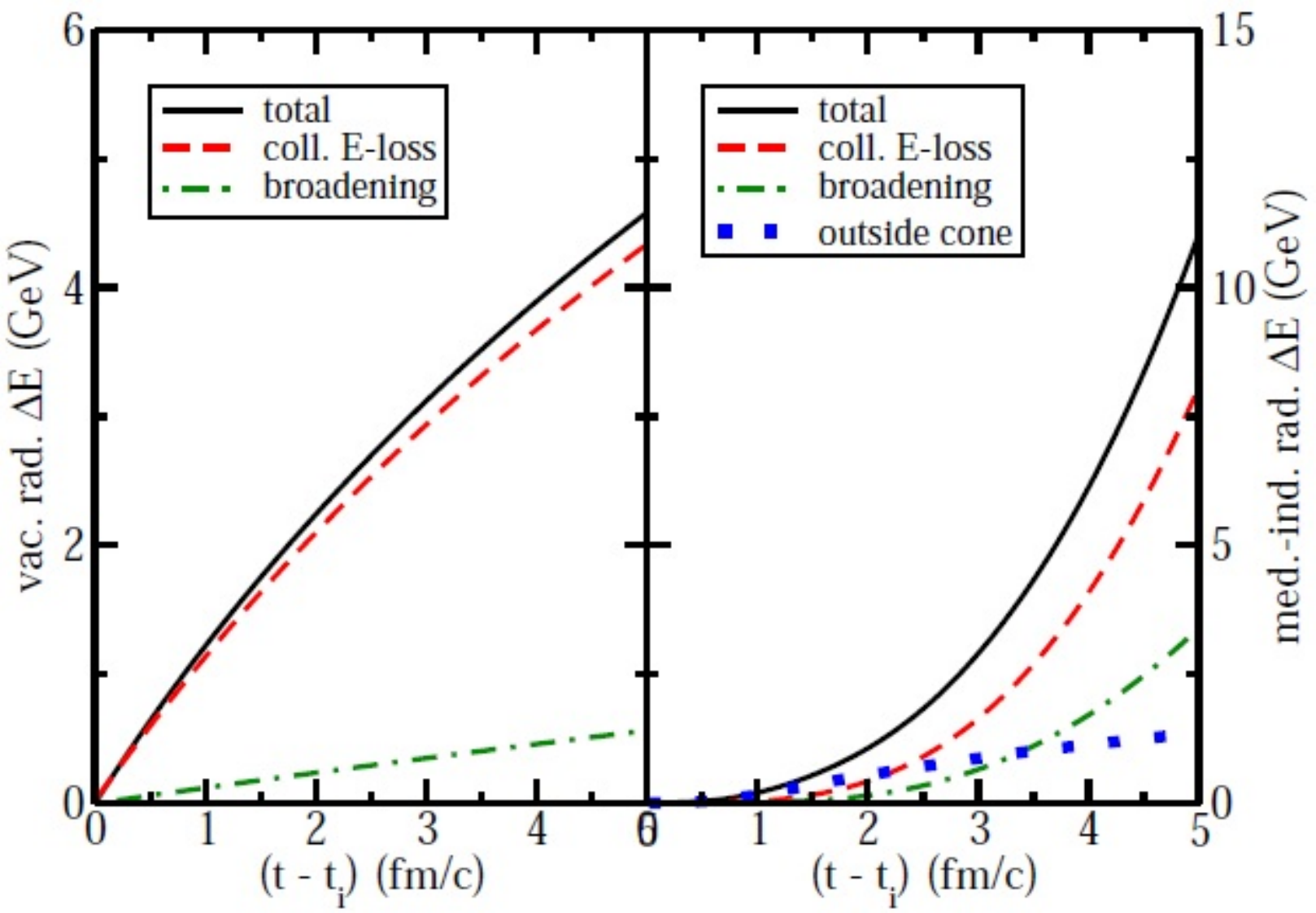}
\caption{(Color online) The energy loss experienced by a quark jet with initial energy $E=100$~GeV traversing a quark-gluon plasma with constant temperature $T=250$~MeV. Contributions from different energy loss mechanisms are compared. The left for the medium effect on vacuum radiation, and the right for the medium effect on medium-induced radiation \cite{Qin:2010mn}.
}
\label{fig_QM_dE}
\end{center}
\end{figure}

Fig. \ref{fig_QM_dE} shows the contributions from vacuum radiated gluons (left) and medium-induced radiated gluons (right) to the total energy loss of the jet defined by a cone angle $R=0.4$.
A quark jet is started with initial energy $E=100$~GeV which radiates gluons in vacuum before $t_i = 1$~fm/c, then the medium modification effect is turned on for the jet shower.
The vacuum radiation is simulated using PYTHIA, and the effect of the medium on the full jet evolution is calculated via solving the above transport equation.
One can see that at early times, the medium-induced energy loss from jet cone is dominated by the medium effect on vacuum radiation, and at later times it is dominated by the medium effect on medium-induced radiation.
One interesting observation is that the most significant contribution originates from the collisional energy loss experienced by the radiated gluons $E_{\rm th, coll}$.
The transverse momentum broadening of shower gluons $E_{\rm out, brd}$ also gives a sizable contribution.
The contribution from the radiation directly outside the jet cone $E_{\rm out, rad}$ is small, since the radiation is mainly dominated by small angle radiation, i.e., the radiated gluons are mostly inside the jet cone.

\begin{figure}
\begin{center}
\includegraphics*[width=11.0cm,height=6.5cm]{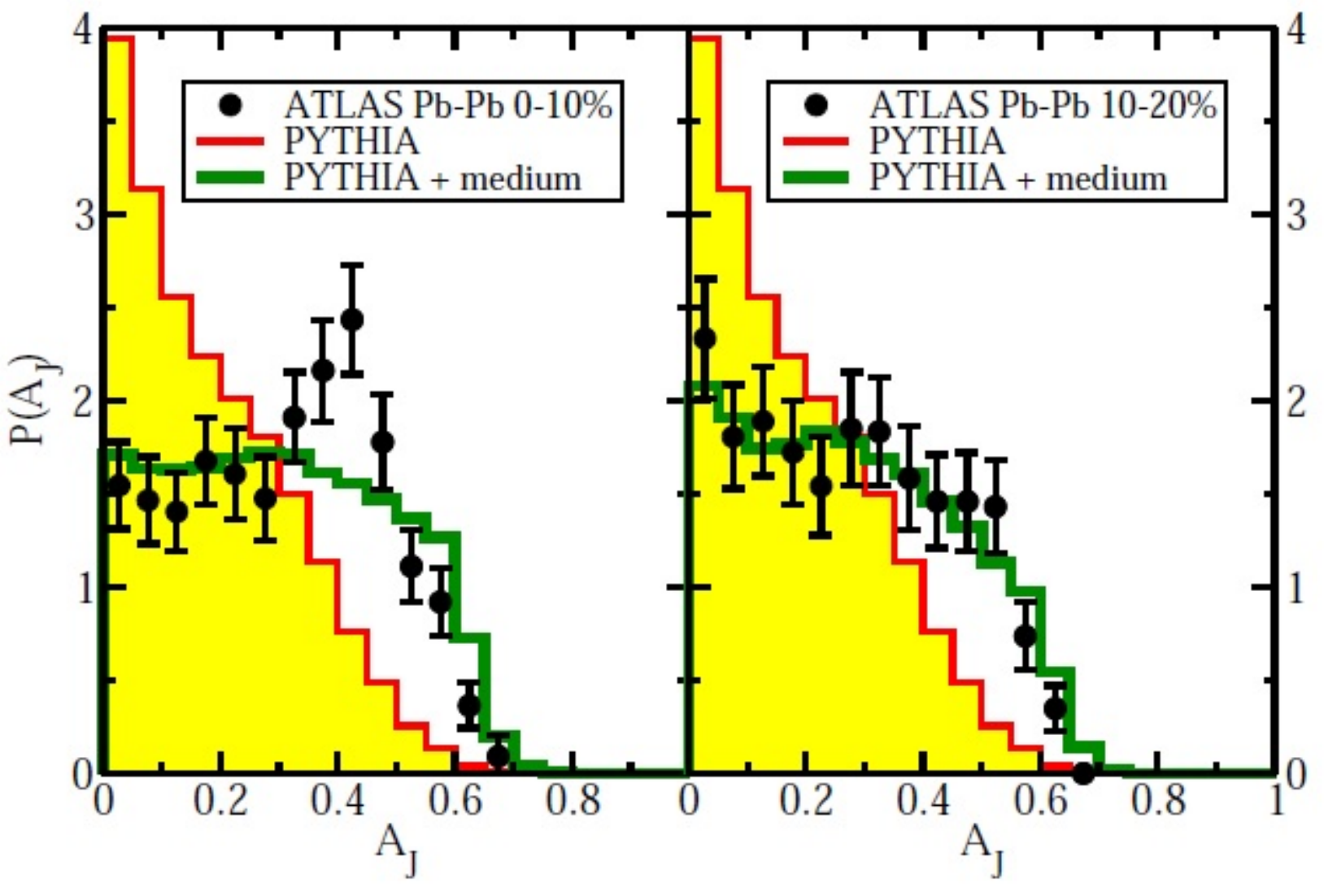}
\caption{(Color online) Di-jet energy asymmetry $A_J$ distribution for proton-proton and Pb-Pb collisions at $\sqrt{s_{NN}} = 2760$~GeV at the LHC. The left panel is for 0-10\% centrality and the right for 10-20\% centrality \cite{Qin:2010mn}.
}
\label{fig_QM_AJ}
\end{center}
\end{figure}

Fig. \ref{fig_QM_AJ} shows the calculation of dijet energy asymmetry factor $A_J$ in proton-proton and Pb-Pb collisions at $\sqrt{s_{NN}} = 2760$~GeV at the LHC.
In proton-proton collisions, $A_J$ distribution for the correlated back-to-back dijets is peaked at $A_J = 0$.
After the medium effect is turned on, the dijet energy asymmetry $A_J$ distribution is significantly shifted to the large $A_J$ direction.
The asymmetry of dijet energies is more prominent in the most central Pb+Pb collisions (left panel) than less central events (right).
The result indicate that the sub-leading jets experience a significant amount of energy loss from the jet cone.
The total energy loss is a combination of collisional energy loss experienced by all shower partons, the radiation outside jet cone and the scattering of radiated gluons out of jet cone.

The above finding that the medium absorption or the thermalization of the soft gluon radiation gives the largest contribution to full jet energy loss is not surprising since one expects the soft components of the hard jet or the accompanying gluons at large angles to experience stronger medium modification than the inner hard core of the jet.
Such effect is often referred to as jet collimation mechanism, i.e., when jet shower propagates through the hot and dense nuclear medium, soft partons of the jet are easily stripped off by the interaction with the medium constituents \cite{CasalderreySolana:2010eh}.

Similar picture of parton shower evolution and full jet energy loss has also been obtained in other studies \cite{MehtarTani:2011tz, Blaizot:2013vha, Mehtar-Tani:2014yea}.
The basic idea underlying these studies is that: if the medium color field varies over the jet transverse size, i.e., $\lambda < r_{\perp}^{\rm jet}$, the interaction between the jet and the medium constituents may destroy the color coherence of the shower partons of the full jet.
It is argued that such decoherence effect may greatly increase the phase space for soft and large angle radiation as compared to the traditional BDMPS-Z formalism.
To the lowest order approximation, one may consider the successive gluon emissions as independent of each other.
This will allows to treat multiple gluon emissions as a probabilistic branching process.

Let us consider the evolution of a gluon shower originating from successive gluon branchings.
The differential probability for a gluon with energy $\omega$ (or the energy fraction $x=\omega/E$) to split into two gluons is given by
\begin{eqnarray}
\frac{dP_{\rm br}}{dz dt} = \frac{\alpha_s}{2\pi} \frac{P_{g \to g}(z)}{\tau_{\rm br}(z, \omega)},
\end{eqnarray}
where $P_{g \to g}(z) = N_c [1-z(1-z)]^2/[z(1-z)]$ is the leading order gluon-gluon splitting function, with $z$ and $1-z$ are energy fractions of two daughter gluons with respect to their parent gluon. The time scale of gluon branching process $\tau_{\rm br}$ is given by
\begin{eqnarray}
\tau_{\rm br}(z, \omega) =  \sqrt{\frac{z(1-z)\omega}{[1-z(1-z)]\hat{q}}},
\end{eqnarray}
where $\hat{q}$ is the jet quenching parameter. To simplify the following analysis, one may replace the light-cone time $t$ by the dimensionless variable $\tau$, defined as
\begin{eqnarray}
\tau = \frac{\alpha_s N_c}{\pi} \sqrt{\frac{\hat{q}}{E}} t = \frac{\alpha_s N_c}{\pi} \sqrt{\frac{\hat{q}L^2}{E}} \frac{t}{L} = \frac{\alpha_s N_c}{\pi} \sqrt{\frac{2\omega_c}{E}} \frac{t}{L},
\end{eqnarray}
where $\omega_c = \hat{q} L^2 / 2$ is the maximum energy that can be taken away by a single radiated gluon. The branching probability may now be written as
\begin{eqnarray}
\frac{dP_{\rm br}}{dx d\tau} = \frac{\kappa(z)}{2\sqrt{x}} = \frac{1}{2\sqrt{x}} \frac{f(z)}{[z(1-z)]^{3/2}} = \frac{1}{2\sqrt{x}} \frac{[1-z(1-z]^{5/2}}{[z(1-z)]^{3/2}},
\end{eqnarray}
where $\kappa(z) = f(z)/[z(1-z)]^{3/2}$ and $f(z) = [1-z(1-z)]^{5/2}$.

To study the multiple gluon branching process, it is convenient to consider the the differential gluon spectra $D(x, \tau) = x dN/dx$, with $N$ the number of gluons in the branching cascade.
Using the fact that the successive branchings are independent of each other, one may solve the following rate equation for the function $D(x, \tau)$,
\begin{eqnarray}
\frac{\partial D(x, \tau)}{\partial \tau} = \int dz \kappa(z) \left[ \sqrt{\frac{z}{z}} D(\frac{x}{z}, \tau) - \frac{z}{\sqrt{x}} D(x, \tau) \right].
\end{eqnarray}
The first term in the equation is the gain term which describes the rise in the number of gluons at $x$ due to the emission from gluons at larger values of $x$.
The second term is the loss term which describes the reduction in the number of gluons at $x$ due to the decay into gluons with smaller values of $x$.
Note that the first integral should be taken as $x<z<1$ since the function $D(x, \tau)$ has support only for $0 \leq x \leq 1$.
Both gain and loss terms have singularities at the ending point $z=1$, but the singularities exactly cancel each other when combining two terms together.

One may simplify the above evolution equation by changing the kernel to be $\kappa(z) = {1}/{[z(1-z)]^{3/2}}$; such replacement is equivalent to taking $f(z) = 1$.
This simplification allows us to solve the equation analytically, but does not affect the singular behavior of the kernel near $z=0$ and $z=1$.
Furthermore, if one is only interested in the small-$x$ behavior, one may further take $\kappa_0(z) = 1$.
For $\tau << 1$, one may use iteration method to solve the above equation.
Given the initial condition $D^{(0)}(x, \tau=0) = \delta(x - 1)$, which means that a single gluon carries all the energy, the first interaction gives:
\begin{eqnarray}
D^{(1)}(x, \tau) = \frac{\tau}{\sqrt{x}(1-x)^{3/2}} \approx \frac{\tau}{\sqrt{x}}.
\end{eqnarray}
In the above equation (and equations below in this subsection), the last approximation is taken for $x \ll 1$, i.e., one is interested in the soft components of the shower gluons.
Taking $t=L$, the first iteration coincides with the traditional BDMPS-Z spectrum, $\omega {dN}/{d\omega} = {\alpha_s N_c}/{\pi} \sqrt{{2\omega_c}/{\omega}}$.
By going beyond the first iteration, one may obtain the complete solution,
\begin{eqnarray}
D(x, \tau) = \frac{\tau}{\sqrt{x}(1-x)^{3/2}} e^{-\pi[(1-x)\tau^2]} \approx \frac{\tau}{\sqrt{x}} e^{-\pi \tau^2}.
\end{eqnarray}

\begin{figure}
\begin{center}
\includegraphics*[width=6.35cm,height=4.6cm]{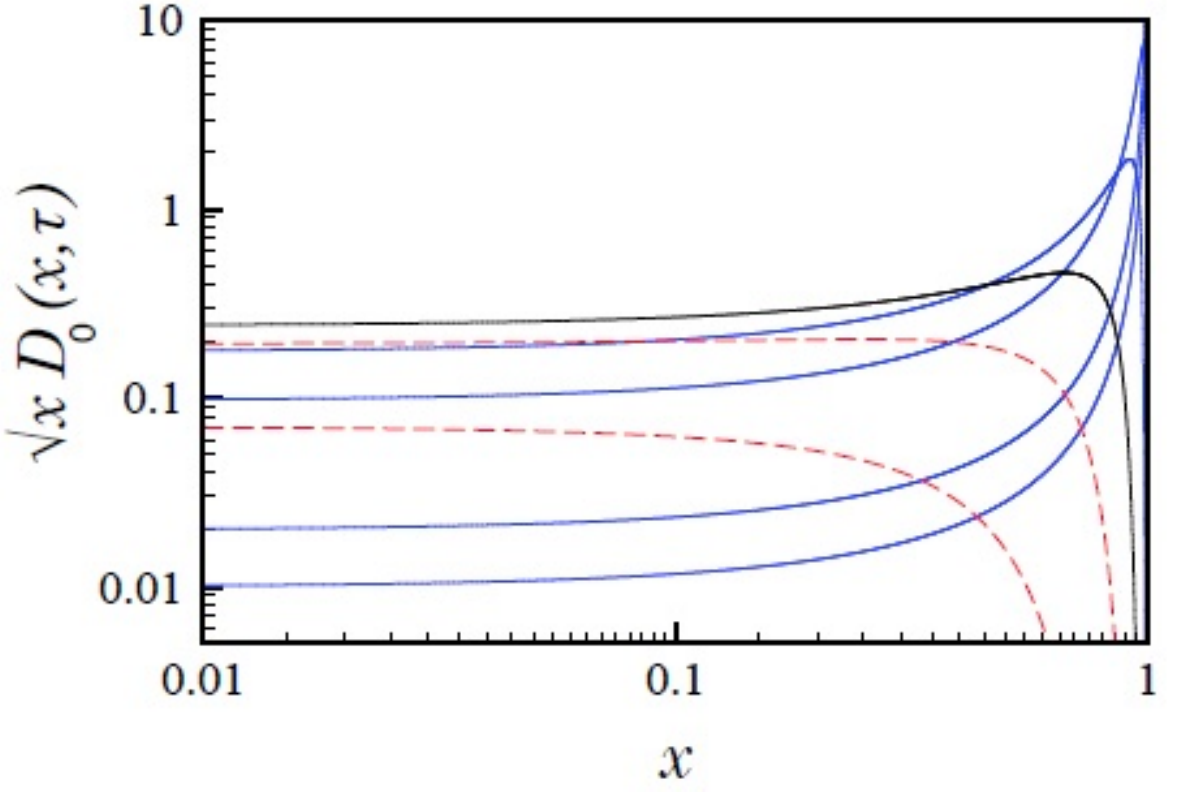}
\includegraphics*[width=6.15cm,height=4.6cm]{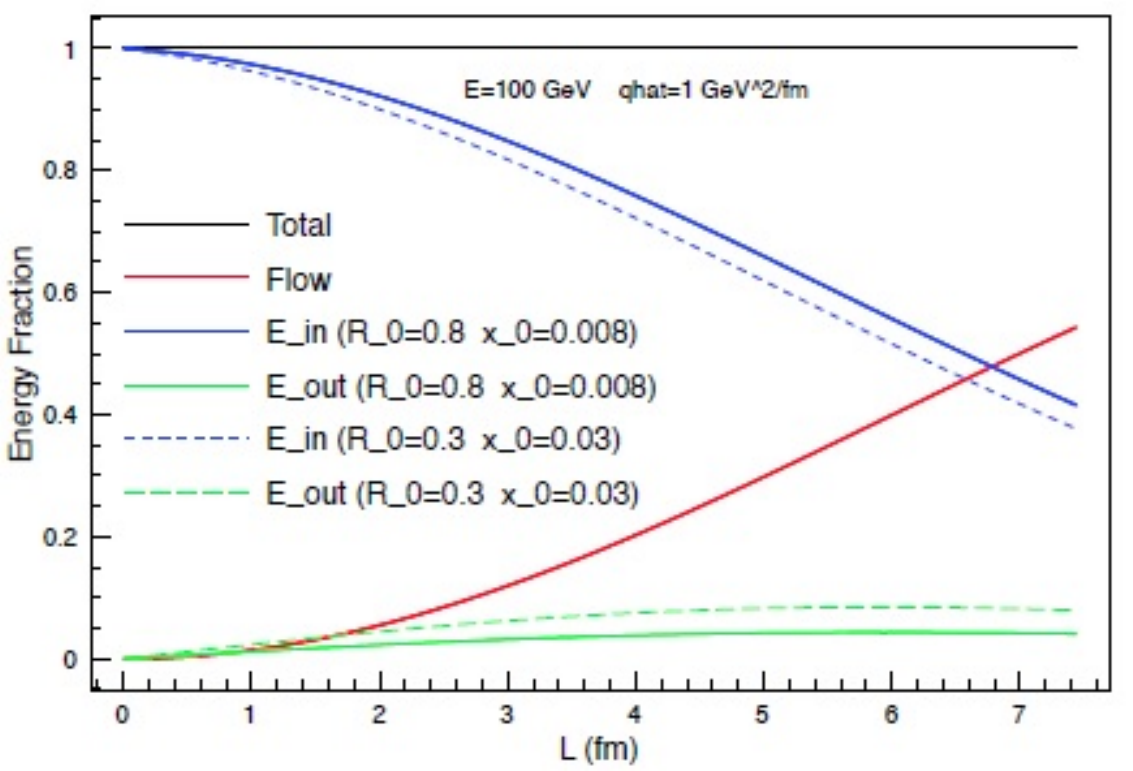}
\caption{(Color online) Left: The $\tau$ evolution of $\sqrt{x} D_0{x, \tau}$, where $\tau=0.01, 0.02, 0.1, 0.2, 0.4$ for solid lines from bottom to top and $\tau=0.6, 0.9$ for dashed lines from top to bottom. Right: The energy fraction for different components $E_{\rm in}$, $E_{\rm flow}$ and $E_{\rm out}$ as a function of the medium size $L$ for two values of jet size $R_0 = 0.3$ and $R_0 = 0.8$.
Figures are taken from Ref. \cite{Mehtar-Tani:2014yea, Iancu:2013ura}.
}
\label{fig_Iancu_gluon_branching}
\end{center}
\end{figure}

The above solution is shown in Fig. \ref{fig_Iancu_gluon_branching} (left panel), where $\sqrt{x} D(x, \tau)$ is plotted as a function of $x$ for different values of $\tau$: for solid lines $\tau=0.01, 0.02, 0.1, 0.2, 0.4$ from bottom to top, and for dashed lines $\tau=0.6, 0.9$ from top to bottom.
One may see that for small times, the spectrum at small $x$ increases linearly with $\tau$, besides the exponential factor.
This is very different from the DGLAP evolution, in which the spectrum get steeper and steeper at small $x$ with increasing evolution time.

One very interesting and important feature of such branching dynamics is that the total energy contained in the spectrum is not conserved but instead decreases with the evolution time,
$\epsilon(\tau) = \int_0^1 dx D(x, \tau) \approx e^{- \pi \tau^2}$.
This means that as the energy flows from the higher to lower values of $x$, there is no energy accumulated at any value at $x>0$.
As the time evolves, the gluon spectrum will eventually disappear into the left moving ``shock wave",
and the energy will be accumulated into a ``condensate" at $x=0$.
Therefore, with increasing evolution time, there could be a substantial fraction of the total jet energy flowing outside the spectrum.
The amount of such energy flow is given by
\begin{eqnarray}
\epsilon_{\rm flow}(\tau) = 1 - \int_0^1 dx D(x, \tau) \approx 1- e^{- \pi \tau^2}.
\end{eqnarray}

Besides the energy flow outside the spectrum that contributes to full jet energy loss, there is also contribution from the radiated energy that is emitted directly at large angles.
To estimate such contribution, one may introduce a scale $x_0$ which is related to the radiation angle $\theta_0$ as: $\theta_0 = [(2\hat{q})/(x_0^3 E)]^{1/4}$.
The energy fraction transported outside jet cone via the gluon emission at large angles can be calculated as follows:
\begin{eqnarray}
\epsilon_{\rm out}(\tau) = \int_0^{x_0} D(x, \tau) = \frac{2 \tau \sqrt{x_0}}{\sqrt{1-x_0}}  e^{-\pi \tau^2} \approx 2 \tau \sqrt{x_0} e^{-\pi \tau^2}.
\end{eqnarray}
One can see that this quantity rapidly decreases with decreasing $x_0$, therefore the gluon radiation directly outside the jet cone angle seems not be able explain the observed full jet energy loss and dijet energy asymmetry.

Another scale $x_{\rm th} = T/E$ may be introduced to take into account the fact that when the gluon energies becomes as low as as the typical energy scale in the medium, the radiated gluons may ``thermalize" and disappear from the jet.
Combining all these effects, the total phase space for the branching gluons may be divided into three different parts separated by two scales, $x_0$ and $x_{\rm th}$.
One may decompose the total energy of the original jet as follows:
\begin{eqnarray}
E_{\rm jet} = E_{\rm in} + E_{\rm lost} = E_{\rm in}(x>x_0) + E_{\rm out}(x_{\rm th}<x<x_0) + E_{\rm flow}(x<x_{\rm th}).
\end{eqnarray}
For $x>x_0$, the branching gluons are inside the jet cone and can be recovered by jet reconstruction.
For the radiated gluons with $x_{\rm th} < x < x_0$, they are radiation directly outside the jet cone.
For $x<x_{\rm th}$, the branching gluons thermalize and flow into the medium (thus disappear from the jet).

Fig. \ref{fig_Iancu_gluon_branching} (right panel) shows the relative contributions to the total jet energy (loss) from these three different components, $E_{\rm in}$, $E_{\rm flow}$ and $E_{\rm out}$, as a function of the medium size $L$.
One can see that with increasing jet cone angle from $R_0 = 0.3$ to $R_0 = 0.8$, the energy contained inside the jet cone increases only slightly.
For the medium length $L \ge 4$~fm, more than $20\%$ of the total jet energy is lost from the jet cone.
The largest energy loss component is found to be the energy flow out of the gluon spectrum (which is independent of $R_0$).
The radiation directly outside the jet cone gives very small contribution to full jet energy loss.
This full jet energy loss picture is very similar to what has been found in Ref. \cite{Qin:2010mn}.

\subsection{Effects of medium recoils}

\begin{figure}
\begin{center}
\includegraphics[width=7.5 cm,bb=15 150 585 687]{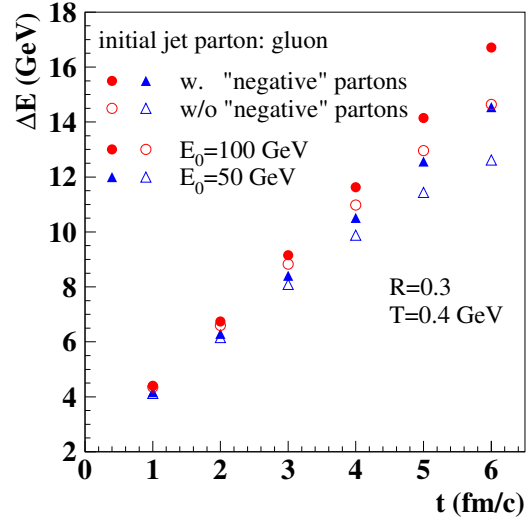}\\
\caption{(Color online) Energy loss of reconstructed jets due to elastic scattering from an initial gluon with energy $E_0=50$ and 100 GeV in a uniform QGP medium at a temperature $T=400$ MeV as a function of propagation time. Jets are reconstructed with all partons (leading and jet-induced medium partons) (solid symbols) or without ``negative" partons (open symbols) \cite{He:2015pra}. }
\label{jet-eloss}
\end{center}
\end{figure}

During the propagation jet shower partons, the leading parton loses energy and experiences transverse momentum broadening due to multiple scattering. The lost energy and momentum transfer will propagate through the medium via recoiled thermal partons, which lead to jet-induced medium excitations.  Some of these recoiled medium partons will be distributed within the jet cone while an increasing fraction will be transported outside the jet cone over the course of the jet transport. Inclusion of the jet-induced medium excitation in the jet reconstruction will affect the final jet energy loss and jet profiles. Such effects of medium recoil can be studied within the LBT model \cite{Li:2010ts,Wang:2013cia,He:2015pra}.  Shown in Fig.~\ref{jet-eloss} is the energy loss of reconstructed jets using the FASTJET algorithm with a cone size $R=0.3$ as a function of time for an initial gluon with energy $E_0=50$ and 100 GeV propagating through a uniform QGP medium with a constant temperature $T=400$ MeV. Only elastic scattering with a constant $\alpha_{s}=0.3$ is considered. The open symbols represent jets without ``negative" partons due to back-reaction while
solid symbols represent jets in which energy of the ``negative" partons are subtracted.

Th effect of ``negative" partons from back-reaction is negligible during the early stage of the parton propagation.  At later times, as the number of ``negative" partons grow, they deplete significantly the thermal medium behind leading partons and effectively modify the background underlying the jet. When such modified background is taken into account via the subtraction of ``negative" partons from the jet cone, the effective jet energy loss becomes bigger as shown in Fig.~\ref{jet-eloss}. The linear distance dependence of the effective jet energy loss is restored only after the energy of ``negative" partons is subtracted.

\begin{figure}
\begin{center}
\centerline{\includegraphics[width=12.5cm]{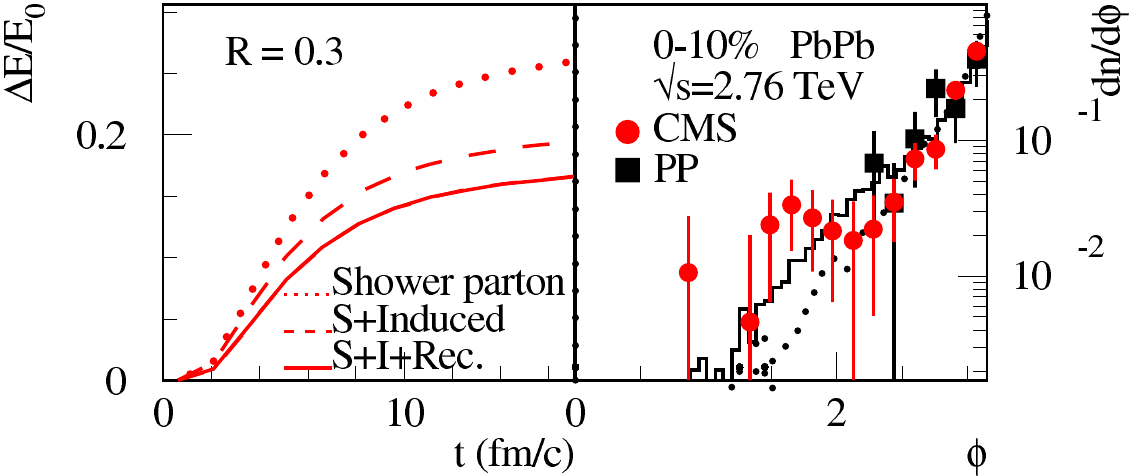}}
 \caption{(Color online) Averaged total energy loss as a function of time (left) and azimuthal jet distribution relative to the $\gamma$ (right) for $\gamma$-tagged jets
 in  central (0-10\%) Pb+Pb collisions at $\sqrt{s}=2.76$ TeV \cite{Wang:2013cia}.
  \label{hydro-jet}}
  \end{center}
\end{figure}

In a realistic case of heavy-ion collisions, one has to include both elastic scattering and induced gluon radiation in an expanding medium as described by relativistic hydrodynamic evolution.
In Ref. \cite{Li:2010ts,Wang:2013cia,He:2015pra}, the case of $\gamma$-tagged jets in heavy-ion collisions is considered.
The initial $\gamma$-jet production from {\footnotesize HIJING} is distributed according to the overlap function of two colliding nuclei with a Wood-Saxon nuclear geometry.
The same kinematic cuts for $\gamma$-jets are applied as in experiments at LHC. For CMS data \cite{Chatrchyan:2012gt},  $p_T^{\gamma}>60$ GeV, $|\eta^{\gamma}|<1.44$, $p_T^\text{jet}>30$ GeV, $|\eta^\text{jet}|<1.6$, and $\Delta\phi=|\phi^\text{jet}-\phi^\gamma|>7\pi/8$, and for ATLAS data \cite{ATLAS:2012cna}, $60< p_T^{\gamma} < 90$ GeV, $|\eta^{\gamma}|<1.3$, $p_T^\text{jet}>25$ GeV, $|\eta^\text{jet}|<2.1$, and $\Delta\phi>7\pi/8$.

Shown in Fig.~\ref{hydro-jet} (left) is the average total jet energy loss with a cone-size $R=0.3$ as a function of time in the most 10\% central Pb+Pb collisions at $\sqrt{s}=2.76$ TeV from LBT simulations with $\alpha_{\rm s}=0.2$. The space-time evolution of the medium is given by Hirano et al \cite{Hirano:2005xf}. Because of rapid cooling due to 3+1D expansion, the total jet energy loss flattens out approximately 10 fm/$c$ after an initial linear rise (solid line).
One can see that inclusion of recoiled partons significantly reduces the net jet energy loss as compared to the case when only shower partons (dotted) and radiated gluons (dashed) are included in the jet reconstruction.

Energy loss for a propagating parton is always accompanied by transverse momentum broadening. This should also be reflected in the $\gamma$-jet azimuthal correlation. As a consequence of the rapid medium expansion, the $\gamma$-jet azimuthal correlation in central Pb+Pb collisions (solid) as shown in Fig.~\ref{hydro-jet} (right) change very little as compared to that in p+p collisions (dotted). This is in agreement with CMS data \cite{Chatrchyan:2012gt} within errors. There is still, however, significant broadening at large values of azimuthal angle difference $\Delta\phi=\phi-\pi\equiv \phi_{\rm jet}-\phi_\gamma-\pi$. Future precision data are therefore necessary to measure $p_T$-broadening of reconstructed jets.

\begin{figure}
\begin{center}
\centerline{\includegraphics[width=8.0cm]{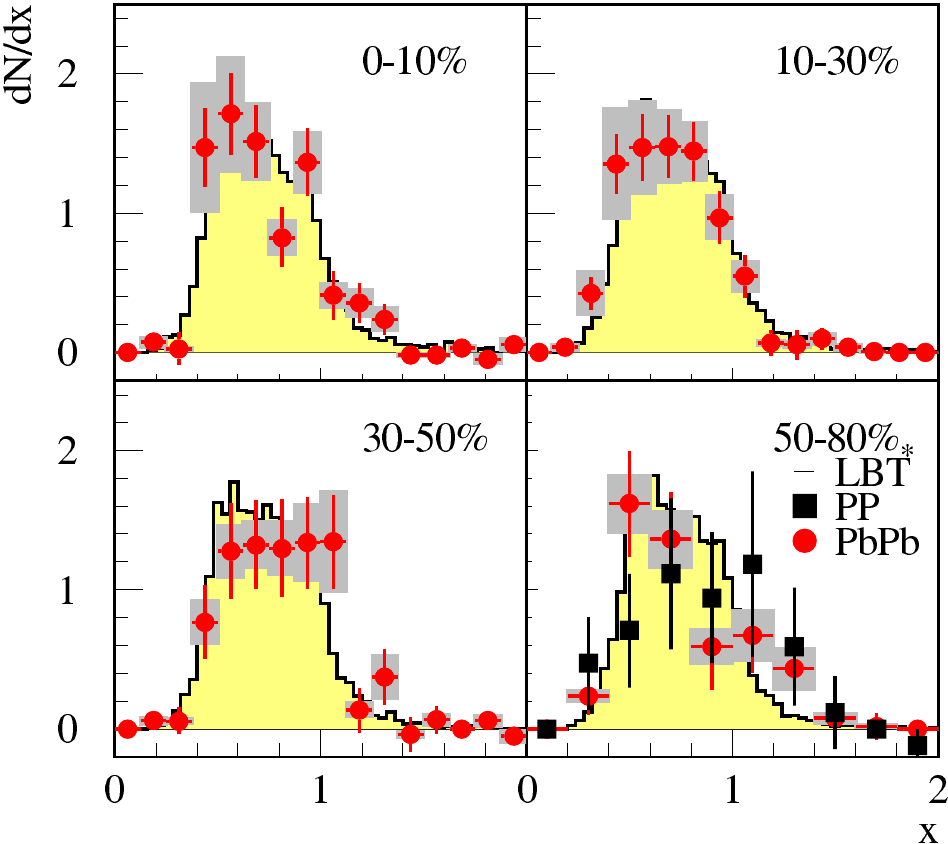}}
 \caption{(Color online) $\gamma$-jet asymmetry distribution in $x=p_T^\text{jet}/p_T^\gamma$ in Pb+Pb collisions  at $\sqrt{s}=2.76$ TeV from LBT with $\alpha_{ s}=0.2$
  as compared to CMS data \cite{Chatrchyan:2012gt} (from Ref. \cite{Wang:2013cia}).
 \label{asym}}
 \end{center}
\end{figure}

Shown in Fig.~\ref{asym} are $\gamma$-jet asymmetry distributions $dN/dx$ with $x=p_T^\text{jet}/p_T^\gamma$  from LBT model simulations (histogram) as compared to CMS  data \cite{Chatrchyan:2012gt} (solid circles)  in Pb+Pb collisions at $\sqrt{s}=2.76$ TeV with four different centralities. $\gamma$ jets are produced with a large momentum asymmetry in $p+p$ and peripheral Pb+Pb collisions  due to initial state radiations. In central Pb+Pb collisions,
the asymmetry distributions are skewed to smaller values of $x$ due to jet energy loss.  The LBT results can fit the experimental data of both $p+p$ and Pb+Pb with different centralities quite well with a fixed value of $\alpha_{s}=0.2$.

\subsection{Full jet substructure}

The elastic and inelastic interactions between the hard jet and the dense nuclear medium not only reduce the total energy of the reconstructed jet, but also change the momentum distribution among the constituents inside the jet. In particular, the scatterings experienced by quarks and gluons in the jet shower may result in both energy loss and deflection of jet constituents. The deflection and additional medium-induced radiation may broaden the parton shower and kick the propagating partons out of jet cone. To fully characterize the effects of jet quenching, one needs not only the modification of total full jet energy, but also the detailed information about the substructures of full jets.
There are two popular observables for the characterization of jet substructures: jet fragmentation function $D(z)$ and jet shape function $\rho(r)$ (sometimes written as $\psi(r)$).
Jet fragmentation function measures the momentum spectrum of hadrons inside the full jet while jet shape function measures the momentum distribution of the jet constituents in the radial direction.
Both jet shape function and jet fragmentation function have been measured for Pb-Pb collisions at the LHC by the ATLAS and CMS Collaborations \cite{Chatrchyan:2012gw, Chatrchyan:2013kwa, Aad:2014wha}.

\begin{figure}
\begin{center}
\includegraphics*[width=12.5cm]{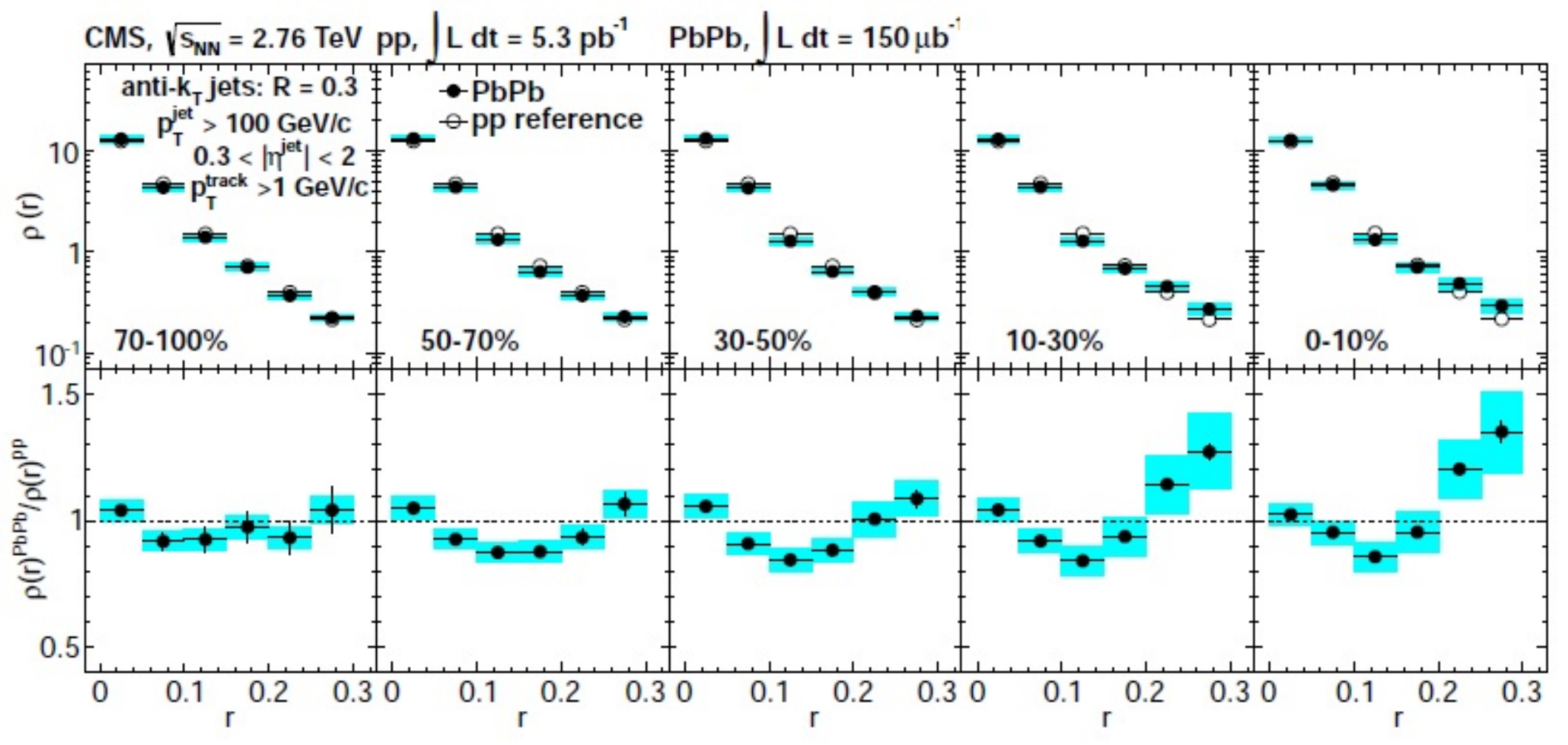}
\caption{(Color online) Top: CMS measurements of the differential jet shapes in Pb-Pb collisions (filled circles) as a function of distance from the jet axis for inclusive jets with $p_T^{\rm jet} > 100$~GeV/c and $0.3 < \eta < 2$~GeV/c in five Pb-Pb centrality intervals. Charged particles with
$p_T^{\rm track} > 1$~GeV/c are used.
Bottom: Jet shape nuclear modification factors, $\rho(r)^{PbPb}/\rho(r)^{pp}$.
Figures are taken from Ref. \cite{Chatrchyan:2012gw}.
}
\label{fig_CMS_jet_shape}
\end{center}
\end{figure}

Jet shape function $\rho(r)$ provides the information about the radial distribution of the momentum carried by the jet constituents (fragments). The differential jet shape function is defined as follows:
\begin{eqnarray}
\rho(r) = \frac{1}{\delta r} \frac{1}{N_{\rm jet}} \sum_{\rm jet}  \sum_{h} \frac{ p_T^h}{p_T^{\rm jet}} \theta[r_h - (r-\delta r/2)] \theta[(r+\delta r/2)-r_h],
\end{eqnarray}
where $r_h = \sqrt{(\eta_h - \eta_{\rm jet})^2 + (\phi_h - \phi_{\rm jet})^2}$, and $\delta r$ is the bin size.
By definition, the differential jet shape function is normalized to unity, $\int \rho(r) dr = 1$.
One may define the integrated jet shape function as: $R(r) = \int_0^r \rho(r) dr$.
One of the important consequences from jet quenching is the broadening of full jets as compared to vacuum jets, due to a combination of medium-induced radiation and the deflection of shower partons.
This effect is usually reflected as an enhancement at large values of $r$ in differential jet shape function $\rho(r)$.
Due to the normalization of jet shape function to unity, and an enhancement at large $r$ must be compensated by a depletion at smaller values of $r$.

Fig. \ref{fig_CMS_jet_shape} shows the nuclear modification of differential jet shape function $\rho(r)$ in Pb-Pb collisions with respect to proton-proton collisions at the LHC measured by CMS Collaborations \cite{Chatrchyan:2012gw}.
A strong centrality dependence is observed for the jet shape functions in Pb-Pb collisions.
In most peripheral collisions, the jet shape functions in Pb-Pb collisions are similar to those in proton-proton collisions, which indicates the radial distribution of the particle momentum inside the jets is similar in these two collision systems.
In more central Pb-Pb collisions (0-10\% and 10-30\%), one observes an excess of transverse momentum fraction emitted at large radius $r > 0.2$ together with a depletion at intermediate radii, $0.1 < r < 0.2$.
This indicates a moderate broadening of the jets after their traversing through the medium.
There is very little change at small radii, indicating that the energy distribution in the inner core of the jet is not affected by the jet-medium interaction.
The result is consistent with previous CMS measurements in which the lost energy from the jets to the medium is found at large distances from the jet axis outside the jet cone \cite{Chatrchyan:2011sx}.
It is expected from jet quenching picture and consistent with some theoretical calculations \cite{Vitev:2008rz, Ramos:2014mba}.

\begin{figure}
\begin{center}
\includegraphics*[width=12.5cm]{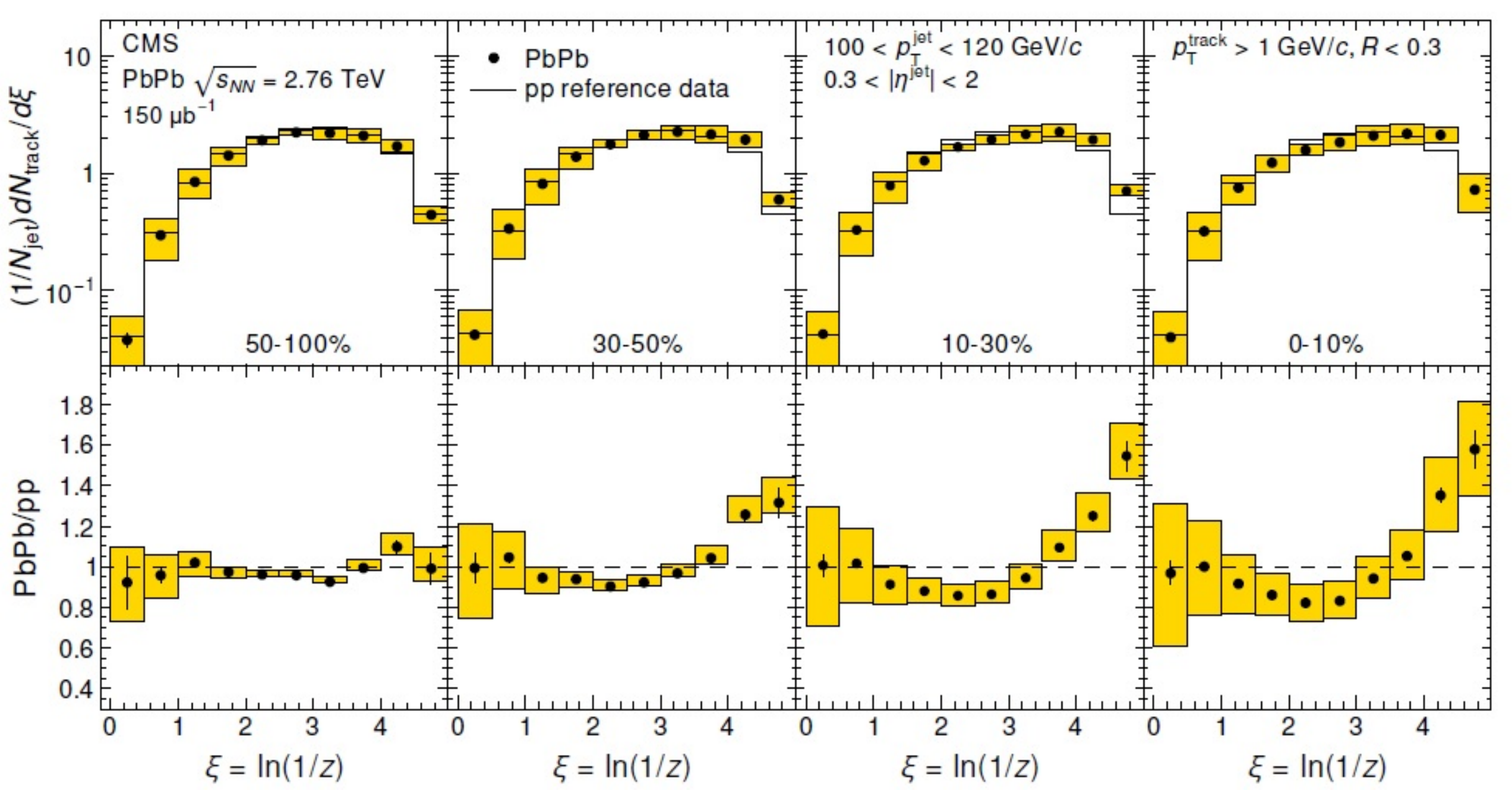}
\caption{(Color online) Top: CMS measurements of the fragmentation functions in bins of centrality (increasing from left to right) in Pb-Pb collisions overlaid with proton-proton reference data. Jets have $100 < p_T < 120$~GeV/c, and tracks have $p_T > 1$~GeV/c.
Bottom: The ratio of each Pb-Pb fragmentation function to its pp reference.
Figures are taken from Ref. \cite{Chatrchyan:2013kwa}.
}
\label{fig_CMS_jet_ff}
\end{center}
\end{figure}

\begin{figure}
\begin{center}
\includegraphics*[width=12.5cm]{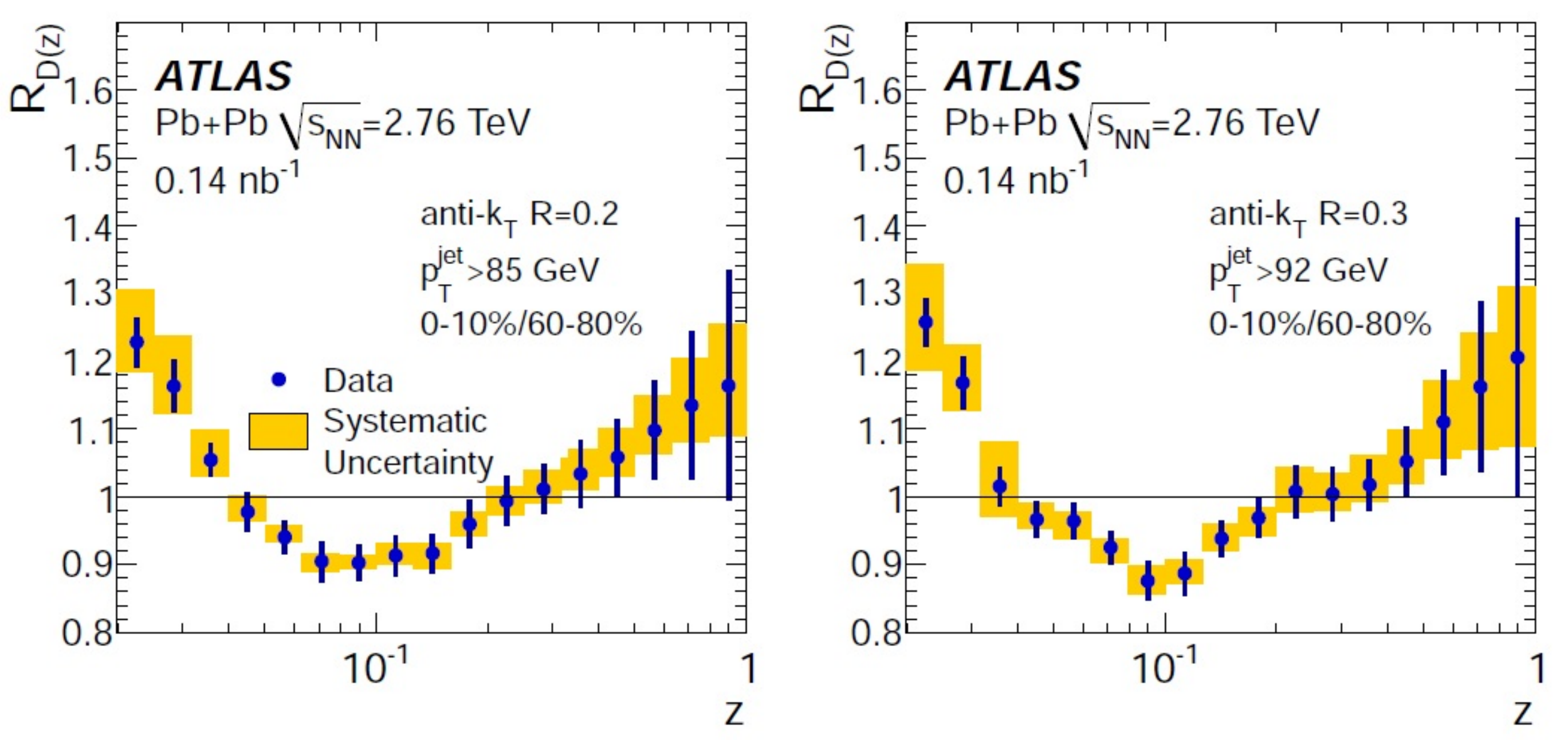}
\caption{(Color online) ATLAS measurements of the ratios of fragmentation functions, D(z) for central (0-10\%) collisions to those in peripheral (60-80\%) collisions for R = 0.2 (bottom left) and R = 0.3 (bottom right) jets.
Figures are taken from Ref. \cite{Aad:2014wha}
}
\label{fig_ATLAS_jet_ff}
\end{center}
\end{figure}

Jet fragmentation function $D(z)$ provides the information about the spectrum of the momentum carried by the jet fragments, where $z =\mathbf{p}_T^{h}\cdot \mathbf{p}_T^{\rm jet} / |\mathbf{p}_T^{\rm jet}|^2$ is the momentum fraction of the jet fragment. Jet fragmentation function is defined as follows:
\begin{eqnarray}
D(z) = \frac{1}{N_{\rm jet}} \frac{dN_h}{dz}.
\end{eqnarray}
The interaction between the hard jet and the dense medium usually leads to a soften of jet fragmentation function due to additional medium-induced radiation.
This will be reflected in the jet fragmentation function as an enhancement at small values of $z$.
Due to the energy conservation, an enhancement at small $z$ must be accompanied by a depletion at higher values of $z$.

The first CMS measurement of jet substructure showed little nuclear modification to the jet fragmentation function (with large uncertainties) in Pb-Pb collisions compared to proton-proton collisions \cite{Chatrchyan:2012gw}.
Later, CMS Collaboration included the charged particles with lower values of $p_T$ in the reconstruction of full jets and a clear nuclear modification of the inclusive jet fragmentation function was observed \cite{Chatrchyan:2013kwa}.
Fig. \ref{fig_CMS_jet_ff} shows the latest CMS measurement of jet fragmentation function (the top panels show jet fragmentation function in Pb-Pb and proton-proton collisions and the lower panels show their ratios, PbPb/pp).
One can see that the nuclear modification of the fragmentation function of jets in Pb-Pb collisions grows with increasing collision centrality.
The ratio of PbPb/pp is almost flat at unity in the peripheral collisions (50-100\% centrality bin), which indicates very little nuclear modification to particle spectrum inside the jet in peripheral Pb-Pb collisions.
For the most central 0-10\% collisions, a significant excess is observed at high $\xi=\ln(1/z)$ (low $z$), together with a depletion in the intermediate $\xi$ regime.
These results indicate that the spectrum of particles inside the full jet has an enhanced contribution in the soft regime, as compared to that from proton-proton collisions.
CMS has also measured jet fragmentation function $D(z)$ for different values of jet transverse momentum $p_T^{\rm jet}$ (up to $300$~GeV/c).
No significant $p_T^{\rm jet}$ dependence was observed for the modification pattern within statistical and systematic uncertainties.

ATLAS Collaboration has also measured the nuclear modification of jet fragmentation function for Pb-Pb collisions at the LHC.
Fig. \ref{fig_ATLAS_jet_ff} shows the ratios of fragmentation functions, $D(z)$, for central (0-10\%) collisions to those in peripheral (60-80\%) collisions, for two different values of jet cone size ($R = 0.2$ and $R=0.3$).
Similar to CMS results, an enhanced yield of low $z$ and large $z$ fragments together with a suppressed yield of fragments at intermediate $z$ was observed in more central collisions compared to more peripheral collisions.

\begin{figure}
\centerline{\includegraphics[width=12.5cm]{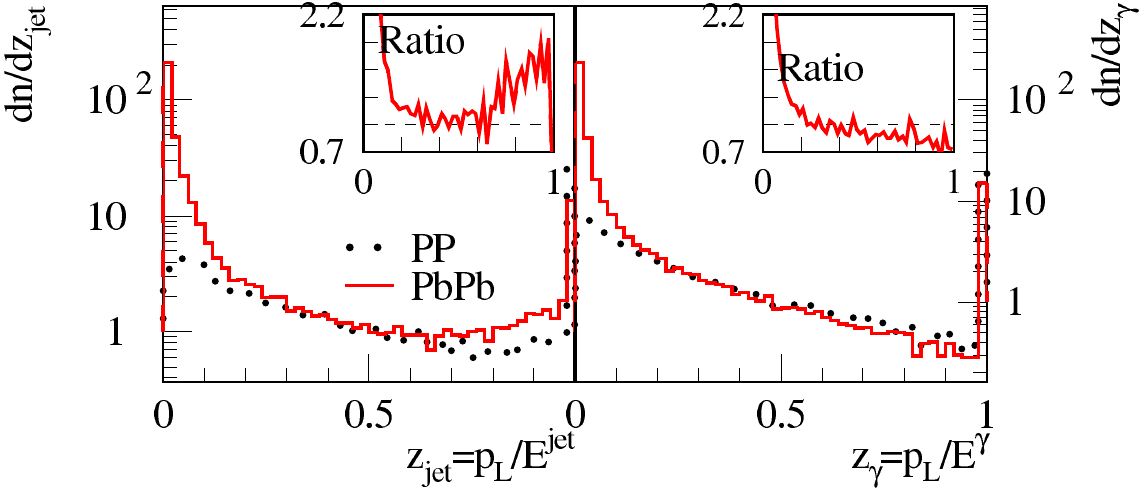}}
 \caption{(Color online) The reconstructed (left) and $\gamma$-tagged jet fragmentation functions (right)
 within a jet cone $R=0.3$ of $\gamma$-tagged jets in central (0\%--10\%) Pb+Pb collisions at $\sqrt{s}=2.76$ TeV \cite{Wang:2013cia}.
 \label{hydro-frag}}
\end{figure}

Jet substructure and their medium modification have also been studied in some details by some parton energy loss models \cite{Qin:2012gp,Wang:2013cia,Ma:2013gga,Majumder:2013re, Perez-Ramos:2014mna}. For example, features of the observed medium modification of jet substructure can
be qualitatively reproduced and understood by the LBT model for jet propagation.
Shown in Fig.~\ref{hydro-frag} (left) is the jet fragmentation function for partons within the jet cone as a function of momentum fraction $z_{\rm jet}$ that is defined by the reconstructed jet energy $z_{\rm jet}=p_{L}/E_{\rm jet}$ as in the LHC experimental data on single jets \cite{Aad:2014wha}.
A large enhancement at small $z_{\rm jet}$  is due to soft partons from the bremsstrahlung and medium recoil within the jet-cone. There is also almost no medium modification at intermediate $z_{\rm jet}$. Similarly to the experimental data, there is an enhancement
at large $z_{\rm jet}$. This enhancement  is caused by the dominance of leading partons in the reconstructed jet energy whose
energy is reduced because a large fraction of nonleading partons is transported outside the jet cone.
To illustrate this point, we plot in Fig.~\ref{hydro-frag} (right) $\gamma$-tagged jet fragmentation function \cite{Wang:1996yh} whose momentum fraction is defined as $z_\gamma=p_{L}/E_\gamma$ by the $\gamma$ energy which more or less reflects the initial jet energy before medium modification.
As a consequence of parton energy loss, the $\gamma$-tagged jet fragmentation function at large $z_\gamma$ is
significantly suppressed. In addition, there is also an enhancement at small $z_\gamma$. Therefore, such $\gamma$-tagged jet fragmentation
function should be more sensitive to parton energy loss and can be used to extract jet transport parameters of the hot medium in heavy-ion collisions.

In general, there are several possible sources that are helpful to explain the observed nuclear modifications and the redistribution of jet energy inside the cone.
First, the change in the fraction of jets originating from quarks and gluons may give different energy distribution inside jet cone.
The extra radiation associated with inelastic collision process may lead to the broadening of the parton shower, and the softening of the spectrum of the jet fragments.
Particles resulting from the medium response to jet transport may also contribute to the energy redistribution in the final reconstructed jets.
For example, different ways of treating the recoiled partons may affect the final jet substructure and the energy distribution inside the jet \cite{Wang:2013cia}.
Fragmentation and recombination mechanisms may lead to different jet fragmentation profiles \cite{Ma:2013gga}.
In short, the understanding of the interplay between the nuclear modifications in different $r$ of jet shape function and different $z$ of jet fragmentation function can put very tight constraints on modeling jet-medium interaction and medium-induced energy loss.
This provides a new window to probe various transport properties of the hot and dense QGP created in relativistic nuclear collisions.

\section{Medium response to jet transport}
\label{sec_medium_response}

Hard partonic jets lose energy when propagating through QGP medium; some of the lost energy from the hard jets is deposited into the traversed medium.
The deposited energy and momentum from the jets may thermalize into the medium, which may contribute to many jet-related observables, such as the angular correlations of hadrons with triggered high-$p_T$ hadrons or direct photons.
The space-time evolution and the final anisotropic flow pattern of the bulk matter may be affected by the transport and mementum deposition of the hard jets \cite{Andrade:2014swa,Schulc:2014jma}.
Particles produced from jet-induced medium excitations may flow into the jet cone and contribute to the energy redistribution in the final reconstructed full jets.
Therefore, the study of the medium response to jet-deposited energy and the knowledge of the fate of medium excitations are essential for a complete understanding of jet-medium interaction.

One of the most interesting phenomena in studying the medium response to hard jets is the induced Mach cone structure \cite{CasalderreySolana:2004qm, Ruppert:2005uz}.
If observed in heavy-ion experiments, it provides a direct probe to the speed of sound of the hot and dense nuclear medium.
But in the dynamically evolving media as produced in relativistic heavy-ion collisions, the Mach cone pattern may be strongly distorted by the large collective flow due to the hydrodynamic expansion of the bulk matter \cite{Renk:2005si, Ma:2010dv, Bouras:2014rea}.
Jet-induced Mach cone structure is also sensitive to many transport properties of the traversed medium, such as the shear viscosity to entropy density ratio \cite{Neufeld:2008dx, Bouras:2014rea}.

Similar to jet quenching and energy loss studies, the study of the medium response to hard jets may be performed in two different frameworks: full Boltzmann transport approach and jet plus hydrodynamics approach.
In the full Boltzmann transport approach such as the AMPT and BAMPS models \cite{Lin:2004en, Xu:2004mz}, both the propagation of the jets in the medium and the medium response to jet transport are simulated at the same time in one Monte-Carlo transport package.
In these transport models, the bulk matter is usually described by a collection of quasi-classical quarks and gluons.
The interaction between jets and the medium constituents has the same cross section as the interaction among medium constituents.
In order to describe the strong collective flow behavior of the bulk matter as indicated by relativistic hydrodynamics calculation and experimental data, usually a very large partonic cross section (or large coupling constant) is used in these calculations .

\begin{figure}
\begin{center}
\includegraphics*[width=12.5cm]{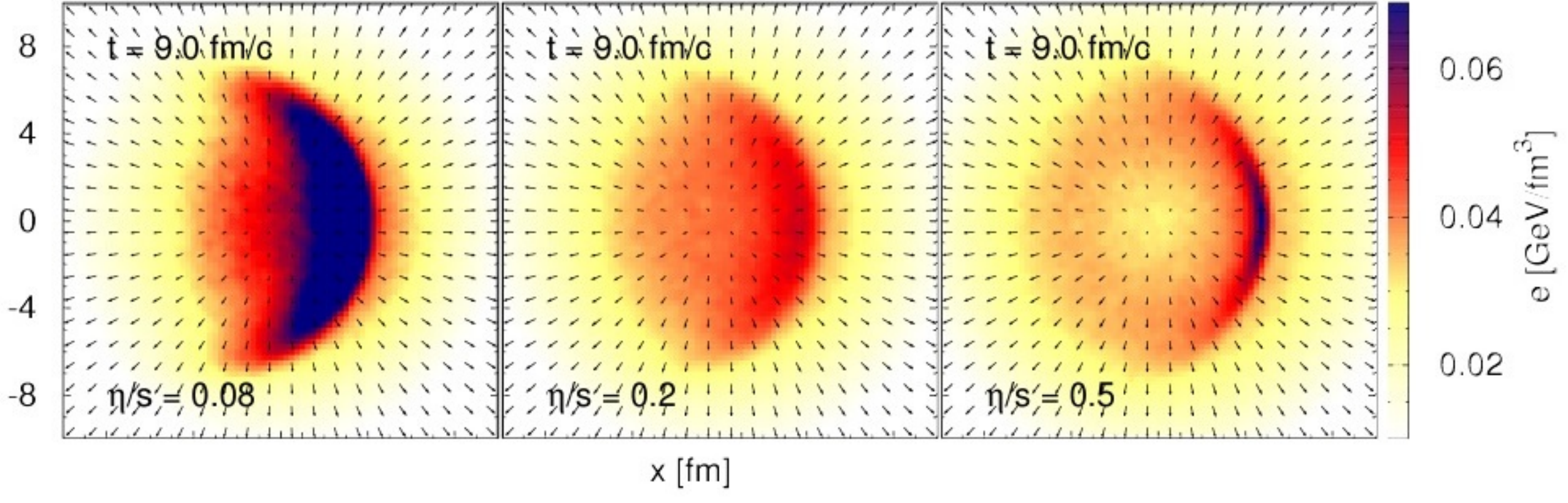}
\includegraphics*[width=12.5cm]{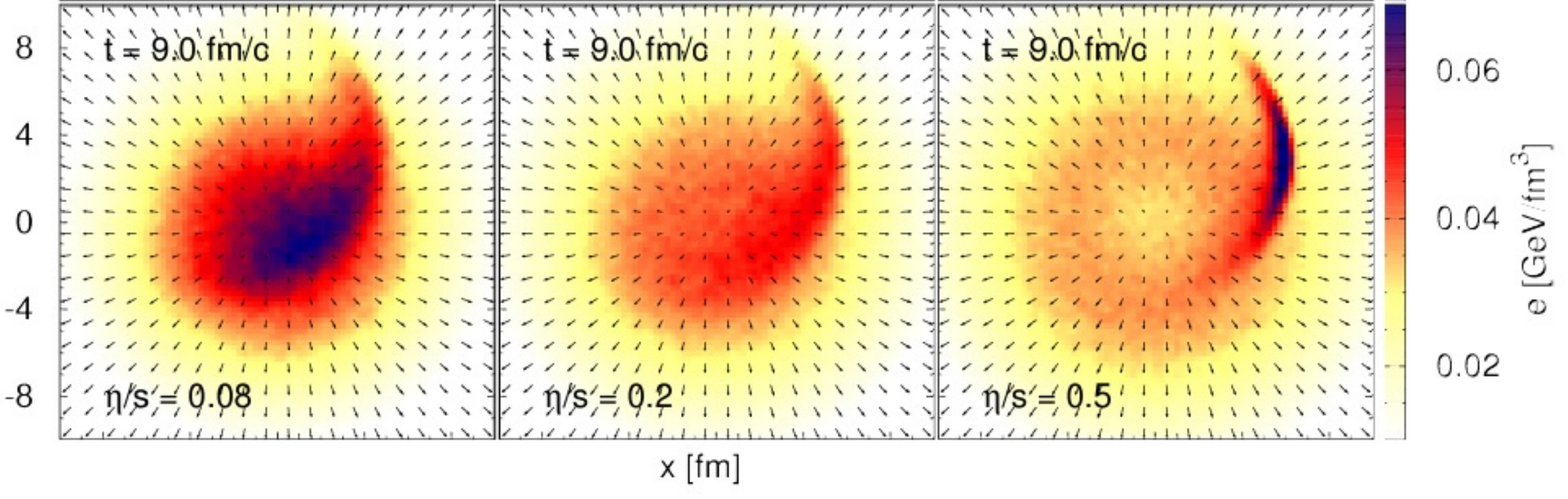}
\caption{
(Color online) Energy density distribution in the transverse plane around mid-rapidity at $\tau=9$~fm/c, overlaid by the velocity profile indicated by arrows.
The results are shown three different values of shear viscosity to entropy density ratio $\eta/s$.
The parton is initialized at a fixed-angle position of $(r, \phi) = (4~{\rm fm}, \pi)$ (upper panels) and $(r, \phi) = (4~{\rm fm}, 3\pi/4)$ (lower panels) with an initial momentum along the direction of $x$-axis, $p_x = E = 20$~GeV.
Figures are taken from Ref. \cite{Bouras:2014rea}.
}
\label{fig_Greiner_mach_cone}
\end{center}
\end{figure}

Fig. \ref{fig_Greiner_mach_cone} shows a simulation of the medium response to high-energy partons using the BAMPS parton cascade model \cite{Bouras:2014rea}.
In the calculation, the medium is treated as a gas of massless gluons, which follow the classical Boltzmann equation,
\begin{eqnarray}
\left( \frac{\partial}{\partial t} + \frac{\mathbf{p}}{E} \cdot \frac{\partial}{\partial \mathbf{r}} \right) f(\mathbf{x}, \mathbf{p}, t) = C[f].
\end{eqnarray}
Particles (both jet and medium partons) collide only via binary collisions with an isotropic cross section.
The effect of shear viscosity is investigated by changing the parton cross section as follows:
\begin{eqnarray}
\eta/s = 0.4 e/(n\sigma),
\end{eqnarray}
where $e$ and $n$ are local energy and particle densities.
In the plots, the parton is produced at $(r, \phi) = (4~{\rm fm}, \pi)$ in the upper panels and at $(r, \phi) = (4~{\rm fm}, 3\pi/4)$ in the lower panels, and traveling in the positive $x$-axis direction.
The energy density distribution of the bulk matter in the transverse plane is shown, together with the velocity profiles.
Three different values of shear viscosity to entropy density ratio $\eta/s$ (from left to right) are used.
The Mach cone structures are clearly seen for small viscosities, whereas for larger viscosities the characteristic structures are smeared out.
When jets propagate in the directions not parallel to the radial flow (lower panels), the Mach cone patterns are strongly distorted by the radial flow.

In jet plus hydrodynamics approach, the space-time evolution of the hot and dense nuclear medium is first simulated by relativistic hydrodynamics.
Then one calculates the medium effect on jet transport using the jet-quenching techniques as described in previous sections.
The medium response to jet transport is then simulated via solving the hydrodynamic equations with a source term,
\begin{eqnarray}
\partial_{\mu} T^{\mu \nu}(x) = J^{\nu}(x).
\end{eqnarray}
Here the source term $J^{\nu}(x)$ represents the space-time profiles of the energy and momentum deposited by the propagating hard jets,
\begin{eqnarray}
J^{\mu}(x) = \left(\frac{dE}{dt}, \frac{d\mathbf{p}_\perp}{dt}, \frac{d {p}_\parallel}{dt} \right),
\end{eqnarray}
where $\mathbf{p}_\perp$ ($p_\parallel$) are the momentum components transverse (longitudinal) to the jet propagation direction.
Currently, a parallel simulation of both jet evolution and the medium response at the same time is not yet available as it is time-consuming.
Most studies only calculate the medium response to jet transport, and simply neglect the effect of the medium response (change) on the subsequent evolution of hard jets.

At a given time and space location, the energy and momentum deposited by the hard jet, the source term $J^{\mu}(x)$), can be obtained from jet quenching calculations.
The energy deposited into the medium is equal to the collisional energy loss experienced by the propagating jet.
For a single parton propagating through a QGP medium, the collisional energy loss rate may be calculated within high temperature field theory.
Taking the leading logarithmic approximation, the collisional energy loss (energy deposition) rate is given by
\begin{eqnarray}
\left.\frac{dE}{dt}\right|_{\rm dep} = \left.\frac{dE}{dt}\right|_{\rm coll} = \frac{1}{4} C_s \alpha_s^2 m_D^2 \left(\frac{4ET}{m_D^2}\right),
\end{eqnarray}
where $m_D^2 = 4 \pi \alpha_s (1 + N_f/6) T^2 $ is the Debye screening mass squared and $C_s$ is the color factor for the propagating parton.
The momentum deposition rate can also be obtained in a similar way.

\begin{figure}
\begin{center}
\includegraphics*[width=12.5cm]{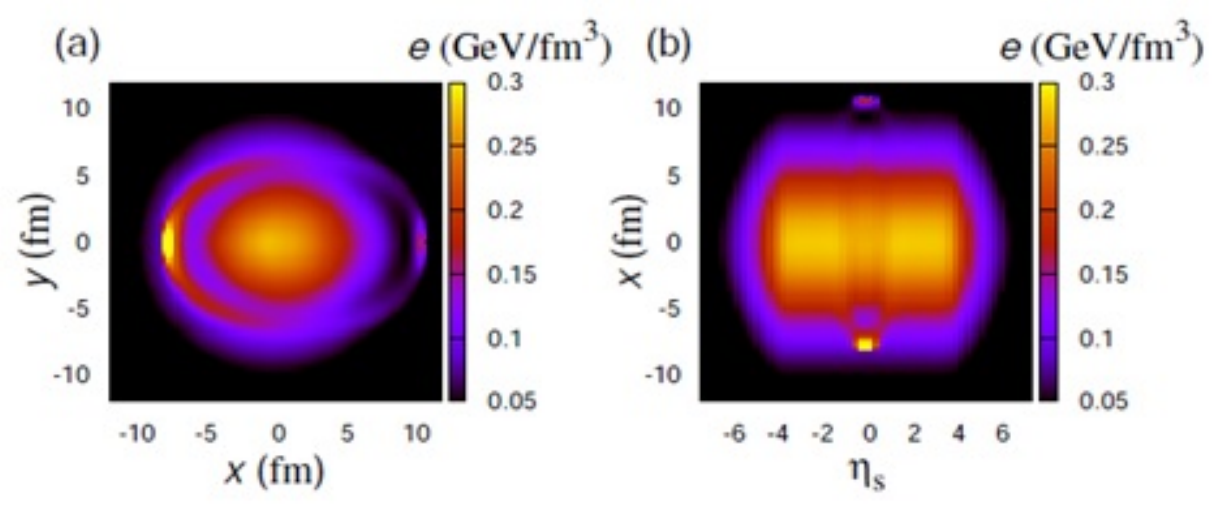}
\caption{(Color online) The energy density distribution of the QGP medium at $\tau = 9.6$~fm/c in the transverse plane
at $\eta_s = 0$ and the reaction plane at $y = 0$.
A pair of partons with the same energy $E_0 = 200$~GeV are produced at ($\tau = 0$, $x = 1.5$~fm, $y = 0$, $\eta_s = 0$) and travel in the opposite (positive and negative $x$-axis) directions.
Figures are taken from Ref. \cite{Tachibana:2014lja}.
}
\label{fig_Tachibana_mach_cone}
\end{center}
\end{figure}

\begin{figure}
\begin{center}
\includegraphics*[width=13.0cm]{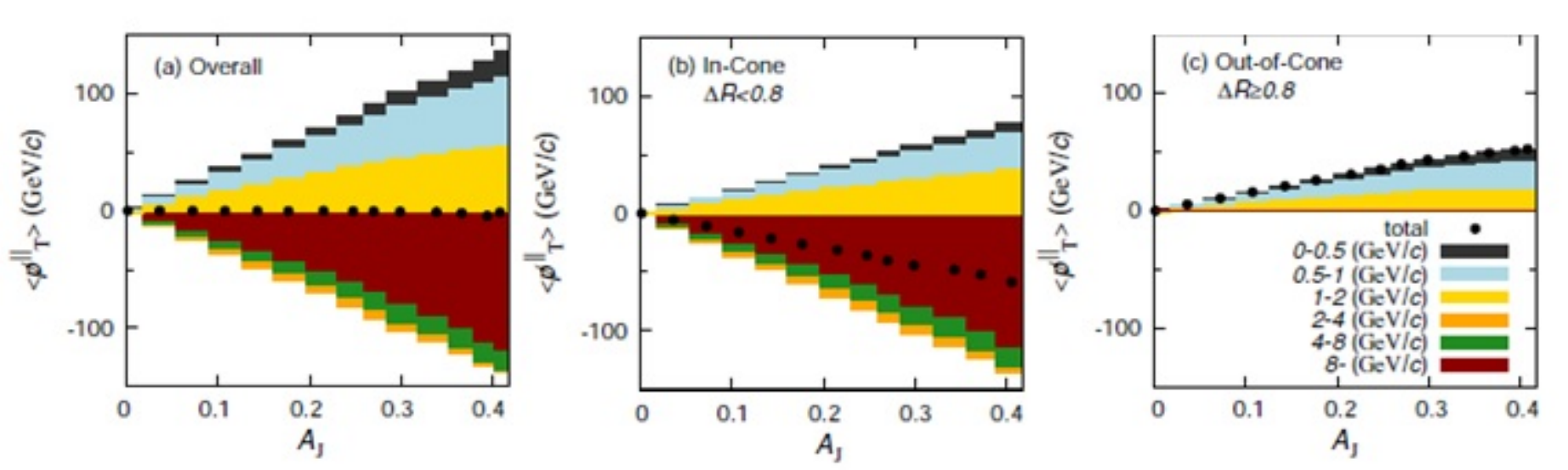}
\caption{(Color online) $p_T^{||}$ as a function of the dijet energy asymmetry variable $A_J$ for the whole region (left panel), inside the jet cone $\Delta R < 0.8$ (middle panel), and outside jet cone $\Delta R > 0.8$ (right panel). Six bands represent the contributions in six $p_T$ ranges: 0-0.5, 0.5-1, 1-2, 2-4, 4-8 GeV/c, and $p_T > 8$~GeV/c. The solid circles include the contributions from all $p_T$ (and the propagating partons).
Figures are taken from Ref. \cite{Tachibana:2014lja}.
}
\label{fig_Tachibana_energy_redistribution}
\end{center}
\end{figure}

After the energy deposition profiles $J^{\mu}(x)$ are obtained from jet quenching calculation, the medium response can be simulated via solving hydrodynamic equation.
Fig. \ref{fig_Tachibana_mach_cone} shows the medium response to the lost energy by a pair of back-to-back partons as simulated via a (3+1)-dimensional ideal hydrodynamics simulation \cite{Tachibana:2014lja}.
In the calculation, a pair of back-to-back partons are created at ($\tau= 0$, $x_0 = 1.5$~fm, $y_0 = 0$, $\eta_s = 0$) with the energy $E_0 = 200$~GeV and travel in the opposite direction along the $x$ axis.
The initial location is chosen as $x_0 = 1.5$~fm to mimic the the different amounts of energy loss between two jets.
The interaction between partons and the medium is switched on at $\tau_0 = 0.6$~fm/c.
The figure shows the energy density profiles of the QGP medium at the time $\tau = 9.6$~fm/c in the transverse ($xy$) plane at $\eta_s = 0$ (left panel) and the reaction plane ($x\eta$ plane) at $y=0$ (right panel).
One may see clearly the medium response and excitations generated by two energetic partons, as well as the distortion of Mach cone patterns by radial flow in the transverse plane.

Given the hydrodynamic evolution of the medium response, it is very interesting to see how the lost energy and medium excitations redistribute in the final state after hydrodynamic evolution.
One may calculate the total $p_T$-balance of the entire system as follows:
\begin{eqnarray}
\langle p_T^{\parallel} \rangle = - \sum_{i} \int dp_T d\phi_p p_T \cos(\phi_p - \phi_1) \frac{dN_i}{dp_T d\phi_p},
\end{eqnarray}
where the sum is taken over all particles in the entire system. Here the projection ``$\parallel$" is taken onto the sub-leading jet axis in the transverse plane.
The momentum distribution of particles may be calculated from hydrodynamics using the Cooper-Fry formula.

In fact, this simulation was motivated by the CMS measurements of dijet momentum imbalance, where it is found that the lost energy from the jet are carried by soft particles at large angles \cite{Chatrchyan:2011sx}.
To mimic dijet momentum imbalance $A_J$, one may change the initial position of the parton pairs $x_0$ such that two partons lose different amounts of energy loss.
The results of the final state momentum distribution are shown in Fig. \ref{fig_Tachibana_energy_redistribution}.
One can see that the total transverse momentum of the whole system is well balanced (left panel).
Inside jet cone (middle panel) the transverse momentum is dominated by hard-$p_T$ particles.
Outside jet cone (right panel) particles with  $p_T<2$~GeV dominate, and these low-$p_T$  particles originate from the jet-deposited energy and momentum which are then transported by hydrodynamic fluid.
This result is qualitatively consistent with CMS measurements of dijet momentum imbalance and the redistribution of lost energy \cite{Chatrchyan:2011sx}.

It should be noted that in the above calculations and many other studies of the medium response to jet transport, a simplified energy deposition deposition profile (for a single parton) is taken as the input to simulate the hydrodynamic medium response.
In fact, jets are collimated showers of partons which may lose energy by a combination of elastic scatterings with medium constituents and inelastic medium-induced emissions.
In fact, both elastic and inelastic collisions play important roles in the study of energy/momentum deposition by the jet shower and the medium response.
The radiative emissions can also deposit energy/momentum into the medium via rescattering with the medium constituents, thus the energy/momentum deposition profile for a jet shower is quite different from a single parton.
Such effect has been investigated in Ref. \cite{Qin:2009uh,Neufeld:2009ep} where it is found that the length dependence of energy deposition rate of a jet shower is very different from that of a single parton.

\begin{figure}
\begin{center}
\includegraphics*[width=6.1cm]{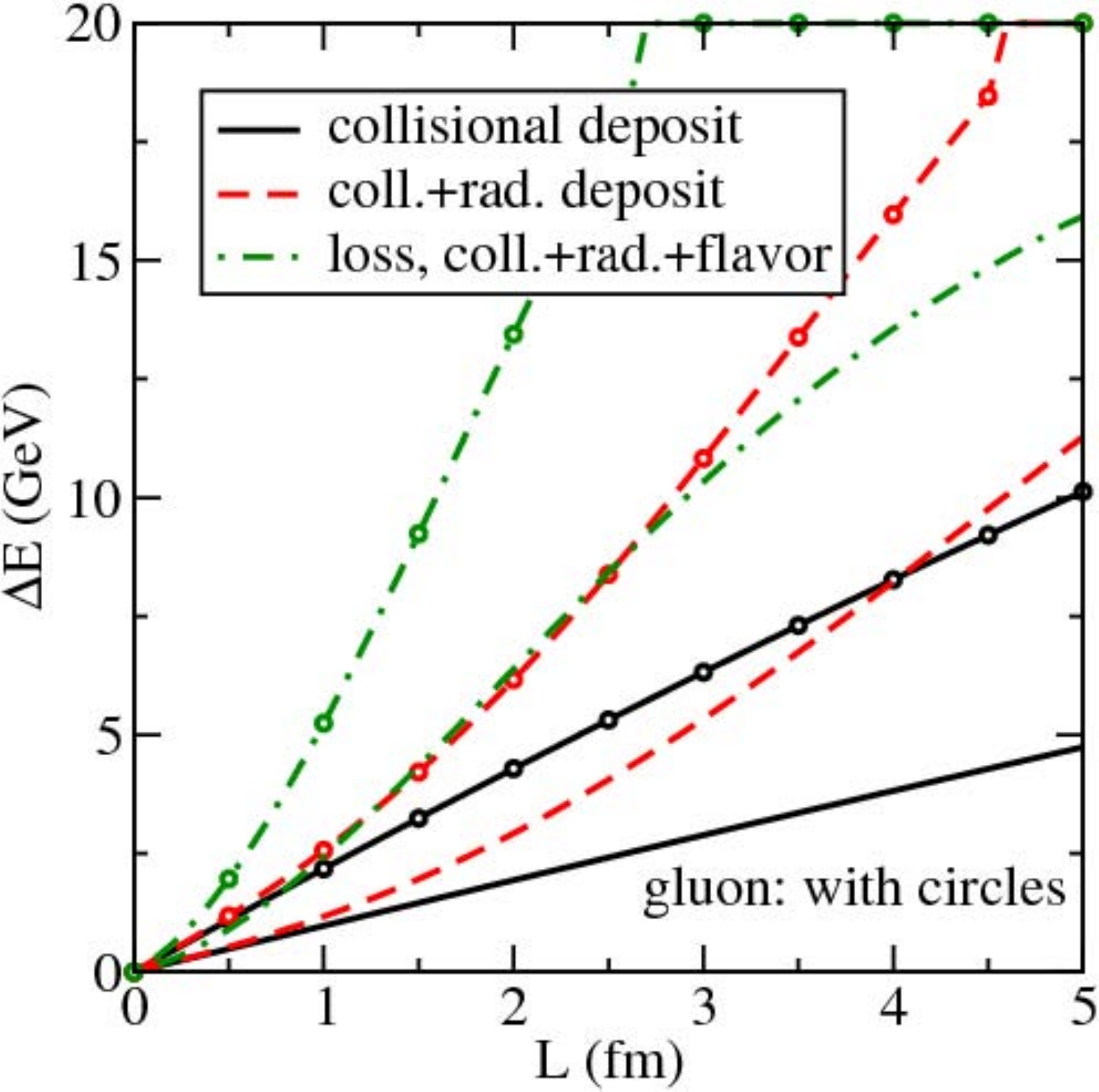}
\includegraphics*[width=6.3cm]{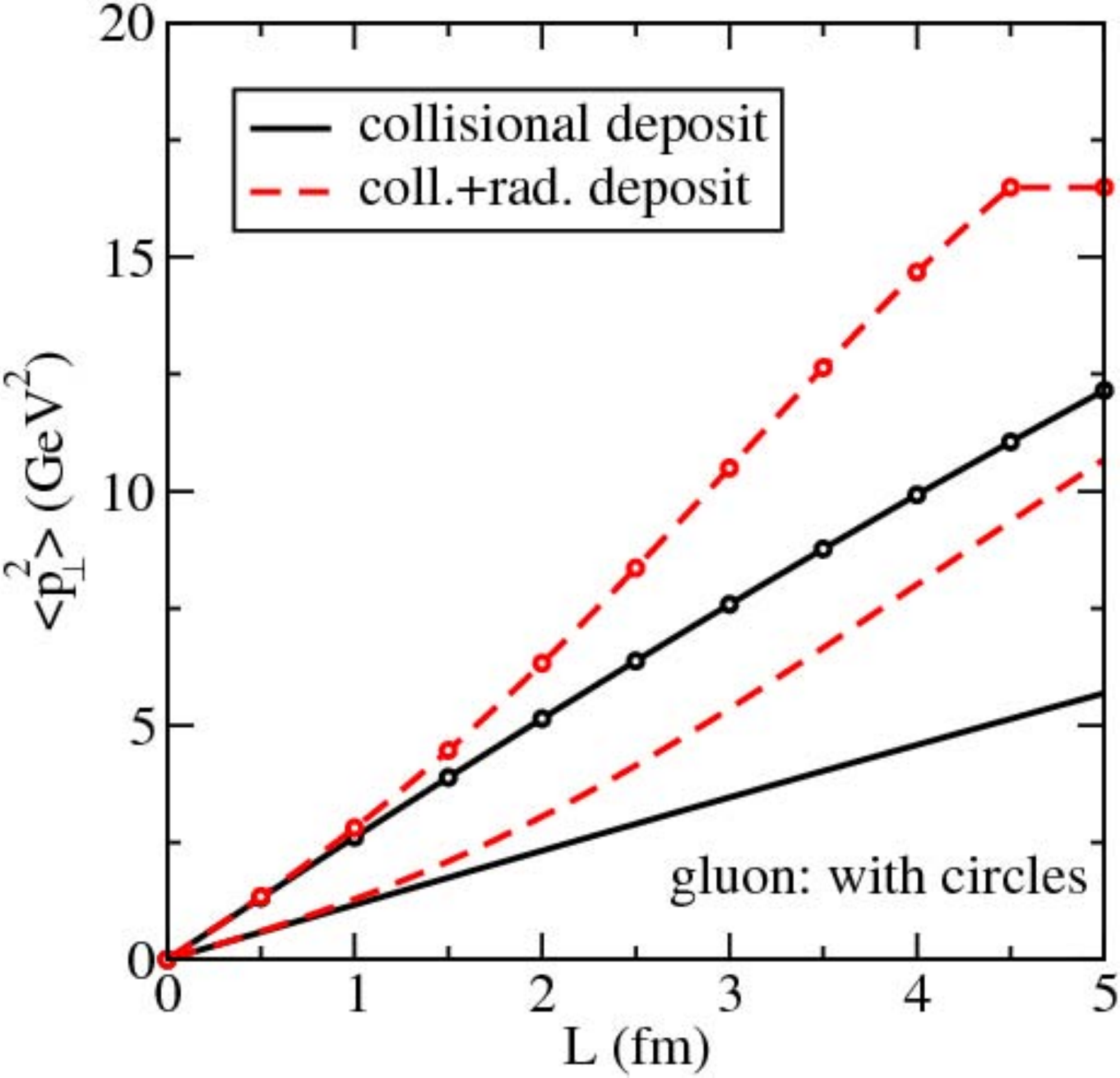}
\caption{(Color online) Left panel: Solid lines denote the energy deposited in the medium by a hard parton which does not radiate. Dashed lines denote the energy deposition by a virtual parton evolving into a partonic shower. Dash-dotted lines denote the total energy lost by a hard parton by radiative, elastic together with flavor changing processes.
Right panel:  Solid lines denote the transverse momentum squared deposited in the medium by a hard parton which does not radiate. Dashed lines denote the transverse momentum squared deposition by a virtual parton evolving into a partonic shower.
Figures are taken from Ref. \cite{Qin:2009uh}.
}
\label{fig_Qin_ep_deposition}
\end{center}
\end{figure}

This can be clearly seen in Fig. \ref{fig_Qin_ep_deposition}, where the energy (left) and transverse momentum squared (right) deposited into the medium by a single parton which does not radiate are compared with those by a virtual parton which evolves into a parton shower.
In the figure, the total energy loss from a single parton is also shown for comparison.
To calculate the energy deposition from a parton shower developed from a virtual parton, one may solve the following DGLAP-like evolution equation \cite{Qin:2009uh}:
\begin{eqnarray}
&&\!\! \frac{\partial \Delta E_j(Q^2, q^-, \zeta_i, \zeta_f)}{\partial \ln Q^2} = \sum_{k,l} \int \frac{dy}{y} P_{j \to kl}(y) \int_{\zeta_i}^{\zeta_j}  d\zeta K_{j \to kl}(Q^2, q^-, y, \zeta) \nonumber \\ &&\!\!
\times \left[\Delta E_j(Q^2, q^-, \zeta_i, \zeta) + \Delta E_k(Q^2, q^-y, \zeta, \zeta_f) + \Delta E_l(Q^2, q^-(1-y) \zeta, \zeta_f) \right],
\end{eqnarray}
where $Q^2$ and $q^-$ are the virtuality and light-cone energy of the propagating jet.
$\Delta E(Q^2, q^-, \zeta_i, \zeta_f)$ is the total energy deposited by a virtual parton traveling from the location $\zeta_i$ to the location $\zeta_f$.
So the total energy deposition $\Delta E$ is now dependent on the virtuality ($Q^2$) of the jet parton which controls how much parton shower can be developed.
The equation describes the change of the total energy deposition due to the change of the virtuality of the parton (which leads to the change of parton shower).
Typically the higher virtuality that the initial parton has, the more shower partons can be developed, and the more energy will be deposited into the medium.
Similar equation can also been obtained for transverse momentum squared $\Delta p_\perp^2$ deposited into the medium by a virtual parton.
The scattering kernel $K_{j \to kl}(Q^2, q^-, y, \zeta)$ in the above equation is the same quantity that controls the energy loss the hard jets as shown in the previous sections.
Therefore, both energy loss and deposition profiles can be calculated self-consistently in the same formalism and they are controlled by the same jet transport coefficients.
Due to the radiated partons which may serve as additional sources depositing energy and momentum into medium, the length dependence of energy deposition rate is significantly enhanced compared to the case of a single parton.

\begin{figure}
\begin{center}
\includegraphics*[width=6.1cm]{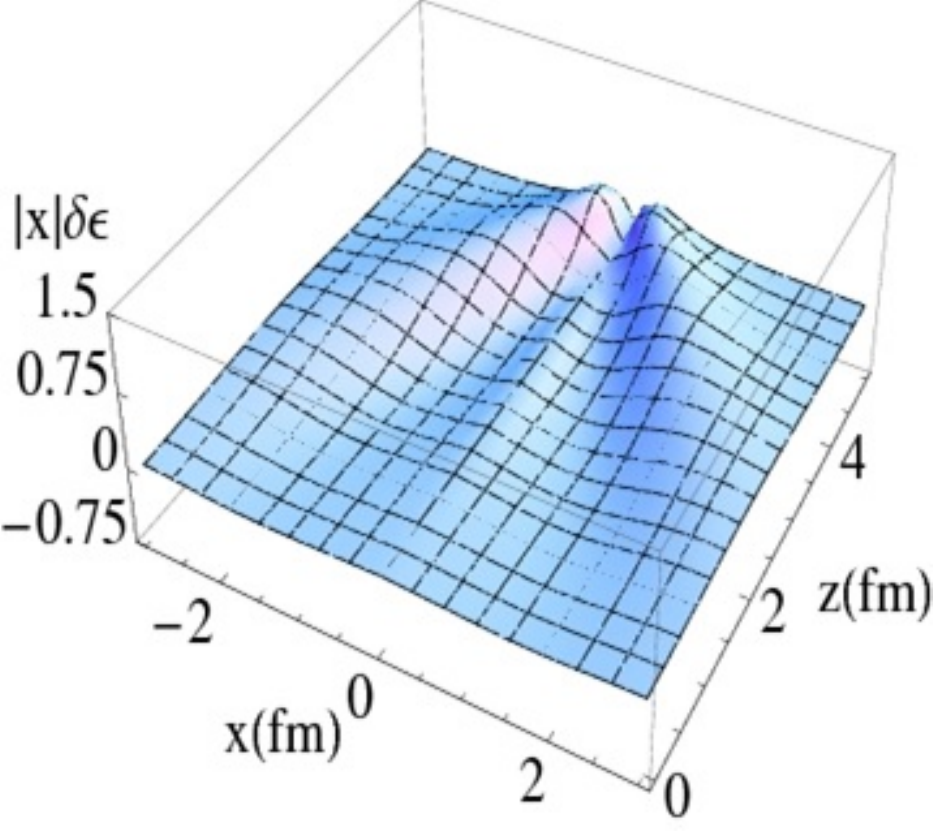}
\includegraphics*[width=6.3cm]{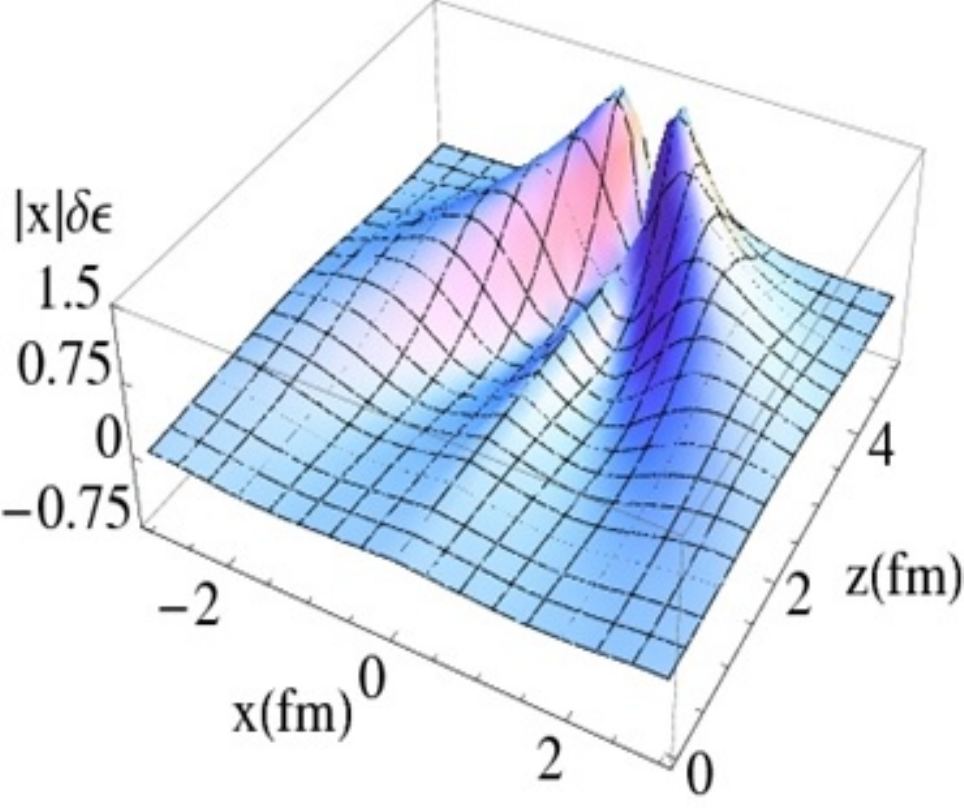}
\caption{(Color online) The fluid dynamical response to the energy deposited by a single quark (left) or by a quark-initiated parton shower (right).
Figures are taken from Ref. \cite{Qin:2009uh}.
}
\label{fig_Qin_mach_cone}
\end{center}
\end{figure}

\begin{figure}
\begin{center}
\includegraphics*[width=12.5cm]{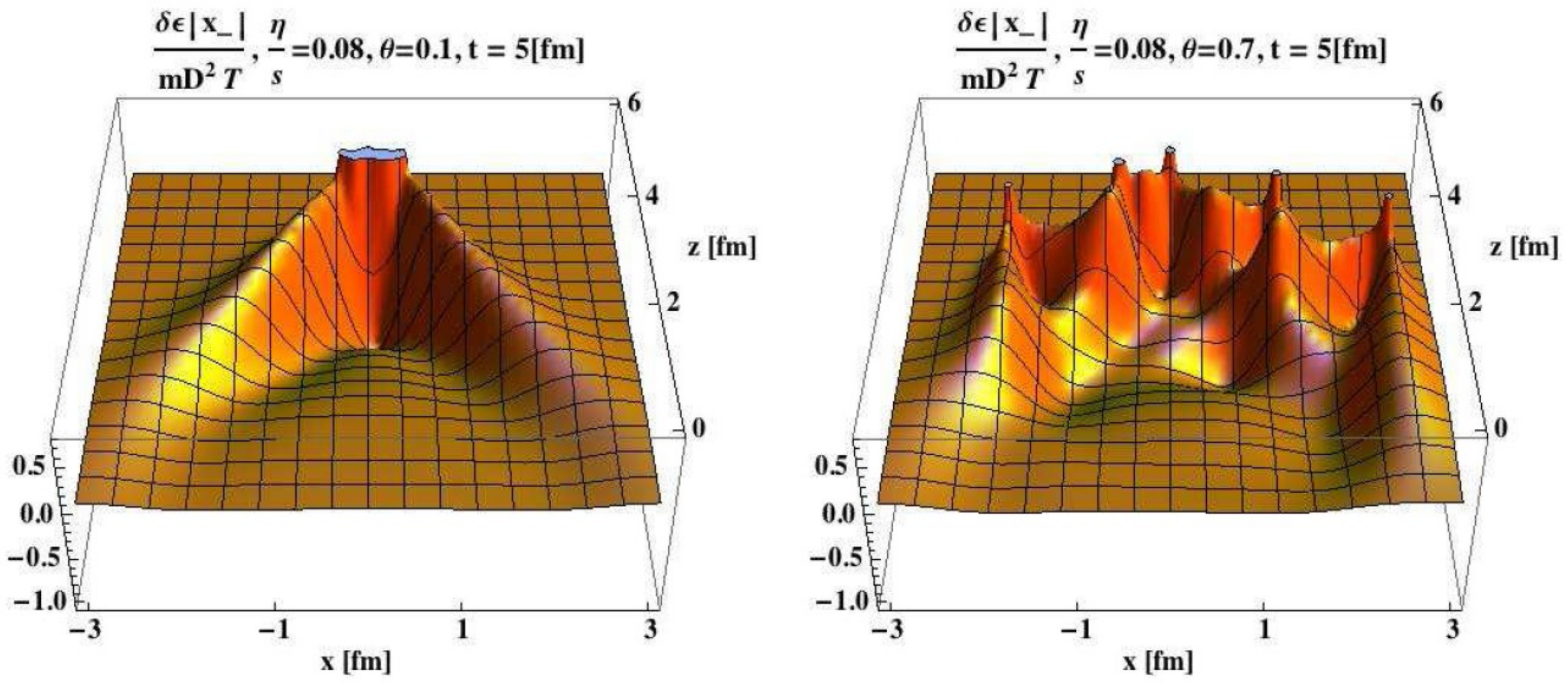}
\caption{(Color online) The energy density disturbance generated by a parton shower originating from a primary quark with average gluon emission angle $\theta = 0.1$ (left panel) and $\theta = 0.7$ (right panel).
Figures are taken from \cite{Neufeld:2011yh}.
}
\label{fig_Neufeld_mach_cone}
\end{center}
\end{figure}

The medium response to a parton shower with the above energy deposition profiles may be calculated as well via solving hydrodynamic equation with a source.
Fig. \ref{fig_Qin_mach_cone} shows the hydrodynamic response to the energy deposited by a single parton (left panel) and by a parton shower (right panel).
Note that in this calculation, the medium is assumed to be static and have constant temperature (density), and the deposited energy-momentum is assumed to equilibrate after a short relaxation time $\sim 1/m_D$.
One can clearly see that the Mach cone pattern is seen for both energy deposition cases, but is strongly enhanced for the virtual parton which develops into a parton shower (almost a fact of three).

However, in the above calculation of the energy deposition profiles from a parton shower, it is assumed that the radiative emissions are collinear, and the spatial distribution of the radiation is neglected.
The effect of spatial distribution of jet energy and momentum deposition profiles is explored in Ref. \cite{Neufeld:2011yh} where it is found that a wide space distribution, such as the case of large angle radiation, may destroy the nice cone-like structure of medium response, as opposed to the collinear radiation.
As an example, Fig. \ref{fig_Neufeld_mach_cone} compares the medium response to a parton shower developed from a primary quark with small average gluon emission angle $\theta = 0.1$ (left panel) and with large emission angle $\theta = 0.7$ (right panel).
One can see that for nearly collinear emissions, the parton shower can generate a nice Mach cone structure.
However, for the larger angle emissions, a well-defined Mach cone is hardly seen; the medium response is more like a superposition of several energy density perturbations.
This is due to the fact that each radiatied parton acts as a separate source which deposits energy and momentum into the medium.

\begin{figure}
\begin{center}
\includegraphics*[width=6.1cm]{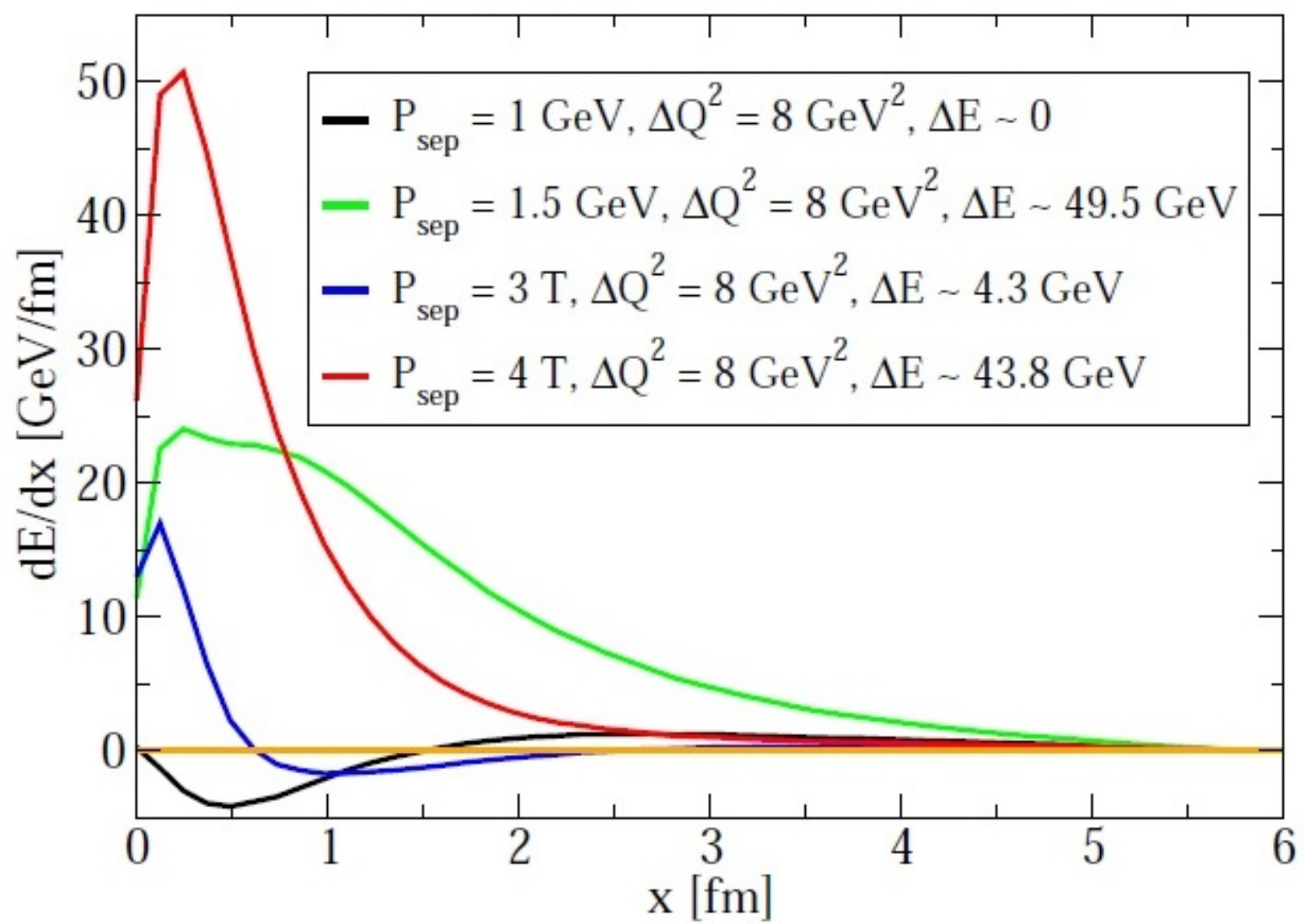}
\includegraphics*[width=6.3cm]{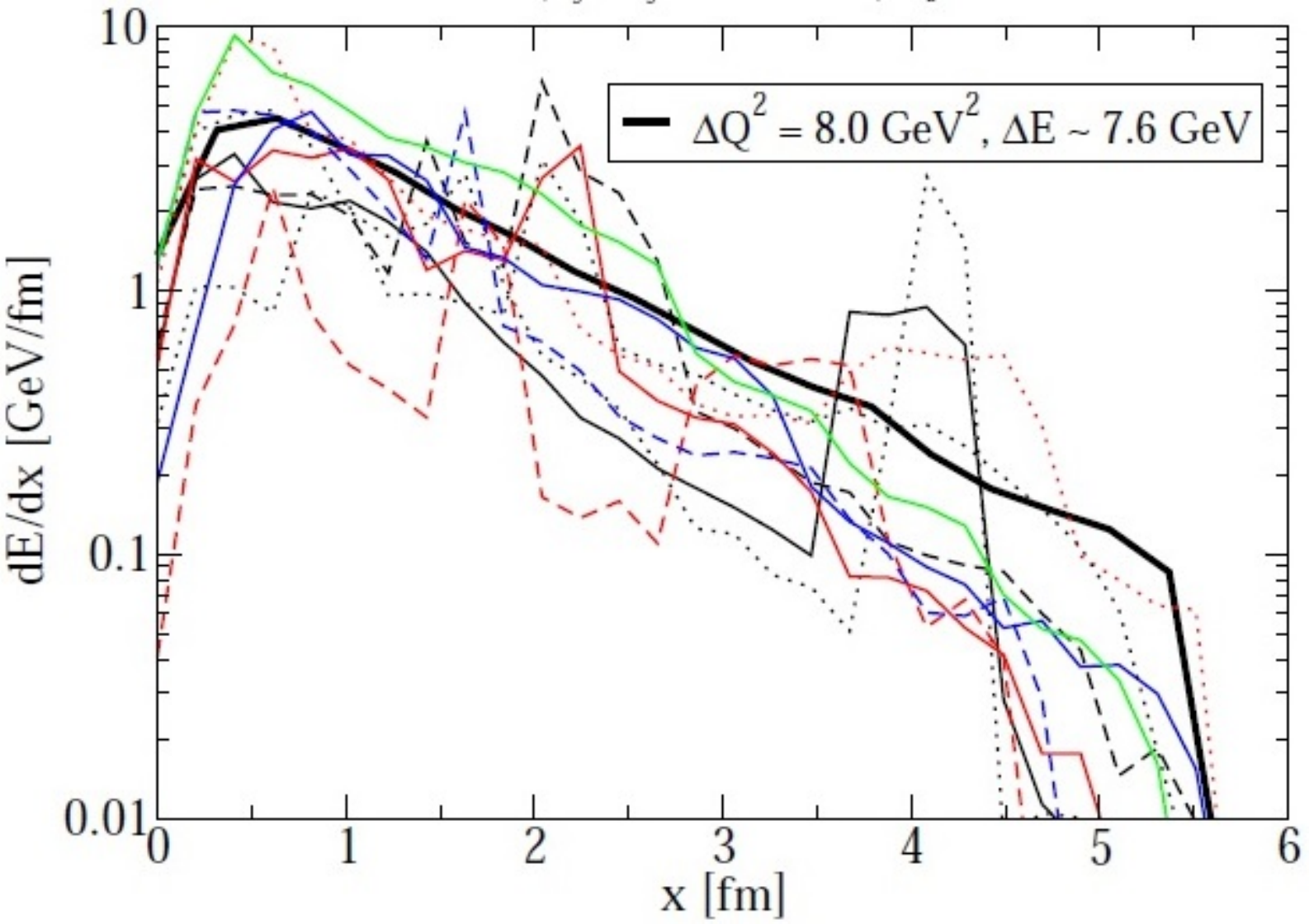}
\caption{(Color online) Left: Energy deposition of a parton shower initiated an 120 GeV gluon propagating through an evolving medium. Results for using different separation scales are compared.
Right: Energy deposition of a parton shower initiated by an 120 GeV gluon produced propagating through an evolving medium. Results for 10 typical individual events are shown.
Figures are taken from Ref. \cite{Renk:2013pua}.
}
\label{fig_Renk_energy_deposition}
\end{center}
\end{figure}

Besides the interplay between elastic collisions and inelastic radiations, and the space-time distributions for the radiation and deposition profiles, many other ingredients need to be taken into account when studying the energy/momentum deposition by the hard jets.
For example, a cutoff energy is often used in many model calculations to determine which part of radiation phase space is treated as thermalization into the medium \cite{Renk:2013pua,Floerchinger:2014yqa}.
Fig. \ref{fig_Renk_energy_deposition} (left panel) shows the effect of different choices of such energy cutoff.
One can see that the change of such separation scale from $T$ to $4T$ may lead to very different energy loss/deposition profiles for the propagating jet.
Another important fact is that the medium probed by the hard jet is dynamically evolving (expanding and cooling) in heavy-ion collisions, so the energy deposition rate is first increasing and then decreasing.
Also in realistic event-by-event simulation which includes the fluctuations of the initial states and jet energy loss, the energy deposition profiles will be very different from one event to another.
This is illustrated in Fig. \ref{fig_Renk_energy_deposition} (right panel) which compares the energy deposition profiles for $10$ typical individual events which are initiated from a $120$~GeV gluon.

\section{Summary}

In this chapter, we have provided a review on the recent progress in theoretical and phenomenological studies of jet quenching in ultra-relativistic heavy-ion collisions at RHIC and the LHC.
After a short introduction, we present the basic framework for studying jet production in elementary proton-proton collisions and the nuclear modification of jets in nucleus-nucleus collisions.
Different parton energy loss mechanisms and jet quenching formalisms are discussed in some detail.
The recent progress on the phenomenological studies of jet quenching in heavy-ion collisions at RHIC and the LHC is presented.
In particular, we have elaborated the effort on the quantitative extraction of jet transport coefficient $\hat{q}$, followed by lattice QCD calculation of $\hat{q}$ and the renormalization of $\hat{q}$.
The study of full jets at RHIC and the LHC is then presented, including energy loss  of reconstructed jets and the nuclear modification of the jet substructure.
In the last section, the progress and some issues about the medium response to jet transport are discussed.

During last years, significant progresses have been achieved in various directions in the studies of jet quenching in heavy-ion collisions, as has been presented in this report.
However, there still exist many complications and open questions.
For example, in order to narrow down the systematic uncertainties in leading-order jet quenching studies, a fully next-to-leading order framework for studying jet evolution and energy loss needs to be developed.
A full understanding of jet substructures and their nuclear modification requires more detailed and systematic investigation.
A general framework, with the incorporation of both realistic hydrodynamic models and realistic jet energy loss/deposition calculations, should be developed for simulating the transport of hard jets and the medium response simultaneously at the same time.
How exactly the energy and momentum deposited by hard jets thermalize into the medium is not well understood.
Future studies along these and other directions will further our understanding of jet-medium interaction, and help to achieve more accurate knowledge about the nature of the hot and dense QCD matter produced in ultra-relativistic nuclear collisions.

\section{Acknowledgments}

This work is supported in part by Natural Science Foundation of China under Grant Nos. 11375072, 11221504, and 11175232, Chinese Ministry of Science and Technology under Grant No. 2014DFG02050, by the Director, Office of Energy Research, Office of High Energy and Nuclear Physics, Division of Nuclear Physics, of the U.S. Department of Energy under Contract Nos. DE-AC02-05CH11231 and within the framework of the JET Collaboration.

\bibliography{GYQ_refs}

\printindex                         
\end{document}